\documentclass[11pt,a4paper]{article}

\usepackage[utf8x]{inputenc}

\usepackage{amsmath}
\usepackage{amsfonts}
\usepackage{amssymb}
\usepackage{amsthm}
\usepackage{mathptmx}
\usepackage{stmaryrd}
\usepackage[english]{babel}

\usepackage{subfiles}
\usepackage{hyperref}

\usepackage{authblk}

\usepackage{tikz}
\usepackage{mathdots}
\usepackage{yhmath}
\usepackage{cancel}
\usepackage{color}
\usepackage{siunitx}
\usepackage{array}
\usepackage{multirow}
\usepackage{gensymb}
\usepackage{tabularx}
\usepackage{booktabs}
\usetikzlibrary{fadings}

\usepackage{mdframed}

\theoremstyle{definition}
\newtheorem{definition}{Definition}%[section]
\newenvironment{defin}{\begin{mdframed}
[hidealllines=true,innerleftmargin=28pt,innerrightmargin=0pt,innertopmargin=-0.5\baselineskip, innerbottommargin=0.1\baselineskip, linewidth=0.7pt] 
\begin{definition}}{\end{definition}\end{mdframed}}

\newtheorem{mydef}{Notion}%[section]
\newenvironment{MyDef}{\begin{mdframed}[hidealllines=true,leftline=true,innerleftmargin=28pt,innerrightmargin=28pt,innertopmargin=-0.5\baselineskip, innerbottommargin=0.1\baselineskip, linewidth=0.7pt] 
\begin{mydef}}{\end{mydef}\end{mdframed}}

\newtheorem{myreq}{Axiom}%[section]
\newenvironment{MyReq}{\begin{mdframed}[hidealllines=true,leftline=true,innerleftmargin=28pt,innerrightmargin=28pt,innertopmargin=-0.5\baselineskip, innerbottommargin=0.1\baselineskip, linewidth=0.7pt] 
\begin{myreq}}{\end{myreq}\end{mdframed}}

\newtheorem{myconj}{Conjecture}%[section]
\newenvironment{MyConj}{\begin{mdframed}[hidealllines=true,innerleftmargin=28pt,innerrightmargin=28pt,innertopmargin=-0.5\baselineskip, innerbottommargin=0.1\baselineskip, linewidth=0.7pt] 
\begin{myconj}}{\end{myconj}\end{mdframed}}

\newtheorem{myaxiom}{A}%[section]

\newtheorem{theorem}{Theorem}%[section]
\newenvironment{theoreme}{\begin{mdframed}
[hidealllines=true,innerleftmargin=28pt,innerrightmargin=0pt,innertopmargin=-0.5\baselineskip, innerbottommargin=0.1\baselineskip, linewidth=0.7pt] 
\begin{theorem}}{\hfill $\blacksquare$ \end{theorem} \end{mdframed}}

\newtheorem{lemma}{\bf Lemma}%[section]
\newenvironment{lemme}{\begin{mdframed}
[hidealllines=true,innerleftmargin=28pt,innerrightmargin=0pt,innertopmargin=-0.5\baselineskip, innerbottommargin=0.1\baselineskip, linewidth=0.7pt] 
\begin{lemma}}{\hfill $\blacksquare$\end{lemma}\end{mdframed}} 

\newtheorem{corollary}{Corollary}%[theorem]
\newenvironment{corollaire}{\begin{mdframed}
[hidealllines=true,innerleftmargin=28pt,innerrightmargin=0pt,innertopmargin=-0.5\baselineskip, innerbottommargin=0.1\baselineskip, linewidth=0.7pt] 
\begin{corollary}}{\hfill $\blacksquare$\end{corollary}\end{mdframed}}

\theoremstyle{remark}
\newtheorem{remark}{Remark}%[section]

\newcommand{\sqcoversupset}{\raisebox{1.3ex}{\,\rotatebox{180}{$\sqsubseteq$}}\,}
\newcommand{\sqcoversubset}{\raisebox{1.3ex}{\,\rotatebox{180}{$\sqsupseteq$}}\,}

\newlength{\textlarg}

\newcommand{\antiwidehat}[1]{\settowidth{\textlarg}{$#1$} {\buildrel {\rotatebox{180}{$\widehat{\hspace{\textlarg \;}}$}} \over {#1}}}

\renewcommand{\textbf}[1]{\begingroup\bfseries\mathversion{bold}#1\endgroup}

\newcommand{\existunique}{\exists \textit{\bf !}}
\renewcommand{\footnote}[1]{\textsuperscript{\addtocounter{footnote}{1}$\lfloor$\thefootnote$\rfloor$}\footnotetext{#1}}
\makeatletter
\newcommand{\subjclass}[2][2020]{ \let\@oldtitle\@title\gdef\@title{\@oldtitle\footnotetext{#1 \emph{\href{https://zbmath.org/static/msc2020.pdf}{Mathematics subject classification}.} #2}}}
\newcommand{\keywords}[1]{\let\@@oldtitle\@title\gdef\@title{\@@oldtitle\footnotetext{\emph{Keywords:} #1.}}}
\makeatother

\DeclareFontFamily{U}{mathx}{\hyphenchar\font45}
\DeclareFontShape{U}{mathx}{m}{n}{
      <5> <6> <7> <8> <9> <10>
      <10.95> <12> <14.4> <17.28> <20.74> <24.88>
      mathx10
      }{}
\DeclareSymbolFont{mathx}{U}{mathx}{m}{n}
\DeclareFontSubstitution{U}{mathx}{m}{n}
\DeclareMathAccent{\widecheck}{0}{mathx}{"71}
\DeclareMathAccent{\wideparen}{0}{mathx}{"75}

\begin{document}

\title{Reconstructing quantum theory\\ from its possibilistic operational formalism}
\author{Eric Buffenoir\footnote{Email: \href{mailto:eric.buffenoir@cnrs.fr}{eric.buffenoir@cnrs.fr}}}
\affil{\href{http://univ-cotedazur.fr}{Universit\'e de la C\^ote d'Azur}, \href{http://www.cnrs.fr}{CNRS}, \href{https://inphyni.cnrs.fr/fr}{InPhyNi}, FRANCE}

\maketitle

\begin{abstract}
We develop a possibilistic semantic formalism for quantum phenomena from an operational perspective. This semantic system is based on a Chu duality between preparation processes and yes/no tests, the target space being a three-valued set equipped with an informational interpretation. \\
A basic set of axioms is introduced for the space of states. This basic set of axioms suffices to constrain the space of states to be a projective domain. The subset of pure states is then characterized within this domain structure.\\
After having specified the notions of properties and measurements, we explore the notion of compatibility between measurements and of minimally disturbing measurements. \\
We achieve the characterization of the domain structure on the space of states by requiring the existence of a scheme of discriminating yes/no tests,  necessary condition for the construction of an orthogonality relation on the space of states.  This last requirement about the space of states constrain the corresponding projective domain to be ortho-complemented. An orthogonality relation is then defined on the space of states and its properties are studied. Equipped with this relation, the ortho-poset of ortho-closed subsets of pure states inherits naturally a structure of Hilbert lattice.\\
Finally, the symmetries of the system are characterized as a general subclass of Chu morphisms. We prove that these Chu symmetries preserve the class of minimally disturbing measurements and the orthogonality relation between states. These symmetries lead naturally to the ortho-morphisms of Hilbert lattice defined on the set of ortho-closed subsets of pure states. 
\end{abstract}

\newpage

\section{Introduction}

The basic description of an 'experimental act' relies generally on (i) a description of a given {\em preparation setting} {{}}{that produces} 
{\em samples} of a given physical object, {{}}{through} some well-established procedures,  and (ii) a particular set of {\em operations/tests} that can be realized by the {\em observer} on these prepared samples. Each prepared sample is associated {{}}{with} a set of {\em information}, checked {{}}{throughout} the preparation process and recorded on devices, and to a set of {\em instructions}, processed by a computer or {{}}{monitored} 
by the human operator. These information/instructions are naturally attached to macroscopically observable events {{}}{that occur} during the preparation process and to macroscopically distinguishable modes of the preparation apparatuses used {{}}{throughout} this preparation process. Analogously, the operations/tests, realized on the prepared samples, must be attached unambiguously to some {\em knowledge/belief} of the observer about the outcomes of this experimental step and to a new {\em state of facts} concerning the outcomes of the whole experiment, in order for these operations/tests to give{{}}{ } sense 
to any 'sorting' of these outcomes. 
{{}}{If} these operations/tests do not 'destroy' or 'alter irremediably' the samples under study, we can then consider the whole experimental protocol\footnote{i.e.{{}}{,} the initial preparation followed by the operation/test step as a global preparation process for subsequent tests} as a completed preparation procedure for forthcoming experiments. The information characterizing the chosen operation, and the information retrieved by the observer during these tests, should then be recorded as new information/instructions {{}}{in} this {{}}{completed} preparation procedure.\footnote{We note that the description of {{}}{the} preparation/measurement process should then exploit some tools of recursion theory.} This description of preparation procedures and operations/tests is a fundamental ingredient of the physical description. It is a {\it sine qua non condition} for the experimental protocol to satisfy a basic requirement{{}}{:} {\em the reproducibility}.   
As noted by Kraus \cite{Kraus}, there exist macroscopic devices undergoing macroscopic changes when interacting with micro-systems, and the observation of micro-objects always requires this inter-mediation. This fundamental empirical fact justifies the attempt to establish such an {\em operational description} for quantum experiments as well. {{}}{Explicitly}, the different procedures designed to prepare collection of similar quantum micro-systems may combine measurements and filtering operations (associated {{}}{with} the unambiguously measured properties) in order to produce collections of samples, that may be {{}}{subjected to subsequent} measurements.\\ 

Obviously, different preparation procedures may be used to produce distinct samples, to which the observer would nevertheless attach the same {\em informational content}. {{}}{This} is the case, {{}}{in particular},
if this observer {{}}{does} not know {{}}{of} any experiments{{}}{ }that could be realized conjointly on these differently prepared samples{{}}{ and that} would produce 'unambiguously-incompatible' logical conclusions. A physical description (of the objects {{}}{subjected}
to experiment) is an attempt to establish {\em a semantic perspective} adapted to previous description{{}}{s} of the process of preparation/measurement. \footnote{Here, and in the following, we will adopt the following basic definition of the word 'semantic' recalled by Reichenbach: "Modern logic distinguishes between object language and metalanguage; the first speaks about physical objects, the second about statements, which in turn are referred to objects. The first part of the meta-language, syntax, concerns only statements, without dealing with physical objects; this part formulates the structure of statements. The second part of the metalanguage, semantics, refers to both statements and physical objects. This part formulates, in particular, the rules concerning truth and meaning of statements, since these rules include a reference to physical objects. The third part of the meta-language, pragmatics, includes a reference to persons who use the object language."\cite{Reichenbach}} The notion of {\em physical state} occupies a central position in this semantic construction. The physical states are abstract names for the different possible realizations of the object under study. Adopting this ontological perspective, the observer may associate an element of the {\em space of states} {{}}{with} any preparation process, which will {\it a priori} characterize any particular sample 'prepared' {{}}{through} this process. The logical truth of a proposition about the 'similarity' of two given samples should then be directly linked to the logical truth of the proposition regarding the identity of the associated states. However, the ontological notion of state has to be faced with its epistemological counterpart. From an ontological perspective, we consider that a given physical system is necessarily in a particular realization, but from an epistemological perspective, the observer should test the hypothesis that this system may indeed be described by this state.\footnote{Note that, {{}}{in practice}, the observer is rather led to {{}}{infer} a 'mixture' {{}}{on the basis of} his limited knowledge about this sample.} Adopting this perspective, it seems that a given physical state could also be meant as a {\em denotation} for the set of preparations{{}}{ }that the observer is led to identify empirically. From a strictly operational point of view, the observer will always establish the equivalence between different preparation procedures, by testing conjointly the corresponding prepared samples through a 'well-chosen' collection of control{{}}{ }tests. In other words, the state {{}}{may} be defined by the set of common facts{{}}{ }that could be established by realizing these control-tests on the corresponding samples. Adopting another perspective, we could also consider that the state should encode the determined aspects relative to the possible results of forthcoming experiments. There should be no problem with such a 'versatile' perspective in the operational description of classical experiments, as {{}}{long} as 
the properties, established as 'actual' during the preparation process, characterize the sample in a way that will be questioned neither by any control{{}}{ }test realized on it, nor by any future non-destructive experiments{{}}{.}
The operational description of quantum experiments is in fact significantly more intricate, due to some {{}}{inconvenient} features of the measurement operation. Indeed, it is a fundamental fact of quantum experiments that, whichever set of properties has been checked as 'actual' by the preparation process, the outcomes of an irreducible part of the measurements{{}}{ }that can be made on these prepared samples, remain completely indeterminate. More than that, if some of these measurements are realized in order to establish some new properties, it generically occurs that the measured samples {{}}{no longer exhibit}, afterwards, some of the properties that had been {{}}{previously} established{{}}{ }on the prepared samples. \\ 

Despite this indeterministic character of quantum theory, it is an empirical fact that the distinct outcomes of these measurements, operated on a large collection of samples, prepared according to the same experimental procedure, have reproducible relative frequencies. This fundamental fact has led physicists to consider{{}}{ } large collections of statistically independent experimental sequences {{}}{as the basic objects of}  {{}}{physical description,} rather than a single experiment on a singular realization of the object under study (see \cite{Peres1993} for a reference book). 
According to {\em Generalized Probabilistic Theory} {{}}{(GPT)}\footnote{This formalism has its origins in the pioneering works of Mackey \cite{MacKey63}, Ludwig \cite{Ludwig1981,ludwig_foundations_1983,Ludwig88} and Kraus \cite{Kraus}. See \cite{Janotta_2014} for a recent review.}, a physical state (corresponding to a class of operationally equivalent preparation procedures) is defined by a vector of probabilities associated {{}}{with} the outcomes of a maximal and irredundant set of fiducial tests that can be effectuated on collections of samples produced by any of these preparation procedures.\footnote{The description of quantum theory in this framework {{}}{then must} deal with the problem {{}}{of defining} the notions of consistency, completeness and irredundancy {{}}{for} the set of control{{}}{ }tests {{}}{that define} an element of the quantum space of states.} In other words, two distinct collections of prepared samples will be considered as operationally equivalent if they lead to the same probabilities for the outcomes of any test on them. The physical description consists{{}}{, therefore,} in a set of prescriptions {{}}{that allows} sophisticated constructs {{}}{to be defined} from elementary ones. In particular, combination rules are defined for the concrete mixtures of states and for the allowed operations/tests. The different attempts to reconstruct quantum mechanics along this path (\cite{Hardy:2001jk,Hardy:2011dm,Hardy:2013faa}\cite{PhysRevA.84.012311}\cite{Mueller:2012ai}\cite{PhysRevA.75.032304}) proceed by the determination of a minimal set of plausible constraints, imposed on the space of states, sufficient to 'derive' the usual Von {{}}{Neumann axiomatic quantum} theory. Although this probabilistic approach is now accepted as a standard conceptual framework for the reconstruction of quantum theory, the adopted perspective appears puzzling for different reasons.
First of all, the observer contributes fundamentally to give an intuitive meaning to the notions of preparation, operation and measurement on physical systems. However, the concrete process of 'acquisition of information' (by the observer / on the system) has no real place in this description. Secondly, the definition of the state has definitively lost its meaning for a singular prepared sample, {{}}{and} the physical state is now intrinsically attached to large collections of similarly prepared samples. This point has concentrated many critics since the original article of Einstein, Podolsky and Rosen \cite{Einstein}, although the empirical testing of quantum theory in EPR experiments has led physicists to definitively accept the traditional probabilistic interpretation. The {{}}{GPT} approach adopts{{}}{ }the probabilistic description of quantum phenomena {{}}{without any discussion or attempt} to explain why {{}}{it is necessary}. Thirdly, in order to clarify the requirements {{}}{of} the basic set of fiducial tests{{}}{} necessary and sufficient to define the space of states, this approach must proceed along a technical analysis which {{}}{fundamentally limits} this description to 'finite dimensional' systems (finite dimensional Hilbert {{}}{spaces} of states). {{}}{Lastly}, the axioms chosen to characterize quantum theory, among other theories encompassed by {{}}{the GPT} formalism, must exhibit {{}}{a} 'naturality' that{{}}{ - from our point of view - }is still missing in the existing proposals.\\

Alternative research programs have tried to overcome some of these conceptual problems. In particular, they try to put the emphasis on the 'informational' relation emerging between the observer and the system, through the concrete set of 'yes/no tests' that can be addressed by this observer to this system, and to characterize quantum theory in these terms through a small number of basic semantic requirements. It must be noted that, although these programs try to clarify the central notion of "information" in the quantum description an observer can {{}}{develop} about the system under study, they basically adopt the 'probabilistic' interpretation {{}}{of} measurements. The fundamental limitation of the information{{}}{ }that an observer can retrieve from a given quantum system through yes-no experiments, has been taken by different physicists as a central principle for the reconstruction of quantum theory (see{{}}{ }reference paper \cite[Chap III]{Rovelli1996}, and see \cite{Zeilinger1999} for another perspective on this basic principle) :
\begin{quote}
{\bf Information Principle} {\it 'Information is a discrete quantity: there is a minimum amount of information exchangeable (a single bit, or the information that distinguishes between just two alternatives).[...] Since information is discrete, any process of acquisition of information can be decomposed into acquisitions of elementary bits of information.'} \cite[p.1655]{Rovelli1996}. 
\end{quote}
We {{}}{intend} to adopt our own version of this fundamental pre-requisite to build the quantum space of states.\\ 
To be more explicit, according to C.Rovelli, quantum mechanics appears to be governed by two seemingly incompatible principles (Postulate 1 and 2 of \cite[Chap III]{Rovelli1996}). {{}}{According to the first},
the amount of independent {{}}{information} that can be retrieved from a 'bounded' quantum system is fundamentally 'finite'. {{}}{According to the second, however,}
the test of any observable property{{}}{ } that has not been stated {{}}{beforehand} as 'actual' {{}}{for} a given state\footnote{In Von Neumann's formalism, this point means: the quantum state of the system is not an eigenstate of the associated operator.}, remains fundamentally {{}}{indeterminate} (of course, this test will establish an actual value for this observable property, valid after the measurement).  Nevertheless, this 'new information' established through this measurement operation will have been {{}}{compensated} by the restoration of an indeterminacy {{}}{in} some of the properties that {{}}{may} have been established as actual beforehand. Another interesting analysis regarding this 'balance' principle 
concerning the knowledge of the observer about the quantum system is given in \cite{Spekkens}. It must be noted that the combinatorics of the 'incompatibilities' between different measurements can be exploited to explore the algebraic structures behind quantum theory and to proceed to the 'reconstruction' of this theory \cite{Hohn2017toolbox,PhysRevA.95.012102}. Nevertheless, these reconstruction programs stay technically imprisoned by a finite dimensional analysis. A third postulate in Rovelli's axiomatic proposal prescribes the nature of automorphisms acting on the {{}}{state space} (the continuous unitary transformations corresponding, in particular, to Schrodinger's dynamics). However, the form of this last postulate is not {{}}{entirely} satisfactory{{}}{, as} it imposes some intricate relations on the probabilities associated {{}}{with} transitions between states. \\ 

Adopting another perspective, the {\em operational quantum logic} {{}}{approach} tries to {{}}{avoid} the introduction of probabilities and explores the relevant categorical structures underlying the space of states and the set of properties of a quantum system. In this description, probabilities appear only as a derived concept.\footnote{For the sake of usual quantum formalism, quantum probabilities can be recovered from this formalism, under the assumption of the existence of some well-behaved measurements{{}}{,} using Gleason's theorem \cite{Gleason1975} (see \cite{Piron1972} for {{}}{a} historical and pedagogical presentation of these elements).} 
Following G. Birkhoff and J. Von Neumann \cite{Birkhoff1936} and G. W. Mackey \cite{MacKey63}, this approach focuses on the structured space of 'testable properties' of a physical system.\footnote{In Von Neumann's quantum mechanics, {{}}{each entity is associated with} a complex Hilbert space $H$. A state $\psi$ of this entity is defined by a ray  $\nu(\psi)$ in $H$, and an observable is defined by a self-adjoint operator on $H$. In particular, a yes/no test ${ \mathfrak{t}}$ is represented by an orthogonal projector $\Pi_{ \mathfrak{t}}$ or equivalently by the closed subspace ${ \mathfrak{A}}_{ \mathfrak{t}}$ defined as the range of $\Pi_{ \mathfrak{t}}$. The answer “yes” {{}}{or "no"} is obtained with certainty for the yes/no test ${ \mathfrak{t}}$, if and only if the state $\psi$ is such that $\nu(\psi)$ is included in ${ \mathfrak{A}}_{ \mathfrak{t}}$ {{}}{or in the orthogonal of ${ \mathfrak{A}}_{ \mathfrak{t}}$, respectively}. 
Birkhoff and von Neumann proposed to focus not on the structure of the Hilbert space itself, but on the structure of the set of closed subspaces of $H$. The mathematical structure associated {{}}{with} the set of quantum propositions defined by the closed subspaces of $H$ is not a Boolean algebra (contrary to the case encountered in classical mechanics). By shifting the attention to the set of closed subspaces of $H$ instead of $H$, the possibility is open to build an operational approach to quantum mechanics, because the basic elements of this description are yes/no experiments.} Mackey identifies axioms on the set of yes/no questions sufficient to relate this set to the set of closed subspaces of a complex Hilbert space. Later, C. Piron proposed a set of axioms that (almost) lead back to the general framework of quantum mechanics (see \cite{Coecke2000} for {{}}{a} historical perspective of the abundant literature inherited from {{}}{Piron's original} works  \cite{Piron1972,Piron1976}.). Piron’s framework has been developed into a full operational approach and the categories underlying this approach were analyzed (see \cite{MOORE199961,moore1995categories} for a detailed account of this categorical perspective). It must be {{}}{noted} that these constructions are established in reference to some general results of projective geometry \cite{faure_projective_2000} and are not {{}}{restricted to}
a finite{{}}{-}dimensional perspective. Despite some beautiful results (in particular the restriction of the division ring associated {{}}{with} Piron's reconstruction of the Hilbert space from Piron's propositional lattices \cite{Holland:447786}\cite{aerts2000}) and the attractiveness of a completely categorical approach (see \cite{STUBBE2007477} for an analysis of the main results on propositional systems), {{}}{this approach has encountered many problems}. Among these problems, we may cite the {{}}{difficulty of building} a consistent description of compound systems due to no-go results {{}}{related} to the existence of a tensor product of Piron's propositional systems \cite{randall1979tensor,Foulis1981}\cite{aerts1984construction,Aerts2004}. These works have cast doubts on the adequacy of Piron's choice of an "orthomodular complete lattice" structure for the set of properties of the system. D. Foulis, C. Piron and C. Randall \cite{Foulis1983} {{}}{ } produced a {{}}{jointly} refined version of their respective {{}}{approaches} in order to rule out these problems (see also \cite{Wilce2000}). This description emphasizes the centrality of the treatment of the incompatibilities between measurements.\footnote{{\it 'It is our contention that the realistic view implicit in classical physics need not be abandoned to accommodate the contemporary conceptions of quantum physics. All that must be abandoned is the presumption that each set of experiments possesses a common refinement (that is, the experiments are compatible). As we shall argue, this in no way excludes the notion of physical systems existing exterior to an observer, nor does it imply that the properties of such systems depend on the knowledge of the observer.'} \cite[p.813]{Foulis1983}} Our work will emphasize the necessity {{}}{of replacing} the lattice structures, introduced to describe the set of propositions about the quantum system, {{}}{with} domains. 
Another central problem with the logico-algebraic approach was its inability to describe the dynamic aspect of the measurement operation. The operational quantum logic {{}}{approach} has then been developed later on in different categorical perspectives in order to clarify the links between quantum logic, {{}}{ } and modal/dynamic{{}}{ } \cite{Baltag2005,baltag_smets_2006,Baltag2008}\cite{CoeckeMooreSmets,Coecke2001,Coecke2004}\cite{Zhong2018,Bergfeld2015}, or linear {{}}{ } \cite{CoeckeSmets2000}\cite{GIRARD19871,GirardProofsTypes} {{}}{logic}.\\ 

Other categorical formalisms, adapted to the axiomatic study of quantum theory, have been developed more recently \cite{ABRAMSKY2009261} and their relation with the 'operational approach' {{}}{has} been partly explored \cite{Abramsky2012,Abramsky2013, abramsky_heunen_2016}. In \cite[Theorem 3.15]{Abramsky2012}, S. Abramsky makes explicit the fact that the {\em Projective quantum symmetry groupoid} $PSymmH$\footnote{The objects of this category are the natural space of states in quantum mechanics, i.e., the Hilbert spaces of dimension greater than two, and the morphisms are the orbits under the $U(1)$ group action %($U\sim V \Leftrightarrow \exists \lambda \in U(1), U=\lambda V$) 
on semi-unitary maps (i.e. unitary or anti-unitary), which are the relevant symmetries of Hilbert spaces from the point of view of {{}}{quantum mechanics}.} is fully and faithfully represented by the category $bmChu_{[0,1]}$, i.e., by the sub-category of the category of bi-extensional Chu spaces associated {{}}{with} the evaluation set $[0,1]$ obtained by restricting {{}}{it} to Chu morphisms $(f_\ast,f^\ast)$ for which $f_\ast$ is injective. 
This result {{}}{suggests that Chu categories could have a} 
central role in the construction of axiomatic quantum mechanics\footnote{The centrality of the notion of Chu {{}}{spaces} for quantum foundations had been noted already by V. Pratt \cite{PrattQuantum}{{}}{.}}, as{{}}{ }they provide a natural characterization of the automorphisms of the theory.  More surprisingly, and interestingly for us, in \cite{Abramsky2012} S. Abramsky shows that the previously mentioned representation of $PSymmH$ is {{}}{'already'} full and {{}}{faithful} if we replace the evaluation space of the Chu category by a three{{}}{-}element set, where the three values represent "definitely yes", "definitely no" and "maybe" \cite[Theorem 4.4]{Abramsky2012} : 
\begin{quote}
{\it 'The results on reduction to finite value sets are also intriguing. Not only is the bare Chu condition on morphisms sufficient to whittle them down to the {{}}{semiunitaries}, this is even the case when the discriminations on which the condition is based are reduced to three values. The general case for two values remains open, but we have shown that the two standard possibilistic reductions both \emph{fail to preserve fullness}. A negative answer for two-valued semantics in general would suggest an unexpected r\^ole for three-valued logic in the foundations of Quantum Mechanics.'}
\end{quote}

The introduction of a three-valued logic in the foundational studies of quantum mechanics goes back to H. Reichenbach's work \cite{Reichenbach}. H. Reichenbach's formalism has been introduced as an alternative to the quantum logic approaches developed to continue the seminal work of G. Birkhoff and J. von Neumann \cite{Birkhoff1936}.\footnote{In order to clarify the fundamental difference in nature between Reichenbach Quantum Logic and Mainstream Quantum Logic the reader is invited to consult \cite{Hardegree}.} This extension of standard logic was designed as a means of dealing with certain conceptual tensions arising in the quantum mechanical description of the world. Briefly, the third truth value, called 'indeterminate', was introduced to capture 'meaningless' statements associated with unmeasured entities during the experimental process.  It is important to note that H. Reichenbach did not clarify how this three-valued logic may lead to results 'comparable' to those obtained using the usual probabilistic formalism, and why this extension of classical logic could be the right setting for quantum theory. \\ 
S. Abramsky's work adopts a radically different perspective.  His study begins with continuous-valued Chu space. He shows that quantum symmetries are elegantly captured by Chu morphisms and proves finally that three values are actually sufficient for this characterization of quantum symmetries.  Despite the suggestive similarities pointed to by S. Abramsky between such a three-valued Chu space description and the Geneva school's formalism, no bridge has been built between these two formalisms until now.  Thus, S. Abramsky did not  affirm that these three values are sufficient to found a complete axiomatic quantum theory, close to Piron's program or alternative to it, and allowing a complete reconstruction of the usual Hilbert formalism.  It is the purpose of the present paper to achieve this goal. We intend to present the basic elements of this 'possibilistic'\footnote{In the rest of this paper we refer to this construction, based on a three-valued Chu space, as a 'possibilistic' approach to distinguish it from the 'probabilistic' one.} semantic formalism, and to give the precise axiomatics that leads to a reconstruction of a generalized Hilbert space structure on the space of states. \\ 

Our formalism is based on a Chu duality between preparation processes and quantum tests. This Chu duality refers to a three-valued target space. This three-valued target space is equipped with a 'possibilist' semantic {{}}{formalism} which leads to an 'informational' interpretation {{}}{of} the set of preparations. 
In the first part of our study, we formulate a precise semantic description of the space of states. The 'Information Principle' introduced by C. Rovelli plays a central role in this formalism. After having introduced a basic set of axioms about the space of states, it will be shown that the space of states is a projective domain.  The space of pure states will then be characterized.\\
In a second part of the study, we clarify successively the notion of 'property' and the notion of {{}}{a} 'measurement' associated {{}}{with} a given property of the system. We explore the consequences of the incompatibilities existing between measurements. \\
An orthogonality relation is then defined on the space of states and its properties are studied using the domain structure obtained on the space of states.The axiomatics on the space of states is achieved by the addition of some conditions relative to the existence of the orthogonality relation. 
Equipped with this relation, the ortho-poset of ortho-closed subsets of pure states inherits a structure of Hilbert lattice. This result is the first part of our reconstruction theorem.\\
In the third part of this paper, we build the set of symmetries of the system as a particular sub-algebra of Chu morphisms. These symmetries appear to {{}}{leave} this subset of minimally disturbing measurement operations {{}}{stable} and preserves the orthogonality relation between states.  Endly, it is shown that these symmetries lead naturally to the ortho-morphisms of Hilbert lattice defined on the set of ortho-closed subsets of pure states.\\ 

Throughout the paper we clearly distinguish (1) elements formulated in the 'material mode of speech' and concerning the structure of the language in which the experimental setting can be described operationally (these elements are designated as 'Notions' {{}}{throughout} the text), although these notions are the occasion to introduce the corresponding mathematical elements, from (2) purely mathematical definitions (these elements are classically designated as 'Definitions').\\
We intentionally emphasize the different requirements of our reconstruction program. Every requirement is introduced accompanied with an analysis of its motivation and summarized under the term 'Axiom'; the analysis of the mathematical consequences of each axiom is declined along Lemmas and Theorems. 

\section{Preparations and states}
\label{sectionpreparationsandstates}
\subsection{Operational formalism}

Adopting the operational perspective on quantum experiments, we will introduce the following definitions :
\begin{MyDef}
A {\em preparation process} is an objectively defined, and {{}}{thus}
'repeatable', experimental sequence {{}}{that allows} singular samples of a certain physical system {{}}{to be produced}, in such a way that we are able to submit them to tests. 
We will denote by ${ \mathfrak{P}}$ the set of preparation processes (each element of ${ \mathfrak{P}}$ can be equivalently considered as the collection of samples produced {{}}{through} this preparation procedure). The {{}}{information} corresponding to macroscopic events/operations {{}}{describing} the procedure depend on an observer $O$. 
If this dependence has to be made explicit, we will adopt the notation ${ \mathfrak{P}}^{{}^{(O)}}$ to denote the set of preparation processes defined by the observer $O$.
\end{MyDef}
\begin{MyDef}
For each  {\em property}, that the observer {{}}{aims}
to test macroscopically on {\em any particular sample} of the considered micro-system, it will be assumed that the observer is able to define (i) some detailed 'procedure', in reference to the modes of use of some experimental apparatuses chosen to {{}}{perform} the operation/test, and (ii) a 'rule' allowing {{}}{ }the answer 'yes' {{}}{to be extracted} if the macroscopic outcome of the experiment {{}}{conforms} with the expectation of the observer, when the test is performed on any input sample (as soon as this experimental procedure can be opportunely applied to this particular sample). 
These operations/tests, {{}}{designed} to determine the occurrence of a given property for a given sample, will be called {\em yes/no tests} {\em associated {{}}{with} this property} ({{}}{also called a} {\em definite experimental project} in \cite{Piron1977}). 
If a yes/no test, associated {{}}{with} a given property, is effectuated according to the established procedure, and if {{}}{a} positive result is actually obtained for a given sample, we will say that {\em this property has been measured} for this sample. The set of 'yes/no tests' at the disposal of the observer will be denoted by ${ \mathfrak{T}}$. If the dependence with respect to the observer $O$ has to be made explicit, we will adopt the notation ${ \mathfrak{T}}^{{}^{(O)}}$ to denote the set of tests defined by the observer $O$.
\end{MyDef}
We {{}}{are essentially} interested {{}}{in the information} gathered by the observer through the implementation of some yes/no tests, designated by elements in ${ \mathfrak{T}}$, on finite collections of samples prepared similarly {{}}{through}
any of the preparation procedures, given as elements of ${ \mathfrak{P}}$. {{}}{With} this perspective, we have to abandon any reference to the probabilistic interpretation\footnote{{{}}{The} finite character of the tested collection of prepared samples {{}}{renders} any notion of relative frequency of the outcomes 'meaningless'{{}}{.}}. Nevertheless, the observer is still able to distinguish the situations where {{}}{one} can pronounce a statement with 'certainty', from the situations where {{}}{one} can judge the result as '{{}}{indeterminate}', on the basis of the knowledge {{}}{ }gathered beforehand. 
\begin{MyDef}
A yes/no test ${ \mathfrak{t}}\in { \mathfrak{T}}$ will be said to be {\em positive with certainty} (resp. {\em negative with certainty}) relatively to a preparation process ${ \mathfrak{p}}\in { \mathfrak{P}}$ iff the observer is led to affirm that the result of this test, realized on any of the particular samples that could be prepared according to this preparation process, would be 'positive with certainty' (resp. would be 'negative with certainty'), 'should' this test be effectuated. If the yes/no test can not be stated as 'certain', this yes/no test will be said to be {\em {{}}{indeterminate}}. Concretely, the observer can establish {{}}{the} 'certainty' {{}}{of} the result of a given yes/no test on any given sample issued from a given preparation procedure, by running the same test on a sufficiently large (but finite) collection of samples issued from this same preparation {{}}{process:} if the outcome is always the same, {{}}{the observer} will be led to claim that similarly prepared 'new' samples would also produce the same result, if the experiment was effectuated.\\ 
To summarize, for any preparation process ${ \mathfrak{p}}$ and any yes/no test ${ \mathfrak{t}}$, the element ${ \mathfrak{e}}({ \mathfrak{p}},{ \mathfrak{t}})\in { \mathfrak{B} } := \{{ \bot}, { \rm \bf Y}, { \rm \bf N}\}$ will be defined to be ${ \bot}$ ({{}}{alternatively,} ${ \rm \bf Y}$ {{}}{or} ${ \rm \bf N}$) if the outcome of the yes/no test ${ \mathfrak{t}}$ on any sample prepared according to the preparation procedure ${ \mathfrak{p}}$ is judged as '{{}}{indeterminate}' ('positive with certainty' {{}}{or} 'negative with certainty'{{}}{, respectively}) by the observer. 
\begin{eqnarray}
\begin{array}{rcrcl}
{ \mathfrak{e}} & : &{ \mathfrak{P}} \times { \mathfrak{T}} & \longrightarrow & { \mathfrak{B} } :=\{{ \bot}, { \rm \bf Y}, { \rm \bf N}\} \\
& &({ \mathfrak{p}},{ \mathfrak{t}}) & \mapsto &{ \mathfrak{e}}({ \mathfrak{p}},{ \mathfrak{t}}).
\end{array}
\end{eqnarray}
\end{MyDef} 

\noindent Several remarks {{}}{should} be {{}}{made regarding the above} definitions.\\ 
\begin{remark} It is essential to note the {\em {{}}{counterfactual}} aspect of these {{}}{definitions:} in the 'determinate' case, the observer is {{}}{asked} to predict the result of {{}}{this} test {\em before {{}}{the} test} and {\em {{}}{regardless of whether the} test {{}}{is} effectuated}. 
Of course, any 'determinate statement' (positive or negative) produced by the observer, about the result of any forthcoming yes/no test {{}}{relative} to a given preparation process, is a strictly falsifiable {{}}{statement:} it may be proved to be false after some test realized on a {\em finite collection} of new{{}}{,} similarly prepared {{}}{samples}\footnote{{{}}{If} the observer is certain of the positive result after having {{}}{performed}
a given yes/no test on a finite number of similarly prepared samples, a negative result obtained for any newly tested sample will lead the observer to revise {{}}{that} prediction and to consider this yes/no test as being  '{{}}{indeterminate}' for this preparation{{}}{.}}.
\end{remark}
\begin{remark}
The 'certainty' of the observer about the occurrence of the considered 'property' is intrinsically attached to {\em any singular sample} prepared {{}}{through} this preparation process and can be falsified as a property of this sample. In other words, it is not necessary to consider {\em a statistical ensemble} of similarly prepared samples to give a meaning to these notions and to the logical perspective adopted to confront these statements {{}}{with} the measurable state of facts. 
\end{remark} 
\begin{remark} 
When the determinacy of a yes/no test is established for an observer, we can consider that this observer {{}}{possesses} some elementary 
'information' about the state of the system, {{}}{whereas}, in the '{{}}{indeterminate} case', {{}}{the observer} has none (relatively to the occurrence of the considered property). 
\end{remark}

\begin{MyDef} 
The set ${ \mathfrak{B} }$ will be equipped with the following poset structure, characterizing the 'information' gathered by the observer: 
\begin{eqnarray}
\forall u,v\in { \mathfrak{B} },&& (u\leq v)\; :\Leftrightarrow\; (u={ \bot}\;\;\textit{\rm or}\;\; u=v).
\end{eqnarray} 
$({ \mathfrak{B} },\leq)$ will be called {{}}{a} {\em flat boolean domain}  in the rest of this paper.\\ 
\end{MyDef} 

\begin{MyDef} 
$({ \mathfrak{B} },\leq)$ is equipped with the following involution map :
\begin{eqnarray}
\overline{{ \bot}}:={ \bot} \;\;\;\;\;\;\;\; \overline{ \rm \bf Y}:={ \rm \bf N}\;\;\;\;\;\;\;\; \overline{ \rm \bf N}:={ \rm \bf Y}.
\end{eqnarray} 
The conjugate of a yes/no test ${{ \mathfrak{t}}}\in { \mathfrak{T}}$ is the yes/no test denoted $\overline{{ \mathfrak{t}}}$ and defined from ${{ \mathfrak{t}}}$ by exchanging the roles of {\bf Y} and {\bf N} in every {{}}{result} obtained by applying ${{ \mathfrak{t}}}$ to any given input sample. In other words,
 \begin{eqnarray}
\forall  { \mathfrak{t}}\in { \mathfrak{T}}, \forall  { \mathfrak{p}}\in { \mathfrak{P}},&& { \mathfrak{e}}({ \mathfrak{p}},\overline{{ \mathfrak{t}}}):=\overline{ { \mathfrak{e}}({ \mathfrak{p}},{{ \mathfrak{t}}}) }.\label{etbar}
\end{eqnarray} 
\end{MyDef} 

\begin{MyDef} 
For any yes/no test ${ \mathfrak{t}}$, the set of preparation processes ${ \mathfrak{p}}$ for which this test is established as {\em actual}, i.e.{{}}{,} 'positive with certainty', will be denoted ${ \mathfrak{A}}_{ \mathfrak{t}}$.
\begin{eqnarray}
\forall { \mathfrak{t}}\in { \mathfrak{T}}, && { \mathfrak{A}}_{ \mathfrak{t}} := \{\, { \mathfrak{p}} \in { \mathfrak{P}}\;\vert\; { \mathfrak{e}}({ \mathfrak{p}},{ \mathfrak{t}})={\rm \bf Y}\,\}.
\end{eqnarray} 
\end{MyDef}

\begin{MyDef} 
For a given yes/no test ${ \mathfrak{t}}$, we define the subset ${ \mathfrak{Q}}_{ \mathfrak{t}}$ of preparation processes that are known by the observer to produce collections of samples leading {{}}{to} positive results to the yes/no test ${ \mathfrak{t}}$. {{}}{Regarding}
these prepared samples, the observer is then {{}}{asked}
to pronounce a statement about any future result of this test on similarly prepared new {{}}{samples:} 'positive' or '{{}}{indeterminate}'. The collections of samples, {{}}{resulting} from these preparation processes, may then be filtered in order to select collections of samples that are known by the observer to have 'passed the yes/no test ${ \mathfrak{t}}$ {{}}{positively}'. 
If a preparation process ${ \mathfrak{p}}$ is in ${ \mathfrak{Q}}_{ \mathfrak{t}}$, we will say that the property associated {{}}{with} the yes/no test ${ \mathfrak{t}}$ is {\em potential} for the samples produced {{}}{through} ${ \mathfrak{p}}$ (or ${ \mathfrak{p}}$ is {\em questionable} by ${ \mathfrak{t}}$). The subset ${ \mathfrak{Q}}_{ \mathfrak{t}}$ is given by
\begin{eqnarray}
{ \mathfrak{Q}}_{ \mathfrak{t}}:= \{\, { \mathfrak{p}} \in { \mathfrak{P}}\;\vert\;  { \mathfrak{e}}({ \mathfrak{p}},{ \mathfrak{t}}) \leq \textit{\rm \bf Y}\,\} \subseteq { \mathfrak{P}}.
\end{eqnarray}
\end{MyDef} 

The evaluation map ${ \mathfrak{e}}$ defines a particular 'duality' between the spaces ${ \mathfrak{P}}$ and ${ \mathfrak{T}}$. Formally, $({ \mathfrak{P}}, { \mathfrak{T}}, { \mathfrak{e}})$ defines a Chu {{}}{space}\footnote{See \cite{Barr1991AutonomousCA} for a reference {{}}{paper} and \cite{Pratt1999}\cite{PRATT1999319} for a basic presentation{{}}{.} }\footnote{In this section, we are concerned with the duality aspect and the situation of Chu morphisms will be treated later{{}}{.}}. 
The set of preparations ${ \mathfrak{P}}$ ({{}}{or} the set of yes/no tests ${ \mathfrak{T}}$) will be {\em a priori} interpreted as the set of {\em points} {{}}{(the} set of {\em opens}) of this Chu space.\footnote{These designations are reminiscent of the basic fact that Chu spaces are generalizations of topological spaces. However this distinction is largely obsolete, as soon as the Chu space construction establishes a duality between these two sets.} Indeed, the preparation processes are naturally considered as 'coexisting entities' distinguished by the properties they {{}}{posses}, whereas the yes/no tests are naturally interpreted as 'alternative predicates' relative to the properties attached to {{}}{the} prepared samples.\\

According to the perspective adopted by \cite{Peres1993}, we will define the {\em states} of the physical system as follows:
\begin{MyDef} \label{defstate}
An equivalence relation, denoted $\sim_{{}_{ \mathfrak{P}}}$, is defined on the set of preparations ${ \mathfrak{P}}$:\\ 
\indent Two preparation processes are identified iff the statements established by the observer about the corresponding prepared samples are identical.\\ 
\indent A {\em state} of the physical system is an equivalence class of preparation processes corresponding to the same informational content, i.e., a class of preparation processes that are not distinguished by the statements established by the observer in reference to the tests{{}}{ }realized on finite collections of samples produced {{}}{through} these preparation processes.\\ 
The set of equivalence classes, modulo the relation $\sim_{{}_{ \mathfrak{P}}}$, will be denoted ${ { \mathfrak{S}}}$. In other words,
\begin{eqnarray}
\forall { \mathfrak{p}}_1,{ \mathfrak{p}}_2\in { \mathfrak{P}}, \;\; ({ \mathfrak{p}}_1\sim_{{}_{ \mathfrak{P}}} { \mathfrak{p}}_2)\;  & :\Leftrightarrow & 
(\; \forall { \mathfrak{t}}\in { \mathfrak{T}},\; { \mathfrak{e}}({ \mathfrak{p}}_1,{ \mathfrak{t}})= { \mathfrak{e}}({ \mathfrak{p}}_2,{ \mathfrak{t}}) \;),\label{equivrelpreparations}\\
&& \sim_{{}_{ \mathfrak{P}}} \;\;\textit{\rm is an equivalence relation},\\
\lceil {{ \mathfrak{p}}} \rceil & := & \{\, { \mathfrak{p}}'\in { \mathfrak{P}}\;\vert\; { \mathfrak{p}}'\sim_{{}_{ \mathfrak{P}}} { \mathfrak{p}}\,\},\\
{ { \mathfrak{S}}} & := & \{\, \lceil {{ \mathfrak{p}}} \rceil \;\vert\; { \mathfrak{p}}\in { \mathfrak{P}}\,\}.
\end{eqnarray}
\end{MyDef} 
\begin{remark}
It must be noticed that a given yes/no test ${ \mathfrak{t}}$ can be applied separately on the two distinct collections of samples prepared{{}}{ through} the two distinct preparation procedures ${ \mathfrak{p}}_1$ and ${ \mathfrak{p}}_2$. The corresponding {{}}{counterfactual} statements ${ \mathfrak{e}}({ \mathfrak{p}}_1,{ \mathfrak{t}})$ and ${ \mathfrak{e}}({ \mathfrak{p}}_2,{ \mathfrak{t}})$, established by the observer about ${ \mathfrak{p}}_1$ and ${ \mathfrak{p}}_2$, are then formulated 'consistently' after these two independent experimental sequences.
\end{remark}

We will derive a map $\widetilde{ \mathfrak{e}}$ from the evaluation map ${ \mathfrak{e}}$  according to the following definition :
\begin{eqnarray}
\begin{array}{rcrcl}
\widetilde{ \mathfrak{e}} & : &{ \mathfrak{T}} & \rightarrow & { \mathfrak{B} }{}^ { \mathfrak{S}} \\
& &{ \mathfrak{t}} & \mapsto & \widetilde{ \mathfrak{e}}_{ \mathfrak{t}} \;\;\;\;\vert\;\;\;\;\; \widetilde{ \mathfrak{e}}_{ \mathfrak{t}}(\lceil { \mathfrak{p}} \rceil):={ \mathfrak{e}}({ \mathfrak{p}},{ \mathfrak{t}}),\; \forall { \mathfrak{p}}\in { \mathfrak{P}} .\label{defetilde}
\end{array}
\end{eqnarray}

As a result of this {{}}{quotient} operation on the space of preparation processes, it appears that we have the following natural property of our Chu space.
\begin{lemme}{}\label{chuseparated}
The Chu space $({ { \mathfrak{S}}}, { \mathfrak{T}}, \widetilde{ \mathfrak{e}})$ is {\em separated}. Different preparation procedures are indeed 'identified' by the observer as soon as this observer attributes the same statements to the differently prepared samples. In other words,
\begin{eqnarray}
\forall \sigma_1,\sigma_2\in { { \mathfrak{S}}},&&
(\,\forall { \mathfrak{t}}\in { \mathfrak{T}},\;\; \widetilde{ \mathfrak{e}}_{ \mathfrak{t}}(\sigma_1)=\widetilde{ \mathfrak{e}}_{ \mathfrak{t}}(\sigma_2)\,)\;\;\Rightarrow\;\; (\,\sigma_1=\sigma_2\,).
\end{eqnarray}
\end{lemme}

\subsection{First axioms for the space of states}

A pre-order relation can be defined on the set ${ \mathfrak{P}}$ of preparation processes. 
\begin{MyDef} 
A preparation process ${ \mathfrak{p}}_2\in { \mathfrak{P}}$ is said to be {\em sharper} than another preparation process ${ \mathfrak{p}}_1\in { \mathfrak{P}}$ (this fact will be denoted ${ \mathfrak{p}}_1\sqsubseteq_{{}_{ \mathfrak{P}}} { \mathfrak{p}}_2$) iff any yes/no test ${ \mathfrak{t}}\in { \mathfrak{T}}$ that is 'determinate' for the samples prepared {{}}{through} ${ \mathfrak{p}}_1$ is also necessarily 'determinate' with the same value for the samples prepared {{}}{through} ${ \mathfrak{p}}_2$, i.e.,
\begin{eqnarray}
 && \forall { \mathfrak{p}}_1, { \mathfrak{p}}_2\in { \mathfrak{P}},\;\; (\; { \mathfrak{p}}_1\sqsubseteq_{{}_{ \mathfrak{P}}} { \mathfrak{p}}_2\;) \;\; :\Leftrightarrow \;\; 
(\; \forall { \mathfrak{t}}\in { \mathfrak{T}},\; { \mathfrak{e}}({ \mathfrak{p}}_1,{ \mathfrak{t}})\leq { \mathfrak{e}}({ \mathfrak{p}}_2,{ \mathfrak{t}}) \;),\label{orderP}
\end{eqnarray}
If ${ \mathfrak{p}}_1\sqsubseteq_{{}_{ \mathfrak{P}}} { \mathfrak{p}}_2$ (i.e., ${ \mathfrak{p}}_2$ is 'sharper' than ${ \mathfrak{p}}_1$), ${ \mathfrak{p}}_1$ is said to be 'coarser' than ${ \mathfrak{p}}_2$.
\end{MyDef} 
\begin{lemme}{} $({ \mathfrak{P}}, \sqsubseteq_{{}_{ \mathfrak{P}}})$ is a pre-ordered set.\end{lemme}
The reflexivity and transitivity properties of the binary relation $\sqsubseteq_{{}_{ \mathfrak{P}}}\; \in { \mathfrak{P}} \times { \mathfrak{P}}$ are trivial to check. \\

The equivalence relation defined in {\bf Notion \ref{defstate}} derives from this pre-order:
\begin{eqnarray}  
\forall { \mathfrak{p}}_1, { \mathfrak{p}}_2\in { \mathfrak{P}},&& (\,{ \mathfrak{p}}_1\sqsubseteq_{{}_{ \mathfrak{P}}} { \mathfrak{p}}_2 \;\;\textit{\rm and}\;\; { \mathfrak{p}}_1\sqsupseteq_{{}_{ \mathfrak{P}}} { \mathfrak{p}}_2\;)\; \Rightarrow\; (\, { \mathfrak{p}}_1 \sim_{{}_{ \mathfrak{P}}} { \mathfrak{p}}_2\,).
\end{eqnarray}

We will define an operation of mixtures on the set of preparations.
\begin{MyReq}\label{reqcap}
If we consider a collection of preparation processes $P\subseteq { \mathfrak{P}}$, we can define a new preparation procedure, called {\em mixture} and denoted $\bigsqcap{}_{{}_{ \mathfrak{P}}} P$, as follows: \\
\indent the samples produced from the preparation procedure $\bigsqcap{}_{{}_{ \mathfrak{P}}} P$ are obtained by a random mixing of the samples issued from the preparation processes of the collection $P$ {{}}{indiscriminately}. \\ 
As a consequence, the statements that the observer can establish after a sequence of tests ${ \mathfrak{t}}\in { \mathfrak{T}}$ on these samples produced {{}}{through} the procedure $\bigsqcap{}_{{}_{ \mathfrak{P}}} P$ is given as the infimum of the statements that the observer can establish for the elements of $P$  separately. In other words, 
\begin{eqnarray}
&& \forall P\subseteq { \mathfrak{P}},\;\; \existunique\; \bigsqcap{}_{{}_{ \mathfrak{P}}} P\in { \mathfrak{P}} \;\;\;\vert\;\;\;(\,\forall { \mathfrak{t}}\in { \mathfrak{T}},\; { \mathfrak{e}}(\bigsqcap{}_{{}_{ \mathfrak{P}}} P,{ \mathfrak{t}})=\bigwedge{}_{{}_{{ \mathfrak{p}}\in P}}{ \mathfrak{e}}({ \mathfrak{p}},{ \mathfrak{t}}),\;\;\;\;\;\;\;\;\;\label{propconjunctiveprop}
\end{eqnarray} 
where $\wedge$ denotes the infimum of a collection of elements in the poset $({ \mathfrak{B} },\leq)$. We will adopt the following notation ${ \mathfrak{p}}_1\sqcap_{{}_{ \mathfrak{P}}} { \mathfrak{p}}_2 = \bigsqcap_{{}_{ \mathfrak{P}}} \{{ \mathfrak{p}}_1,{ \mathfrak{p}}_2\}$. 
\end{MyReq}

We note the following obvious properties deduced from the literal definitions of the random mixing operation $\sqcap_{{}_{ \mathfrak{P}}}$ and the equivalence relation $\sim_{{}_{ \mathfrak{P}}}$ defining the space of states.
\begin{lemme}{}
For any ${ \mathfrak{p}}_1, { \mathfrak{p}}_2,{ \mathfrak{p}}_3 \in { \mathfrak{P}}$, we have
\begin{eqnarray}  
\;\;\;\;\;&& { \mathfrak{p}}_1\sqcap_{{}_{ \mathfrak{P}}} ({ \mathfrak{p}}_2\sqcap_{{}_{ \mathfrak{P}}} { \mathfrak{p}}_3) \sim_{{}_{ \mathfrak{P}}} ({ \mathfrak{p}}_1\sqcap_{{}_{ \mathfrak{P}}} { \mathfrak{p}}_2)\sqcap_{{}_{ \mathfrak{P}}} { \mathfrak{p}}_3\label{capaxiom1}\\
&&({ \mathfrak{p}}_1\sqcap_{{}_{ \mathfrak{P}}} { \mathfrak{p}}_1)\sim_{{}_{ \mathfrak{P}}} { \mathfrak{p}}_1,\label{capaxiom2}\\
&&({ \mathfrak{p}}_2\sqcap_{{}_{ \mathfrak{P}}} { \mathfrak{p}}_1) \sim_{{}_{ \mathfrak{P}}} ({ \mathfrak{p}}_1\sqcap_{{}_{ \mathfrak{P}}} { \mathfrak{p}}_2)\label{capaxiom3}
\end{eqnarray}
\end{lemme}

The properties of the equivalence relation $\sim_{{}_{ \mathfrak{P}}}$ with respect to the pre-order $\sqsubseteq_{{}_{ \mathfrak{P}}} $ and the random-mixing binary operation $\sqcap_{{}_{ \mathfrak{P}}}$ leads to the following properties: 
\begin{lemme}{}
The binary operation $\sqcap_{{}_{ \mathfrak{P}}}$  being {{}}{literally} designed to satisfy{{}}{ }properties (\ref{capaxiom1}){{}}{, }(\ref{capaxiom2}) {{}}{and} (\ref{capaxiom3}), the binary relation $\sqsubseteq_{{}_{ \mathfrak{P}}}$ is then equivalently defined by:
\begin{eqnarray}  
\forall { \mathfrak{p}}_1, { \mathfrak{p}}_2\in { \mathfrak{P}},&& ({ \mathfrak{p}}_1\sqsubseteq_{{}_{ \mathfrak{P}}} { \mathfrak{p}}_2)\; \Leftrightarrow\; (\, { \mathfrak{p}}_1 \sim_{{}_{ \mathfrak{P}}} ({ \mathfrak{p}}_1 \sqcap_{{}_{ \mathfrak{P}}} { \mathfrak{p}}_2)\,).\label{lemmabasicsuqsubseteqcapsim}
\end{eqnarray}
The following properties of the pre-order $\sqsubseteq_{{}_{ \mathfrak{P}}}$ are direct consequences of this fact:
\begin{eqnarray}  
\forall  { \mathfrak{p}},{ \mathfrak{p}}_1, { \mathfrak{p}}_2\in { \mathfrak{P}}, &&
({ \mathfrak{p}}_1\sqcap_{{}_{ \mathfrak{P}}} { \mathfrak{p}}_2)\sqsubseteq_{{}_{ \mathfrak{P}}} { \mathfrak{p}}_1,\\
&& ({ \mathfrak{p}} \sqsubseteq_{{}_{ \mathfrak{P}}} { \mathfrak{p}}_1\;\;\textit{\rm and}\;\;{ \mathfrak{p}} \sqsubseteq_{{}_{ \mathfrak{P}}} { \mathfrak{p}}_2)\;\Rightarrow\; (\,{ \mathfrak{p}} \sqsubseteq_{{}_{ \mathfrak{P}}}({ \mathfrak{p}}_1\sqcap_{{}_{ \mathfrak{P}}} { \mathfrak{p}}_2)\,).
\end{eqnarray}
\end{lemme}

\begin{lemme}{}
The space of states ${ { \mathfrak{S}}}$ is partially ordered
\begin{eqnarray}
\forall \sigma_1,\sigma_2\in { { \mathfrak{S}}},&& (\, \sigma_1 \sqsubseteq_{{}_{ { \mathfrak{S}}}} \sigma_2\,) \;\;:\Leftrightarrow\;\; (\, \forall { \mathfrak{p}}_1,{ \mathfrak{p}}_2 \in  { \mathfrak{P}}, \;\; 
(\, \sigma_1=\lceil {{ \mathfrak{p}}_1}\rceil , \sigma_2=\lceil {{ \mathfrak{p}}_2}\rceil \,)\;\;\Rightarrow\;\; (\,{ \mathfrak{p}}_1\sqsubseteq_{{}_{ \mathfrak{P}}} { \mathfrak{p}}_2\,) \,),\;\;\;\;\;\;\;\;\;\;\;\;\\
&&({ { \mathfrak{S}}},\sqsubseteq_{{}_{ { \mathfrak{S}}}}) \;\;\textit{\rm  is a partial order}.\label{axiomsigmapartialorder}
\end{eqnarray}
\end{lemme}
Moreover, the existence of the 'mixed' preparations satisfying{{}}{ }property (\ref{propconjunctiveprop}) leads to the following definition.
\begin{MyDef}
\begin{eqnarray}
\forall P\subseteq { \mathfrak{P}},&&\bigsqcap{}_{{}_{{ \mathfrak{p}}\in P}} \lceil  { \mathfrak{p}}\rceil :=\lceil \bigsqcap{}_{{}_{ \mathfrak{P}}}P \rceil.
\end{eqnarray}
\end{MyDef}

\noindent Quite naturally, we will assume the existence and uniqueness of a bottom element in ${ \mathfrak{P}}$:
\begin{MyReq}\label{reqbot}
 {{}}{There} exists a unique preparation process, that can be interpreted as a 'randomly-selected' collection of 'un-prepared samples'. This element leads to {{}}{ }complete {{}}{indeterminacy} for any yes/no test realized on it. In other words,  the following axiom will be imposed 
\begin{eqnarray} 
&& \existunique \;{ \mathfrak{p}}_{{ \bot}}\in { \mathfrak{P}}\;\;\vert \;\; (\,\forall { \mathfrak{t}}\in { \mathfrak{T}},\;\; { \mathfrak{e}}({ \mathfrak{p}}_{{ \bot}},{ \mathfrak{t}})={ \bot}\,).
\end{eqnarray} 
 \end{MyReq}

\begin{lemme}{}
$({ { \mathfrak{S}}},\sqsubseteq_{{}_{ { \mathfrak{S}}}})$ admits a bottom element{{}}{,} denoted ${ \bot}_{{}_{ { \mathfrak{S}}}}$:
\begin{eqnarray}
{ \bot}_{{}_{ { \mathfrak{S}}}}:= \lceil  {{ \mathfrak{p}}_{ \bot}}\rceil && \textit{\rm is the bottom element of $({ { \mathfrak{S}}},\sqsubseteq_{{}_{ { \mathfrak{S}}}})$}.\label{axiomsigmabottom}
\end{eqnarray}
\end{lemme}

\begin{MyDef}
A collection of preparation processes $P\subseteq { \mathfrak{P}}$ will be said to be {\em consistent} (this fact will be denoted $\widehat{P}^{{}^{ \mathfrak{P}}}$) iff the different elements  can be considered as different incomplete preparations of the same targeted collection of prepared samples, i.e., iff {{}}{there} exists a preparation process ${ \mathfrak{p}}\in { \mathfrak{P}}$ which is simultaneously sharper than any ${ \mathfrak{p}}'$ in $P$ (i.e., {{}}{which} is a common upper-bound in ${ \mathfrak{P}}$). In other words,
\begin{eqnarray}
\forall P\subseteq { \mathfrak{P}},\;\; \widehat{\;P\;}^{{}_{ \mathfrak{P}}} & :\Leftrightarrow &(\, \exists { \mathfrak{p}}\in { \mathfrak{P}} \;\vert\;  { \mathfrak{p}}' \sqsubseteq_{{}_ { \mathfrak{P}}}  { \mathfrak{p}}, \; \forall { \mathfrak{p}}' \in P \,).
\end{eqnarray}
We will also denote $\widehat{{ \mathfrak{p}}_1 { \mathfrak{p}}_2 }^{{}_{ \mathfrak{P}}}:=\widehat{\{{ \mathfrak{p}}_1, { \mathfrak{p}}_2\} }^{{}_{ \mathfrak{P}}}$
This consistency relation is obviously reflexive and symmetric. \\
Due to the following relation $\forall { \mathfrak{p}}_1,{ \mathfrak{p}}_2,{ \mathfrak{p}}_3\in { \mathfrak{P}},\;\; (\,\widehat{{ \mathfrak{p}}_1{ \mathfrak{p}}_2}^{{}_{ \mathfrak{P}}}\;\textit{\rm and}\; { \mathfrak{p}}_2 \sim{}_{{}_{ \mathfrak{P}}} { \mathfrak{p}}_3\,)\;\Rightarrow\; \widehat{{ \mathfrak{p}}_1{ \mathfrak{p}}_3}^{{}_{ \mathfrak{P}}}$, the consistency relation can be defined between states as follows.
%\begin{MyDef}
\begin{eqnarray}
\forall \sigma_1,\sigma_2\in { \mathfrak{S}},\;\;\widehat{\sigma_1\sigma_2}^{{}_{ \mathfrak{S}}} & :\Leftrightarrow & (\, \forall { \mathfrak{p}}_1,{ \mathfrak{p}}_2\in { \mathfrak{P}}, \;\; \sigma_1=\lceil{ \mathfrak{p}}_1\rceil,\sigma_2=\lceil{ \mathfrak{p}}_2\rceil \;\;\Rightarrow\;\; \widehat{{ \mathfrak{p}}_1{ \mathfrak{p}}_2}^{{}_{ \mathfrak{P}}}\,).
\end{eqnarray} 
For any $\sigma_1,\sigma_2\in { \mathfrak{S}}$, we will denote $\antiwidehat{\sigma_1\;\sigma_2}{}^{{}^{ \mathfrak{S}}}$ the property $\neg \,\widehat{\;\sigma_1\sigma_2\;}^{{}_{ \mathfrak{S}}}$.
\end{MyDef} 

\begin{theoreme}\label{theoremeboundedcomplete}
$({ { \mathfrak{S}}},\sqcap_{{}_{ { \mathfrak{S}}}})$  is bounded complete.  In other words,
\begin{eqnarray}
\forall S\subseteq { \mathfrak{S}}\;\vert\; \widehat{S}{}^{{}^{ \mathfrak{S}}},&& (\bigsqcup{}_{{}_{ \mathfrak{S}}} S) \; \textit{\rm exists in ${ \mathfrak{S}}$.}
\end{eqnarray}
\end{theoreme}

The previous construction of the space of states, albeit usual, appears a bit strange from an operational perspective. Indeed, concretely, the observer is never confronted {{}}{with} a given 'state' (i.e., to a generically-infinite class of preparation processes, indistinguishable by the generically-infinite set of tests that can be realized on them) in order to decide if it is consistent (or not) to affirm with 'certainty' the occurrence of a given 'property' for a given sample corresponding to this state. The observer is rather confronted {{}}{with} a restricted set of preparation processes, {{}}{enabling} mixtures {{}}{to be produced}, which generically lead to undetermined results when {{}}{they are} confronted {{}}{with} a family of tests. \\

In order to produce 'determinacy', relative to the occurrence of a given property for a given state of the system, the observer extracts ({{}}{from} the {{}}{selected} family of preparation processes {{}}{available, that} are {{}}{detected} to produce samples corresponding {{}}{more or less} to the chosen state) some sub-families {{}}{that concretely realize} {\em a sharpening} of the parameters {{}}{that define} the preparation setting/procedure, according to a given set of {{}}{prerequisites} concerning the samples that will be submitted to the test. Through each 'sharpening' of its preparation procedures, the observer {{}}{intends} to fix 'unambiguously', but 'inductively', a 'state' of the system. This limit process is understood in terms of the limit taken for every {{}}{statement} that can be {{}}{made} about the selected samples (i.e., the samples prepared according to any of the preparation processes{{}}{ that} are elements of the chosen sharpening family). 
\begin{MyDef} \label{defsharpening1}
A family ${ \mathfrak{Q}}\subseteq { \mathfrak{P}}$ is {\em a sharpening family of preparation processes} {{}}{(denoted} 
${ \mathfrak{Q}}\subseteq_{Chain} { { \mathfrak{P}}}$) iff every {{}}{pair} of elements of ${ \mathfrak{Q}}$ are ordered by $\sqsubseteq_{{}_{ \mathfrak{P}}}$, i.e., for any ${ \mathfrak{p}}_1$ and ${ \mathfrak{p}}_2$ in ${ \mathfrak{Q}}$, we have necessarily ${ \mathfrak{p}}_1\sqsubseteq_{{}_{ \mathfrak{P}}} { \mathfrak{p}}_2$ or ${ \mathfrak{p}}_2\sqsubseteq_{{}_{ \mathfrak{P}}} { \mathfrak{p}}_1$.
\end{MyDef} 
\begin{MyReq} \label{reqsharpening1}
For any family ${ \mathfrak{Q}}$ in ${ \mathfrak{P}}$ defining a 'sharpening', {{}}{there} exists a state $\sigma$ in ${ { \mathfrak{S}}}$ which is the supremum of the chain of states corresponding to the elements of ${ \mathfrak{Q}}$.  
\begin{eqnarray}
 \forall { \mathfrak{Q}}\subseteq_{Chain} { { \mathfrak{P}}},&& \textit{\rm the supremum} \;\; (\bigsqcup{}_{{}_{ { \mathfrak{S}}}}  \lceil { \mathfrak{Q}} \rceil )\;\;\textit{\rm exists in the partially ordered set} \;\;{ { \mathfrak{S}}} % \nonumber \\ && (\, \forall \sigma\in  { \mathfrak{C}}, \sigma \sqsubseteq_{{}_{ { \mathfrak{S}}}}  (\bigsqcup  { \mathfrak{C}})\,) \;\;\textit{\rm and}\;\; (\, \forall \sigma'\in { { \mathfrak{S}}},\,  (\,\forall \sigma \in { \mathfrak{C}},\, \sigma \sqsubseteq_{{}_{ { \mathfrak{S}}}}  \sigma'\,)\Rightarrow  (\, (\bigsqcup  { \mathfrak{C}}) \sqsubseteq_{{}_{ { \mathfrak{S}}}}  \sigma'\,)\,).\;\;\;\;\;\;\;\;\;\;
\label{axiomsigmachaincomplete}
\end{eqnarray}
In other words, ${ \mathfrak{S}}$ will be required to be {\em a chain-complete partial order}. 
\end{MyReq}

\begin{remark}
${ \mathfrak{S}}$ is then also {\em a directed-complete partial order}. 
\end{remark}

Let us then fix a yes/no test ${ \mathfrak{t}}$. If the observer {{}}{intends} to designate the corresponding property  as an 'element of reality' attached to the system itself, and not as a datum depending on the explicit operational requirements used to define the state, the following condition must be satisfied.
\begin{MyReq}\label{reqsharpening2}
The observer is authorized to formulate a 'determinate' statement, about the occurrence of a given property, for the 'limit state' induced from a given sharpening family of preparation processes, iff {{}}{it is possible} to formulate this same statement for {{}}{another} preparation process {{}}{that is an} element of the chosen sharpening family (and {{}}{thus} also for any {{}}{sharper} preparation process{{}}{)}. In other words,  
\begin{eqnarray}
\forall { \mathfrak{t}}\in { \mathfrak{T}},\;\forall { \mathfrak{C}}\subseteq_{Chain} { { \mathfrak{S}}},&& (\,  (\bigsqcup{}_{{}_{ { \mathfrak{S}}}} { \mathfrak{C}})\in \lceil { \mathfrak{A}}_{{ \mathfrak{t}}} \rceil \,)\;\;\Rightarrow\;\; (\, \exists \sigma \in { \mathfrak{C}}\;\vert\; \sigma \in \lceil { \mathfrak{A}}_{{ \mathfrak{t}}} \rceil \,).\label{continuityrequirement}
\end{eqnarray}
The set of states $\lceil { \mathfrak{A}}_{{ \mathfrak{t}}} \rceil$, for which the property tested by ${ \mathfrak{t}}$ is actual, must be a '{{}}{Scott-open} filter' in the {{}}{directed-complete partial order} $({ \mathfrak{S}},  \sqsubseteq_{{}_{ { \mathfrak{S}}}})$.
\end{MyReq}

We can reformulate {{}}{the above} requirement in terms of a continuity property of the evaluation map with respect to the sharpening process. 
\begin{lemme}{}\label{reqsharpening3}
For any ${ \mathfrak{t}}\in { \mathfrak{T}}$, the map ${\widetilde{ \mathfrak{e}} }_{ \mathfrak{t}}$ is chain-continuous, i.e., continuous with respect to the Scott{{}}{ }topology on ${ \mathfrak{S}}$ and ${ \mathfrak{B} }$. In other words, 
\begin{eqnarray}
\forall { \mathfrak{t}}\in { \mathfrak{T}},\;\forall { \mathfrak{C}}\subseteq_{Chain} { { \mathfrak{S}}},&& \bigvee\!\!{}_{{}_{\sigma \in { \mathfrak{C}}}} {\widetilde{ \mathfrak{e}} }_{ \mathfrak{t}} (\sigma) =  {\widetilde{ \mathfrak{e}} }_{ \mathfrak{t}} (\bigsqcup{}_{{}_{ { \mathfrak{S}}}} { \mathfrak{C}}).\label{continuityrequirement2}
\end{eqnarray}
\end{lemme}

\noindent An interesting consequence of the continuity property (\ref{continuityrequirement2}) formalizing{{}}{ } Lemma \ref{reqsharpening3}   
is the following property of $({ \mathfrak{S}},\sqsubseteq_{{}_{ { \mathfrak{S}}}})$.
\begin{lemme}{}\label{lemmemeetcontinuous}
The chain-complete Inf semi-lattice $({ \mathfrak{S}},\sqsubseteq_{{}_{ { \mathfrak{S}}}})$ is meet-continuous, i.e.,
\begin{eqnarray}
\forall { \mathfrak{C}}\subseteq_{Chain} { { \mathfrak{S}}},\; \forall \sigma \in { { \mathfrak{S}}}, &&
\sigma \sqcap_{{}_{ { \mathfrak{S}}}}  (\bigsqcup{}_{{}_{ { \mathfrak{S}}}} { \mathfrak{C}}) \;=\;  \bigsqcup{}_{\!{}_{\sigma'\in { \mathfrak{C}}}} (\sigma\sqcap_{{}_{ { \mathfrak{S}}}} \sigma').\label{axiomsigmameetcontinuous}
\end{eqnarray}
\end{lemme}
\begin{proof} 
This fact is easily established using the continuity of {{}}{ }map ${\widetilde{ \mathfrak{e}} }_{ \mathfrak{t}}$ and the meet-continuity of {{}}{ }dcpo ${ \mathfrak{B} }$. For any ${ \mathfrak{C}}\subseteq_{Chain} { { \mathfrak{S}}}$ and any $\sigma \in { { \mathfrak{S}}}$, $(\,\sigma \sqcap_{{}_{ { \mathfrak{S}}}}  (\bigsqcup { \mathfrak{C}})\,)$ and $\bigsqcup{}_{\!{}_{\sigma'\in { \mathfrak{C}}}} (\sigma\sqcap_{{}_{ { \mathfrak{S}}}} \sigma')$ exist as elements of ${ \mathfrak{S}}$. Moreover, 
\begin{eqnarray*}
\forall { \mathfrak{t}}\in { \mathfrak{T}},\;\; %{\widetilde{ \mathfrak{e}} }_{ \mathfrak{t}}(\sigma \sqcap_{{}_{ { \mathfrak{S}}}}  (\bigsqcup { \mathfrak{C}}))=
{\widetilde{ \mathfrak{e}} }_{ \mathfrak{t}} (\sigma \sqcap_{{}_{ { \mathfrak{S}}}}  (\bigsqcup{}_{{}_{ { \mathfrak{S}}}} { \mathfrak{C}})) &=&  
{\widetilde{ \mathfrak{e}} }_{ \mathfrak{t}} (\sigma) \wedge {\widetilde{ \mathfrak{e}} }_{ \mathfrak{t}} (\, (\bigsqcup{}_{\!{}_{\sigma'\in { \mathfrak{C}}}} \sigma'))\\
&=&  
{\widetilde{ \mathfrak{e}} }_{ \mathfrak{t}} (\sigma) \wedge \bigvee\!\!{}_{\!{}_{\sigma'\in { \mathfrak{C}}}}{\widetilde{ \mathfrak{e}} }_{ \mathfrak{t}} ( \sigma')\\
&=&  \bigvee\!\!{}_{\!{}_{\sigma'\in { \mathfrak{C}}}} (\,
{\widetilde{ \mathfrak{e}} }_{ \mathfrak{t}} (\sigma) \wedge {\widetilde{ \mathfrak{e}} }_{ \mathfrak{t}} ( \sigma')\,)\\
&=&  \bigvee\!\!{}_{\!{}_{\sigma'\in { \mathfrak{C}}}} {\widetilde{ \mathfrak{e}} }_{ \mathfrak{t}} (\sigma\sqcap_{{}_{ { \mathfrak{S}}}} \sigma')\\
&=&  {\widetilde{ \mathfrak{e}} }_{ \mathfrak{t}} (\, \bigsqcup{}_{\!{}_{\sigma'\in { \mathfrak{C}}}} (\sigma\sqcap_{{}_{ { \mathfrak{S}}}} \sigma')).
\end{eqnarray*}
We finally use the separation of the Chu space $({ \mathfrak{P}}, { \mathfrak{T}}, { \mathfrak{e}})$ (i.e., Lemma \ref{chuseparated}) to conclude that $(\,\sigma \sqcap_{{}_{ { \mathfrak{S}}}}  (\bigsqcup { \mathfrak{C}})\,)=\bigsqcup{}_{\!{}_{\sigma'\in { \mathfrak{C}}}} (\sigma\sqcap_{{}_{ { \mathfrak{S}}}} \sigma')$.
\end{proof}

{\bf Axioms \ref{reqcap}, \ref{reqbot}, \ref{reqsharpening1}, \ref{reqsharpening2}} {{}}{shall} be {{}}{complemented} by several new axioms {{}}{that are closely related} to the specific character of quantum systems. Indeed, C.Rovelli {{}}{aims} to reconstruct quantum mechanics from the following conceptual proposal \cite[Chap.III]{Rovelli1996}:
\begin{quote}
{\it 'Information is a discrete quantity: there is a minimum amount of information exchangeable (a single bit, or the information that distinguishes between just two alternatives). 
[...] Since information is discrete, any process of acquisition of information can be decomposed into acquisitions of elementary bits of information.'} \cite[p.1655]{Rovelli1996}. 
\end{quote}
We will translate Rovelli's conceptual proposal as follows.
\begin{MyDef}
A state $\sigma_2$ is said to {\em contain one more bit of information than} another state $\sigma_1$ (this fact will be denoted $\sigma_1 \sqcoversubset_{{}_{ \mathfrak{S}}}\sigma_2$), iff (i) $\sigma_2$ is strictly sharper than $\sigma_1$, and (ii) there is no state strictly separating $\sigma_1$ and $\sigma_2$. In other words, 
\begin{eqnarray}
\forall \sigma_1, \sigma_2\in { \mathfrak{S}},\;\;
(\,\sigma_1 \sqcoversubset_{{}_{ \mathfrak{S}}}\sigma_2\,) \; :\Leftrightarrow\; \left\{
\begin{array}{l}
(\, \sigma_1 \sqsubset_{{}_{ \mathfrak{S}}}\sigma_2\,)\\
(\, \forall \sigma\in { \mathfrak{S}},\;
 (\, \sigma_1 \sqsubseteq_{{}_{ \mathfrak{S}}}\sigma \sqsubseteq_{{}_{ \mathfrak{S}}}\sigma_2\,)\;\Rightarrow\; 
 (\, \sigma_1 =\sigma\;\;\textit{\rm or}\;\; \sigma_2=\sigma \,)
 \,).
 \end{array}\right.
 && 
\end{eqnarray}
\end{MyDef}
The discreteness of the informational content, encoded in the partial order $\sqsubseteq_{{}_{ \mathfrak{S}}}$ defined on the space of states ${ \mathfrak{S}}$, is translated into the following topological condition.
\begin{MyReq}\label{reqrovelli1}
For any state $\sigma_1$ admitting a sharper state $\sigma_2$, {{}}{there} exists another state $\sigma_3$ which contains one more 'bit of information' than $\sigma_1$ and coarser than $\sigma_2$.  In other words, the partial-order $({ \mathfrak{S}},\sqsubseteq_{{}_{ \mathfrak{S}}})$ is strongly atomic, i.e.
\begin{eqnarray}
\forall\, \sigma,\sigma' \in  {\mathfrak{S}}\;\vert\; \sigma \sqsubset \sigma' &\Rightarrow& (\,\exists \, \sigma''\in {\mathfrak{S}},\; \sigma \sqcoversubset_{{}_{ \mathfrak{S}}} \sigma''\sqsubseteq_{{}_{ \mathfrak{S}}} \sigma' \,).\label{axiomsigmastrongatom} 
\end{eqnarray}
\end{MyReq}

However, this requirement is not sufficient to capture all the aspects of our conceptual proposal. The exchangeable information {{}}{comprises} a collection of distinguishable bits of information and the acquisition of {{}}{the} informational content is reduced to the acquisition of these bits of information.  

\begin{MyReq}\label{reqrovelli2}
For any state $\sigma\in { \mathfrak{S}}$ and any other $\sigma'\in { \mathfrak{S}}$ containing more information (i.e., $\sigma \sqsubset_{{}_{ \mathfrak{S}}}\sigma'$), the {pair} 
$(\sigma, \sigma')$ {\em 'encodes' an unambiguous piece of information entering into the 'decomposition' of}  $\sigma'$. \\
More {{}}{explicitly}, for any $\sigma_2\in { \mathfrak{S}}$, any sharper state $\sigma_2'\in { \mathfrak{S}}$, and any $\sigma_1\in { \mathfrak{S}}$ 'coarser' than $\sigma_2$, {{}}{there} exists a state $\sigma'_1$ such that (i) $\sigma'_2$ is the coarsest state sharper than $\sigma'_1$ and $\sigma_2$, (ii) $\sigma_1$ is the sharpest state coarser than $\sigma'_1$ and $\sigma_2$. Moreover,  (i) if $(\sigma_2,\sigma'_2)$ encodes for a 'bit' of information, this is also the case for $(\sigma_1,\sigma'_1)$, and (ii) if $(\sigma_1,\sigma_2)$ encodes for a 'bit' of information, this is also the case for $(\sigma_1',\sigma'_2)$. In other words, 
\begin{eqnarray}
\forall \sigma_1,\sigma_2,\sigma_2'\in {\mathfrak{S}},\;\;\;\;\;
(\,\sigma_1 \sqsubseteq_{{}_{ \mathfrak{S}}} \sigma_2 \sqsubseteq_{{}_{ \mathfrak{S}}} \sigma_2',)&\Rightarrow &(\,\exists\; \sigma_1'\in {\mathfrak{S}},\;\vert \;(\, \sigma_1=\sigma_1' \sqcap_{{}_{ \mathfrak{S}}} \sigma_2\;\;\textit{\rm and}\;\;\sigma_2'=\sigma_1' \sqcup_{{}_{ \mathfrak{S}}} \sigma_2\,)\,),\;\;\;\;\;\;\;\;\;\;\;\;
\label{axiomsigmarelativecomplement2prime}
\end{eqnarray}
and
\begin{eqnarray}
\forall \sigma'_1,\sigma_2,\sigma'_2\in {\mathfrak{S}},\;\;\;\;\; (\,\sigma'_1 \sqsubseteq_{{}_{ \mathfrak{S}}} \sigma'_2,\; \sigma_2\sqcoversubset_{{}_{ \mathfrak{S}}}\sigma'_2\,) &\Rightarrow & (\, (\sigma'_1\sqcap_{{}_{ \mathfrak{S}}}\sigma_2)\sqcoversubset_{{}_{ \mathfrak{S}}}\sigma'_1\,),\label{axiomsigmalowersm}\\
\forall \sigma'_2,\sigma'_1,\sigma_2,\sigma_1\in {\mathfrak{S}},\;\;\;\;\; (\,\sigma'_2\sqsupseteq_{{}_{ \mathfrak{S}}} \sigma'_1 \sqsupseteq_{{}_{ \mathfrak{S}}} \sigma_1,\; \sigma'_2\sqsupseteq_{{}_{ \mathfrak{S}}} \sigma_2\sqcoversupset_{{}_{ \mathfrak{S}}}\sigma_1\,)&\Rightarrow &(\, (\sigma'_1\sqcup_{{}_{ \mathfrak{S}}}\sigma_2)\sqcoversupset_{{}_{ \mathfrak{S}}}\sigma'_1\,).\;\;\;\;\;\;\;\;\;\label{axiomsigmauppersm}
\end{eqnarray}
As a summary, ${ \mathfrak{S}}$ is relatively-complemented (property (\ref{axiomsigmarelativecomplement2prime})) and satisfies lower semi-modularity (property (\ref{axiomsigmalowersm})) and 
conditional upper semi-modularity (property (\ref{axiomsigmauppersm})).
\end{MyReq}

\begin{remark}
Note that, in (\ref{axiomsigmalowersm}), $(\sigma'_1 \sqsubseteq_{{}_{ \mathfrak{S}}} \sigma'_2,\; \sigma_2\sqcoversubset_{{}_{ \mathfrak{S}}}\sigma'_2)$ implies trivially $\sigma'_2=(\sigma'_1\sqcup_{{}_{ \mathfrak{S}}}\sigma_2)$, and, in (\ref{axiomsigmauppersm}), $(\sigma'_1 \sqsupseteq_{{}_{ \mathfrak{S}}} \sigma_1,\;  \sigma_2\sqcoversupset_{{}_{ \mathfrak{S}}}\sigma_1)$ implies trivially $\sigma_1=(\sigma'_1\sqcap_{{}_{ \mathfrak{S}}}\sigma_2)$.
\end{remark}

\subsection{First properties of the space of states as a domain}

{{}}{We} now exploit the {{}}{above} requirements to characterize the structure of the space of states. Let us firstly summarize the collected information. \\

$({ \mathfrak{S}}, \sqsubseteq_{{}_{ \mathfrak{S}}})$ is a partial order (property (\ref{axiomsigmapartialorder})). Due to {\bf Axiom \ref{reqbot}}, this partial order is pointed (property (\ref{axiomsigmabottom})). Due to {\bf Axiom \ref{reqcap}}, $({ \mathfrak{S}}, \sqsubseteq_{{}_{ \mathfrak{S}}})$ is also a complete Inf semi-lattice (it is then also a bounded-complete partial order, see Theorem \ref{theoremeboundedcomplete}). Moreover, due to {\bf Axiom \ref{reqsharpening1}}, this partial order is a directed complete partial order (property (\ref{axiomsigmachaincomplete})).  Due to Lemma \ref{lemmemeetcontinuous}, deriving from {\bf Axiom \ref{reqsharpening2}}, this partial order is meet-continuous. 
Due to{{}}{ }{\bf Axiom \ref{reqrovelli1}}, the space of states $({ \mathfrak{S}}, \sqsubseteq_{{}_{ \mathfrak{S}}})$ satisfies the strong atomicity property (\ref{axiomsigmastrongatom}). {{}}{Lastly}, due to{{}}{ }{\bf Axiom \ref{reqrovelli2}}, the space of states $({ \mathfrak{S}}, \sqsubseteq_{{}_{ \mathfrak{S}}})$ is relatively complemented (property (\ref{axiomsigmarelativecomplement2prime})) and satisfies lower semi-modularity (property (\ref{axiomsigmalowersm})) and conditional upper semi-modularity (property (\ref{axiomsigmauppersm})).\\
  
\begin{lemme}{\bf \cite[Theorem 3.6 p.24]{Crawley}}\label{theoremmodular}
Due to the meet-continuity property (\ref{axiomsigmameetcontinuous}), the strong-atomicity property (\ref{axiomsigmastrongatom}), the lower semi-modularity (property (\ref{axiomsigmalowersm})) and the conditional upper semi-modularity (property (\ref{axiomsigmauppersm})), we deduce that ${ \mathfrak{S}}$ is conditionally-modular.  In other words,
\begin{eqnarray}
\forall \sigma_1,\sigma_2,\sigma_3 \in { \mathfrak{S}}\;\vert\; \widehat{\sigma_2\sigma_3}{}^{{}^{ \mathfrak{S}}} && (\sigma_1 \sqsupseteq_{{}_{ \mathfrak{S}}}\sigma_2)\;\Rightarrow\; \sigma_1 \sqcap_{{}_{ \mathfrak{S}}} (\sigma_2\sqcup_{{}_{ \mathfrak{S}}}\sigma_3)=\sigma_ 2\sqcup_{{}_{ \mathfrak{S}}} (\sigma_1\sqcap_{{}_{ \mathfrak{S}}}\sigma_3),\label{defmodularity}
\end{eqnarray}
or, equivalently,
\begin{eqnarray}
\forall \sigma_1,\sigma_2,\sigma_3 \in { \mathfrak{S}}\;\vert\; \widehat{\sigma_1\sigma_3}{}^{{}^{ \mathfrak{S}}}, (\sigma_1 \sqsupseteq_{{}_{ \mathfrak{S}}}\sigma_2), && (\sigma_1 \sqcap_{{}_{ \mathfrak{S}}} \sigma_3=\sigma_2 \sqcap_{{}_{ \mathfrak{S}}} \sigma_3 \;\;\textit{\rm and}\;\; \sigma_1 \sqcup_{{}_{ \mathfrak{S}}} \sigma_3=\sigma_2 \sqcup_{{}_{ \mathfrak{S}}} \sigma_3)\;\Rightarrow\; (\sigma_1=\sigma_2).\;\;\;\;\;\;\;\;\;\;\label{propmodularity}
\end{eqnarray}
\end{lemme}

\begin{lemme}{\bf \cite[Theorem 4.3 (3)$\Rightarrow$(4)]{Crawley}}\label{theorematomistic}
Due to the meet-continuity property (\ref{axiomsigmameetcontinuous}), the strong-atomicity property (\ref{axiomsigmastrongatom}), the bounded-completeness (Theorem \ref{theoremeboundedcomplete}) and the relative complement property (property (\ref{axiomsigmarelativecomplement2prime})), we deduce that ${ \mathfrak{S}}$ is atomistic.  
\begin{eqnarray}
\forall \sigma \in { \mathfrak{S}},&& \sigma=\bigsqcup{}_{{}_{ \mathfrak{S}}} \{\,a\;\vert\; \sigma \sqsupseteq_{{}_{ \mathfrak{S}}} a \sqcoversupset_{{}_{ \mathfrak{S}}} \bot_{{}_{ \mathfrak{S}}}\,\}. 
\end{eqnarray}
\end{lemme}

\begin{lemme}\label{theoremalgebraic}
Due to the meet-continuity property (\ref{axiomsigmameetcontinuous}), the atoms are compact. Then, using Lemma \ref{theorematomistic}, we deduce that the domain ${ \mathfrak{S}}$ is algebraic (i.e. compactly generated).
\begin{eqnarray}
\forall \sigma \in { \mathfrak{S}},&& \sigma=\bigsqcup{}_{{}_{ \mathfrak{S}}} \{\,c\;\vert\; c \sqsubseteq_{{}_{ \mathfrak{S}}} \sigma,\; c\in { \mathfrak{S}}_c \,\}. 
\end{eqnarray}
\end{lemme}

Let us recall that a lattice is said to be {\it a projective lattice} iff it is complete, atomistic, meet-continuous (or algebraic), complemented and modular \cite[Definition 2.4.7 p.38]{faure_projective_2000}. Moreover,  every complemented modular lattice is relatively-complemented \cite[Theorem 4.2 p.31]{Crawley}.  We introduce the definition :
\begin{defin}
A directed complete, bounded complete, atomistic, meet-continuous (or algebraic), relatively-complemented and conditionally-modular poset will be called {\em projective domain}. 
\end{defin}

As a consequence of previous lemmas, we have

\begin{theoreme}\label{spacestatesprojectivedomain}
The space of states is a projective domain.
\end{theoreme}

\begin{MyConj}\label{conjecturejoincontinuity}
As a consequence of relative-complement property and meet-continuity property we suggest that ${ \mathfrak{S}}$ is join-continuous. In other words,
\begin{eqnarray}
\forall \sigma\in { \mathfrak{S}},\forall { \mathfrak{C}}\subseteq_{Chain} { \mathfrak{S}}\;\vert\; \exists  \sigma'\in { \mathfrak{C}},\; \widehat{\sigma\sigma'}{}^{{}^{ \mathfrak{S}}},&&  \sigma \sqcup_{{}_{ \mathfrak{S}}} \bigsqcap{}^\veebar_{{}_{ \mathfrak{S}}}{ \mathfrak{C}}= \bigsqcap{}^\veebar_{{}_{\sigma''\in { \mathfrak{C}},\;\sigma' \sqsupseteq_{{}_{ \mathfrak{S}}}\sigma''}}(\sigma \sqcup_{{}_{ \mathfrak{S}}}\sigma'').\;\;\;\;\;
\end{eqnarray}
\end{MyConj}
\begin{remark}see \cite[Corollary 6]{Bordalo1998} for a key intermediate step in the proof.\end{remark}

\subsection{Pure states}

\begin{MyDef}
A state is said to be {\em a pure state} if and only if it cannot be {{}}{built} as a mixture of other states (the set of pure states will be denoted ${ \mathfrak{S}}_{{}_{pure}}$). More {{}}{explicitly},
\begin{eqnarray}
\sigma \in { \mathfrak{S}}_{{}_{pure}} & :\Leftrightarrow & (\,\forall S \subseteq^{\not= \varnothing} { \mathfrak{S}},\;\;  (\,\sigma = \bigsqcap{}_{{}_{ { \mathfrak{S}}}}S\,) \;\; \Rightarrow \;\; (\, \sigma \in S\,) \,).\label{purestates1}
\end{eqnarray}
In other words, pure states are associated {{}}{with} complete meet-irreducible elements in the selection structure ${ \mathfrak{S}}$.
\end{MyDef}
\begin{remark}
Complete meet-irreducibility implies meet-irreducibility. In other words,
\begin{eqnarray}
\sigma \in { \mathfrak{S}}_{{}_{pure}} & \Rightarrow &  \forall \sigma_1, \sigma_2\in { \mathfrak{S}}, \;\;(\, \sigma =
\sigma_1 \sqcap_{{}_{ { \mathfrak{S}}}} \sigma_2\,) \;\; \Rightarrow \;\; (\, \sigma =  \sigma_1  \;\;\textit{\rm or}\;\; \sigma =  \sigma_2  \,).
\end{eqnarray}
\end{remark}

A simple characterization of completely meet-irreducible elements within posets is adopted in \cite[Definition I-4.21]{gierz_hofmann_keimel_lawson_mislove_scott_2003}. This characterization is equivalent to {{}}{the above} one for {{}}{a} bounded{{}}{-}complete inf semi-lattice. We have {{}}{explicitly}

\begin{theoreme}{\bf [Characterization of pure states]}\label{characterizationpurestates}
\begin{eqnarray}
\sigma \in { \mathfrak{S}}_{{}_{pure}} & \Leftrightarrow &\left\{\begin{array}{ll}  \sigma\in Max({ \mathfrak{S}})  & \textit{\rm (Type 1)}\\ \textit{\rm or}\\ (\uparrow^{{}^{ { \mathfrak{S}}}}\!\!\!\! \sigma)\!\smallsetminus\! \{\sigma\}  \;\textit{\rm admits a minimum element} & \textit{\rm (Type 2)}\end{array}\right.
\end{eqnarray}
\end{theoreme}

It is clear that 'type 2' pure states have no physical meaning. Indeed, for any 'type 2' pure states, it exists a unique 'type 1' pure state sharper than it (and, then, containing more information than it). The existence of 'type 2' pure states in the space of state leads then to a redundant description of the system.  

\begin{MyReq}\label{Axiomnotype2}
${ \mathfrak{S}}_{{}_{pure}}$ admits no element of type 2.
\end{MyReq}

\begin{theoreme}
Every state can be written as a mixture of pure states. In other words, 
\begin{eqnarray}
\forall \sigma \in { \mathfrak{S}} && \sigma= \bigsqcap{}_{{}_{ { \mathfrak{S}}}}  ({ \mathfrak{S}}_{{}_{pure}} \cap (\uparrow^{{}^{ { \mathfrak{S}}}}\!\!\!\! \sigma) ).\label{completemeetirreducibleordergenerating}
\end{eqnarray}
Moreover, ${ \mathfrak{S}}_{{}_{pure}}$ is the unique smallest subset of states generating any state by mixture. In other words,  ${ \mathfrak{S}}_{{}_{pure}}$ is the unique smallest order-generating subset in ${ \mathfrak{S}}$ (i.e., the unique smallest subset of ${ \mathfrak{S}}$ satisfying property (\ref{completemeetirreducibleordergenerating})).
\end{theoreme}
\begin{proof}
${ \mathfrak{S}}$ being, in particular, a bounded{{}}{-}complete algebraic domain, this result is a direct consequence of \cite[Theorem I-4.26]{gierz_hofmann_keimel_lawson_mislove_scott_2003}.
\end{proof}

\begin{MyDef}
We will introduce the following subset of pure states associated with any state :
\begin{eqnarray}
\forall S\subseteq { \mathfrak{S}}^\ast, && \underline{S} :=({ \mathfrak{S}}_{{}_{pure}} \cap (\uparrow^{{}^{ { \mathfrak{S}}}}\!\!\!\! S)).
\end{eqnarray}
\end{MyDef}

\section{Properties and measurements}
\label{sectionpropertiesandmeasurements}
\subsection{Properties and States}

Let us now focus on the set of yes/no tests. Adopting our perspective on the Chu duality between ${ \mathfrak{P}}$ and ${ \mathfrak{T}}$, it is natural to introduce the following equivalence relation on ${ \mathfrak{T}}$. 
\begin{MyDef} 
An equivalence relation, denoted $\sim_{{}_{ \mathfrak{T}}}$, is defined on the set of yes/no tests ${ \mathfrak{T}}$ :\\ 
\indent Two yes/no tests are identified iff the corresponding statements established by the observer about any given preparation process are the same.\\ 
\indent A {\em property} of the physical system is an equivalence class of yes/no tests, i.e., a class of yes/no tests that are not distinguished from the point of view of the statements that the observer can produce by using these yes/no tests on finite collections of samples.\\ 
The set of equivalence classes of yes/no tests, modulo the relation $\sim_{{}_{ \mathfrak{T}}}$, will be denoted ${ \mathcal{L}}$. In other words,
\begin{eqnarray}
\forall { \mathfrak{t}}_1, { \mathfrak{t}}_2\in { \mathfrak{T}},\;\; (\; { \mathfrak{t}}_1 \sim_{{}_{ \mathfrak{T}}} { \mathfrak{t}}_2\;) & :\Leftrightarrow &
(\; \forall { \mathfrak{p}}\in { \mathfrak{P}},\; { \mathfrak{e}}({ \mathfrak{p}},{ \mathfrak{t}}_1)= { \mathfrak{e}}({ \mathfrak{p}},{ \mathfrak{t}}_2) \;),\label{simTchu}\\
&&\sim_{{}_{ \mathfrak{T}}} \;\;\textit{\rm is an equivalence relation},\\
\lfloor { \mathfrak{t}} \rfloor & := & \{\, { \mathfrak{t}}'\in { \mathfrak{T}}\;\vert\; { \mathfrak{t}}'\sim_{{}_{ \mathfrak{T}}} { \mathfrak{t}}\,\},\\
{ \mathcal{L}} & := & \{\,\lfloor { \mathfrak{t}} \rfloor \;\vert\; { \mathfrak{t}}\in { \mathfrak{T}}\,\}.
\end{eqnarray}
\end{MyDef} 

The following equivalence justifies the use of the notion of 'property' in the literal definitions of 'potentiality' and 'actuality' :
\begin{eqnarray}
\forall  { \mathfrak{t}}_1,t_2\in { \mathfrak{T}},&&
(\, { \mathfrak{t}}_1 \sim_{{}_{ \mathfrak{T}}}  { \mathfrak{t}}_2\,)\;\;\Leftrightarrow\;\; (\,{ \mathfrak{Q}}_{{ \mathfrak{t}}_1}={ \mathfrak{Q}}_{{ \mathfrak{t}}_2}\;\; \textit{\rm and}\;\; { \mathfrak{A}}_{{ \mathfrak{t}}_1}={ \mathfrak{A}}_{{ \mathfrak{t}}_2}\,).\label{simTQandA}
\end{eqnarray} 
Hence, 
\begin{MyDef}
for any ${ \mathfrak{l}}\in { \mathcal{L}}$, we will {{}}{now} denote by  ${ \mathfrak{Q}}_{{ \mathfrak{l}}}$ (resp. ${ \mathfrak{A}}_{{ \mathfrak{l}}}$) the set ${ \mathfrak{Q}}_{{ \mathfrak{t}}}$ (resp. ${ \mathfrak{A}}_{{ \mathfrak{t}}}$) taken for any ${ \mathfrak{t}}$ such that ${ \mathfrak{l}}=\lfloor { \mathfrak{t}} \rfloor$.
\end{MyDef} 
Moreover,
\begin{MyDef}
for any ${ \mathfrak{l}}\in { \mathcal{L}}$, we will {{}}{now} denote by  $\widetilde{ \mathfrak{e}}_{{ \mathfrak{l}}}$ the evaluation map defined on ${ \mathfrak{S}}$ and defined by $\widetilde{ \mathfrak{e}}_{{ \mathfrak{l}}}:=\widetilde{ \mathfrak{e}}_{{ \mathfrak{t}}}$ for any ${ \mathfrak{t}}$ such that ${ \mathfrak{l}}=\lfloor { \mathfrak{t}} \rfloor$.
\end{MyDef}

\begin{MyDef}
A property ${ \mathfrak{l}}\in { \mathcal{L}}$ will be said to be {\em testable} iff it can be revealed as 'actual' at least for some collections of prepared samples.  In other words,  
\begin{eqnarray}
{ \mathfrak{l}} \;\textit{\rm is 'testable'}&:\Leftrightarrow &{ \mathfrak{A}}_{ \mathfrak{l}} \;\textit{\rm is a non-empty  subset of ${ \mathfrak{S}}$}.\;\;\;\;\;\;\;\;\;\label{axiompotential}
\end{eqnarray} 
\end{MyDef}

We check {{}}{immediately} that $\forall { \mathfrak{t}}_1, { \mathfrak{t}}_2\in { \mathfrak{T}},\;\;(\; { \mathfrak{t}}_1 \sim_{{}_{ \mathfrak{T}}} { \mathfrak{t}}_2\;) \;\; \Leftrightarrow \;\; (\; \overline{{ \mathfrak{t}}_2} \sim_{{}_{ \mathfrak{T}}} \overline{{ \mathfrak{t}}_1}\;)$. As a consequence, the bar involution will be defined on the space of properties simply by requiring
\begin{MyDef} 
\begin{eqnarray}
\forall { \mathfrak{l}}\in { \mathcal{L}},&& \overline{{ \mathfrak{l}}}:=\{\, \overline{{ \mathfrak{t}}}\;\vert\; { \mathfrak{l}}=\lfloor{ \mathfrak{t}}\rfloor\,\}.
\end{eqnarray} 
\end{MyDef}

$({ \mathfrak{S}}, \sqsubseteq_{{}_{ \mathfrak{S}}})$ being  a selection structure, we know that $({ \mathfrak{S}}, \sqsubseteq_{{}_{ \mathfrak{S}}})$ is a bounded-complete Inf semi-lattice, and in particular is closed under arbitrary {{}}{infima}. In other words,
\begin{lemme}{}
\begin{eqnarray}
\forall { \mathfrak{Q}} \subseteq^{\not= \varnothing} { \mathfrak{P}},\; 
&&\textit{\rm the infimum}\; (\bigsqcap{}_{{}_{ \mathfrak{S}}} \lceil { \mathfrak{Q}}\rceil )\;\textit{\rm exists in}\; { \mathfrak{S}},
\end{eqnarray}
\end{lemme}
Moreover, we inherit from Lemma \ref{reqsharpening3} the following continuity property :
\begin{lemme}{}
\begin{eqnarray}
\forall { \mathfrak{R}} \subseteq^{\not= \varnothing} { \mathfrak{S}},\;  \forall { \mathfrak{t}}\in { \mathfrak{T}},&&
{\widetilde{ \mathfrak{e}} }_{ \mathfrak{t}}(\bigsqcap{}_{{}_{ \mathfrak{S}}} { \mathfrak{R}}) \;=\; \bigwedge{}_{\!{}_{\sigma\in { \mathfrak{R}}}} {\widetilde{ \mathfrak{e}} }_{ \mathfrak{t}} (\sigma).\label{infcontinuity}
\end{eqnarray}
\end{lemme}
\begin{proof}
$\forall { \mathfrak{R}} \subseteq^{\not= \varnothing} { \mathfrak{S}}$, we define ${ \mathfrak{M}}_{{}_{ \mathfrak{R}}}:=\{\sigma\in { \mathfrak{S}}\;\vert\; \sigma \sqsubseteq_{{}_{ \mathfrak{S}}} { \mathfrak{R}}\,\}$. ${ \mathfrak{M}}_{{}_{ \mathfrak{R}}}$ is obviously directed and $\bigsqcap{}_{{}_{ \mathfrak{S}}} { \mathfrak{R}} = \bigsqcup{}^\barwedge_{{}_{ \mathfrak{S}}} { \mathfrak{M}}_{{}_{ \mathfrak{R}}}$. Now, using the Scott{{}}{ }continuity of ${\widetilde{ \mathfrak{e}} }_{ \mathfrak{t}}$, we have ${\widetilde{ \mathfrak{e}} }_{ \mathfrak{t}}(\bigsqcap{}_{{}_{ \mathfrak{S}}} { \mathfrak{R}}) = \bigvee{}^\barwedge_{{}_{ \sigma \in { \mathfrak{M}}_{{}_{ \mathfrak{R}}}}} {\widetilde{ \mathfrak{e}} }_{ \mathfrak{t}} (\sigma)$. The monotonicity of 
${\widetilde{ \mathfrak{e}} }_{ \mathfrak{t}}$ and the fact that the target space of ${\widetilde{ \mathfrak{e}} }_{ \mathfrak{t}}$ is the boolean domain ${ \mathfrak{B}}$ implies moreover that $\bigvee{}^\barwedge_{{}_{ \sigma \in { \mathfrak{M}}_{{}_{ \mathfrak{R}}}}} {\widetilde{ \mathfrak{e}} }_{ \mathfrak{t}} (\sigma)=\bigwedge{}_{\!{}_{\sigma\in { \mathfrak{R}}}} {\widetilde{ \mathfrak{e}} }_{ \mathfrak{t}} (\sigma)$.
\end{proof}

\begin{theoreme}{}\label{widetildeelawson}
For any property ${ \mathfrak{l}}\in { \mathcal{L}}$, the evaluation map ${\widetilde{ \mathfrak{e}} }_{ \mathfrak{l}}$ is order{{}}{ }preserving and continuous with respect to the Lawson topologies on ${ \mathfrak{S}}$ and ${ \mathfrak{B}}$.
\end{theoreme}
\begin{proof}
From previous lemma we have that, for any ${ \mathfrak{t}}\in { \mathfrak{T}}$, the map ${\widetilde{ \mathfrak{e}} }_{ \mathfrak{t}}$ is continuous with respect to the lower{{}}{ }topologies on ${ \mathfrak{P}}$ and ${ \mathfrak{B} }$.\\
Due to Lemma \ref{reqsharpening3} and property (\ref{infcontinuity}), and using \cite[Theorem III-1.8 p.213]{gierz_hofmann_keimel_lawson_mislove_scott_2003}, we then prove the announced continuity property.
\end{proof}

We can then deduce the form of the subsets $ \lceil { \mathfrak{A}}_{{ \mathfrak{t}}}\rceil$ determining the sub-space of states for which the property tested by ${ \mathfrak{t}}$ is actual. 

\begin{theoreme}{\bf [Property-state]}\label{defUwithsigma}\\
For any ${ \mathfrak{t}}$ in ${ \mathfrak{T}}$, corresponding to a testable property $\lfloor { \mathfrak{t}}\rfloor$,  {{}}{there} exists an element ${\Sigma}_{ \mathfrak{t}}\in { \mathfrak{S}}$ (in fact, a compact element in the algebraic domain ${ \mathfrak{S}}$), such that the Scott-open filter$ \lceil { \mathfrak{A}}_{{ \mathfrak{t}}}\rceil$ is the principal filter associated {{}}{with} ${\Sigma}_{ \mathfrak{t}}$ :
\begin{eqnarray}
\forall { \mathfrak{t}} \in { \mathfrak{T}}\;\vert\; \lfloor { \mathfrak{t}}\rfloor \;\textit{\rm is testable},\;\; \exists {\Sigma}_{ \mathfrak{t}}\in { \mathfrak{S}}_c & \vert & \lceil { \mathfrak{A}}_{{ \mathfrak{t}}}\rceil =  (\uparrow^{\!{}^{ \mathfrak{S}}}\!\!{\Sigma}_{ \mathfrak{t}}).
\end{eqnarray}
In particular, we have
\begin{eqnarray}
\forall { \mathfrak{t}} \in { \mathfrak{T}}, \forall \sigma \in { \mathfrak{S}},&& \widetilde{\mathfrak{e}}_{ \mathfrak{t}}(\sigma)=\textit{\bf Y}\;\; \Leftrightarrow\;\; {\Sigma}_{{ \mathfrak{t}}} \sqsubseteq_{{}_{ \mathfrak{S}}}\sigma.
\end{eqnarray}
If the conjugate test $\overline{ \mathfrak{t}}$ corresponds to a testable property as well, there exists an element an element $\Sigma_{\overline{ \mathfrak{t}}}$ such that 
\begin{eqnarray}
\Sigma_{\overline{ \mathfrak{t}}} &=& \bigsqcap{}_{{}_{ \mathfrak{S}}} \widetilde{\mathfrak{e}}_{ \mathfrak{t}}{}^{-1} (\textit{\bf N}) \\
{ \mathfrak{Q}}_{ \mathfrak{t}} &=& { \mathfrak{S}} \smallsetminus (\uparrow^{\!{}^{ \mathfrak{S}}}\!\!{\Sigma}_{\overline{ \mathfrak{t}}})
\end{eqnarray}
\end{theoreme}
\begin{proof}
We have already remarked that $ \lceil { \mathfrak{A}}_{{ \mathfrak{t}}}\rceil$ is a Scott-open filter. Using the fact that ${ \mathfrak{S}}$ is a complete Inf semi-lattice {{}}{as well as} property (\ref{infcontinuity}), we also note that the element ${\Sigma}_{ \mathfrak{t}}:=\bigsqcap{}_{{}_{ \mathfrak{S}}} \lceil { \mathfrak{A}}_{{ \mathfrak{t}}}\rceil$ obeys ${\widetilde{ \mathfrak{e}} }_{ \mathfrak{t}} ({\Sigma}_{ \mathfrak{t}})=\bigwedge{}_{\!{}_{\sigma\in \lceil { \mathfrak{A}}_{{ \mathfrak{t}}}\rceil}} {\widetilde{ \mathfrak{e}} }_{ \mathfrak{t}} (\sigma)=\textit{\bf Y}$. Then, ${\Sigma}_{ \mathfrak{t}}$ is the 'minimum' of $ \lceil { \mathfrak{A}}_{{ \mathfrak{t}}}\rceil$. As a consequence, the filter $ \lceil { \mathfrak{A}}_{{ \mathfrak{t}}}\rceil$ {{}}{is revealed} to be the principal filter $(\uparrow^{{}^{ \mathfrak{S}}}\!\!\!{\Sigma}_{ \mathfrak{t}})$. From  \cite[Remark I-4.24]{gierz_hofmann_keimel_lawson_mislove_scott_2003}, we deduce that ${\Sigma}_{ \mathfrak{t}}$ is a compact element in ${ \mathfrak{S}}$. \\
The properties associated to $\Sigma_{\overline{ \mathfrak{t}}}$ are derived along the same way.
\end{proof}

\begin{MyDef} \label{reqcontinuityarbitraryinfima}
For any yes/no test ${ \mathfrak{t}} \in { \mathfrak{T}}$, corresponding to the testable property $\lfloor { \mathfrak{t}}\rfloor$,  the state ${\Sigma}_{ \mathfrak{t}}$ is defined to be the minimal element of the principal filter $\lceil { \mathfrak{A}}_{{ \mathfrak{t}}}\rceil$ in ${{({ \mathfrak{S}}, \sqsubseteq_{{}_{ \mathfrak{S}}})}}$.  
\begin{eqnarray}
\forall { \mathfrak{t}} \in { \mathfrak{T}}\;\vert\; \lfloor { \mathfrak{t}}\rfloor \;\textit{\rm is testable},&& {\Sigma}_{ \mathfrak{t}}:=\bigsqcap{}_{{}_{ \mathfrak{S}}} \lceil { \mathfrak{A}}_{{ \mathfrak{t}}}\rceil = \bigsqcap{}_{{}_{ \mathfrak{S}}}  \widetilde{ \mathfrak{e}}_{ \mathfrak{t}}{}^{-1}(\textit{\bf Y}).\label{defSigma}
\end{eqnarray}
The state ${\Sigma}_{ \mathfrak{t}}$ depends only on the testable property $\lfloor { \mathfrak{t}}\rfloor$ associated {{}}{with} ${ \mathfrak{t}}$. This state will then be called the {\em property-state} associated to ${ \mathfrak{t}}$ and we will {{}}{henceforth} adopt the following abuse of notation ${\Sigma}_{\lfloor { \mathfrak{t}}\rfloor}:={\Sigma}_{ \mathfrak{t}}$. 
\end{MyDef}

\subsection{Tests and measurements}

The measurement process offers {{}}{another} perspective on the relation between the spaces ${ \mathfrak{P}}$ and ${ \mathfrak{T}}$ (the first {{}}{being} a duality relation): this {{}}{new} perspective emphasizes the 'recursive' aspect of {{}}{the} preparation process. Indeed, a given yes/no test ${ \mathfrak{t}}\in { \mathfrak{T}}$ can be used to complete a given preparation procedure $ { \mathfrak{p}} \in { \mathfrak{P}}$ in order to produce a new preparation procedure, as {{}}{ }the 'filtering operation' associated {{}}{with} ${ \mathfrak{t}}$ actually operates on a collection of produced samples that can exhibit the desired property.
\begin{MyDef} \label{Defmeasurementmap}
For any yes/no test ${ \mathfrak{t}} \in { \mathfrak{T}}$ and any preparation procedure ${ \mathfrak{p}}\in { \mathfrak{Q}}_{ \mathfrak{t}}$, we can define a preparation procedure denoted ${ \mathfrak{p}} {\centerdot} { \mathfrak{t}}$ and defined as follows :\\ 
\indent the samples, previously prepared {{}}{through} procedure ${ \mathfrak{p}}$, are actually submitted to the measurement operation defined according to the yes/no test ${ \mathfrak{t}}$, the {{}}{resulting (or `outcoming')} samples of the 'whole preparation process' (i.e., the initial preparation ${ \mathfrak{p}}$ followed by the filtering operation defined by ${ \mathfrak{t}}$), denoted $({ \mathfrak{p}}{\centerdot} { \mathfrak{t}})$, are the samples 'actually measured as positive' through the yes/no test ${ \mathfrak{t}}$.\\ 
{{}}{For} any yes/no test ${ \mathfrak{t}} \in { \mathfrak{T}}$, we will then associate {{}}{it with} the partial map denoted ${\centerdot} { \mathfrak{t}}$ and defined by ({{}}{where} its domain will be denoted $Dom^{ \mathfrak{P}}_{{\centerdot} { \mathfrak{t}}}\subseteq { \mathfrak{P}}$, and its range $Im^{ \mathfrak{P}}_{{\centerdot} { \mathfrak{t}}}$): 
\begin{eqnarray}
&&\begin{array}{rcrccr}
{\centerdot} { \mathfrak{t}} & : & { \mathfrak{P}} & \dashrightarrow & { \mathfrak{P}} & \;\;\;\;\;\;\;\;\;\;Dom^{ \mathfrak{P}}_{{\centerdot} { \mathfrak{t}}}:={ \mathfrak{Q}}_{ \mathfrak{t}}\\
& &   { \mathfrak{p}} & \mapsto & { \mathfrak{p}}{\centerdot} { \mathfrak{t}} &
\end{array}
\end{eqnarray} 
These maps define the measurement operation associated {{}}{with} a given property. The map ${\centerdot} { \mathfrak{t}}$ associated {{}}{with} a yes/no test ${ \mathfrak{t}}$ is called {\em the measurement map} associated {{}}{with} ${ \mathfrak{t}}$.
\end{MyDef}

%To begin, let us clarify the notion of 'succession' of measurements. 

\begin{MyDef} \label{Deft1t2}
For any yes/no tests ${ \mathfrak{t}}_1,{ \mathfrak{t}}_2\in { \mathfrak{T}}$, we can build a new yes/no test denoted $({ \mathfrak{t}}_1 {\centerdot} { \mathfrak{t}}_2)\in { \mathfrak{T}}$ and called {\em the succession of ${ \mathfrak{t}}_1$ by ${ \mathfrak{t}}_2$}. It is defined as follows :\\
to begin, the incoming sample is tested {{}}{through} the yes/no test ${ \mathfrak{t}}_1$; if the result of this test is negative, then the whole yes/no test  $({ \mathfrak{t}}_1 {\centerdot} { \mathfrak{t}}_2)$ is declared 'negative'{{}}{, but if the result is positive, } the outcoming sample having been positively measured by ${ \mathfrak{t}}_1$ is submitted to the yes/no test ${ \mathfrak{t}}_2$; the result of this test is then attributed to the whole yes/no test  $({ \mathfrak{t}}_1 {\centerdot} { \mathfrak{t}}_2)$ for the given prepared sample.\\ 
 In other words,
\begin{eqnarray}
\forall { \mathfrak{t}}_1,{ \mathfrak{t}}_2\in { \mathfrak{T}}, \exists ({ \mathfrak{t}}_1 {\centerdot} { \mathfrak{t}}_2)\in { \mathfrak{T}} &\vert & \nonumber \\
&&{ \mathfrak{A}}_{{{ \mathfrak{t}}_1}{\centerdot} { \mathfrak{t}}_2} :=\{\, { \mathfrak{p}} \in { \mathfrak{A}}_{{ \mathfrak{t}}_1}\;\vert\; ({ \mathfrak{p}}{\centerdot} { \mathfrak{t}}_1)\in { \mathfrak{A}}_{{ \mathfrak{t}}_2} \,\}\label{At1dott2}\\
&&{ \mathfrak{Q}}_{{{ \mathfrak{t}}_1}{\centerdot} { \mathfrak{t}}_2} := \{\, { \mathfrak{p}} \in { \mathfrak{Q}}_{{ \mathfrak{t}}_1}\;\vert\; ({ \mathfrak{p}}{\centerdot} { \mathfrak{t}}_1)\in { \mathfrak{Q}}_{{ \mathfrak{t}}_2} \,\}\label{Qt1dott2}\\
&&{ \mathfrak{A}}_{\overline{{{ \mathfrak{t}}_1}{\centerdot} { \mathfrak{t}}_2}} := { \mathfrak{A}}_{\overline{{ \mathfrak{t}}_1}} \cup \{\, { \mathfrak{p}} \in { \mathfrak{Q}}_{{ \mathfrak{t}}_1}\;\vert\; ({ \mathfrak{p}}{\centerdot} { \mathfrak{t}}_1)\in { \mathfrak{A}}_{\overline{{ \mathfrak{t}}_2}} \,\}\label{Abart1dott2}\\
&&\forall { \mathfrak{p}} \in Dom^{ \mathfrak{P}}_{{\centerdot}({{ \mathfrak{t}}_1}{\centerdot} { \mathfrak{t}}_2)} = { \mathfrak{Q}}_{{{ \mathfrak{t}}_1}{\centerdot} { \mathfrak{t}}_2}, \;\;  
 { \mathfrak{p}} {\centerdot} ({ \mathfrak{t}}_1 {\centerdot} { \mathfrak{t}}_2) := ({ \mathfrak{p}} {\centerdot} { \mathfrak{t}}_1) {\centerdot} { \mathfrak{t}}_2.\label{pt1dott2} 
\end{eqnarray}
The 'succession law' satisfies the following associativity properties:
\begin{eqnarray}
\forall{ \mathfrak{p}}\in { \mathfrak{P}}, \forall { \mathfrak{t}}_1,{ \mathfrak{t}}_2,\in { \mathfrak{T}}, && { \mathfrak{p}} {\centerdot} ({ \mathfrak{t}}_1{\centerdot} { \mathfrak{t}}_2)=({ \mathfrak{p}} {\centerdot} { \mathfrak{t}}_1){\centerdot} { \mathfrak{t}}_2,\\
\forall { \mathfrak{t}}_1,{ \mathfrak{t}}_2,{ \mathfrak{t}}_3\in { \mathfrak{T}}, && ({ \mathfrak{t}}_1 {\centerdot} { \mathfrak{t}}_2){\centerdot} { \mathfrak{t}}_3={ \mathfrak{t}}_1 {\centerdot} ({ \mathfrak{t}}_2{\centerdot} { \mathfrak{t}}_3).
\end{eqnarray}
\end{MyDef}

We note the following natural relations.

\begin{lemme}{}
\begin{eqnarray}
\forall { \mathfrak{t}}_1,{ \mathfrak{t}}_2\in { \mathfrak{T}},&&\forall { \mathfrak{p}} \in { \mathfrak{Q}}_{{ \mathfrak{t}}_1},\;\;\; { \mathfrak{e}} ({ \mathfrak{p}}{\centerdot} { \mathfrak{t}}_1, { \mathfrak{t}}_2) \geq { \mathfrak{e}} ({ \mathfrak{p}},{ \mathfrak{t}}_1 {\centerdot} { \mathfrak{t}}_2),\\
&& \overline{\;{ \mathfrak{t}}_1 {\centerdot} { \mathfrak{t}}_2\;} = \overline{{ \mathfrak{t}}_1} \; {\centerdot}\; \overline{{ \mathfrak{t}}_2}
\end{eqnarray}
\end{lemme}

As a basic requirement {{}}{of} measurement maps, we will impose that they are monotone maps on their domain. The following simple analysis justifies this requirement.

\begin{theoreme}\label{Axiommeasurementmap1}
A measurement operation associated {{}}{with} any yes/no test{{}}{, ${ \mathfrak{t}}$,} respects the ordering of information established by the observer about the collection of samples on which it is realized.
\begin{eqnarray}
\forall { \mathfrak{t}} \in { \mathfrak{T}},\;\;\forall { \mathfrak{p}}_1, { \mathfrak{p}}_2\in  { \mathfrak{Q}}_{ \mathfrak{t}},&& (\, { \mathfrak{p}}_1 \sqsubseteq_{{}_{ \mathfrak{P}}} { \mathfrak{p}}_2\,) \;\; \Rightarrow\;\; (\, ({ \mathfrak{p}}_1 {\centerdot} { \mathfrak{t}}) \sqsubseteq_{{}_{ \mathfrak{P}}} ({ \mathfrak{p}}_2 {\centerdot} { \mathfrak{t}})\,).
\end{eqnarray}    
The measurement map  $({\centerdot} { \mathfrak{t}})$ is an order-preserving map on $( { \mathfrak{P}},\sqsubseteq_{{}_{ \mathfrak{P}}})$.
\end{theoreme}
\begin{proof}
If the preparation processes ${ \mathfrak{p}}_1$ and ${ \mathfrak{p}}_2$ are ordered by $\sqsubseteq_{{}_{ \mathfrak{P}}}$ (${ \mathfrak{p}}_2$ being sharper than ${ \mathfrak{p}}_1$), every {{}}{statement made} by the observer about ${ \mathfrak{p}}_1$ {{}}{is} also necessarily {{}}{made} about ${ \mathfrak{p}}_2$, i.e., $\forall { \mathfrak{u}}\in { \mathfrak{T}}, { \mathfrak{e}}({ \mathfrak{p}}_1,{ \mathfrak{u}})\leq { \mathfrak{e}}({ \mathfrak{p}}_2,{ \mathfrak{u}})$. This is true in particular for the statements that can be {{}}{made} by the observer about the corresponding collections of samples after having separately measured the property associated {{}}{with} a given yes/no test{{}}{, ${ \mathfrak{t}}$,} on each collection of samples beforehand. 
More precisely, we must then have $\forall { \mathfrak{v}}\in { \mathfrak{T}}, { \mathfrak{e}}({ \mathfrak{p}}_1{\centerdot} { \mathfrak{t}},{ \mathfrak{v}})\leq { \mathfrak{e}}({ \mathfrak{p}}_2{\centerdot} { \mathfrak{t}},{ \mathfrak{v}})$. 
\end{proof}

\begin{corollaire}{}
As a consequence, the measurement operation $({\centerdot} { \mathfrak{t}})$ associated {{}}{with} a given yes/no test{{}}{, ${ \mathfrak{t}}$,} {{}}{cannot} distinguish different {{}}{collections} of samples on which it acts, {{}}{when} these collections of samples correspond to the same state of the system, i.e.,
\begin{eqnarray}
\forall { \mathfrak{t}} \in { \mathfrak{T}},\;\; \forall { \mathfrak{p}}_1, { \mathfrak{p}}_2\in   { \mathfrak{Q}}_{ \mathfrak{t}},&& (\, { \mathfrak{p}}_1 \sim_{{}_{ \mathfrak{P}}} { \mathfrak{p}}_2\,)\;\;\Rightarrow\;\; (\, ({ \mathfrak{p}}_1 {\centerdot} { \mathfrak{t}}) \sim_{{}_{ \mathfrak{P}}} ({ \mathfrak{p}}_2 {\centerdot} { \mathfrak{t}})\,).\label{tsursim}
\end{eqnarray}
\end{corollaire}

\begin{MyDef}\label{actionmeasurementsonstates}
The measurement operation will then be defined to act on states as follows:
\begin{eqnarray}
\forall { \mathfrak{t}} \in { \mathfrak{T}}, \;\;\forall { \mathfrak{p}} \in  { \mathfrak{Q}}_{ \mathfrak{t}},&& \lceil { \mathfrak{p}}  \rceil {\centerdot} { \mathfrak{t}} := \lceil { \mathfrak{p}} {\centerdot} { \mathfrak{t}} \rceil \label{tsursimdef}
\end{eqnarray}
We will adopt the following notations $Dom^{ \mathfrak{S}}_{{\centerdot} { \mathfrak{t}}}:=\lceil Dom^{ \mathfrak{P}}_{{\centerdot} { \mathfrak{t}}}\rceil$ and $Im^{ \mathfrak{S}}_{{\centerdot} { \mathfrak{t}}}:=\lceil Im^{ \mathfrak{P}}_{{\centerdot} { \mathfrak{t}}}\rceil$.
\end{MyDef}
\begin{corollaire}{}
The measurement map  $({\centerdot} { \mathfrak{t}})$ is an order-preserving map on $({ \mathfrak{S}},\sqsubseteq_{{}_{ \mathfrak{S}}})$
\begin{eqnarray}
\forall { \mathfrak{t}} \in { \mathfrak{T}}, \forall \sigma_1,\sigma_2 \in { \mathfrak{S}}_{ \mathfrak{t}},&& (\, \sigma_1 \sqsubseteq_{{}_{ \mathfrak{S}}} \sigma_2\,) \;\; \Rightarrow\;\; (\, (\sigma_1 {\centerdot} { \mathfrak{t}}) \sqsubseteq_{{}_{ \mathfrak{S}}} (\sigma_2 {\centerdot} { \mathfrak{t}})\,).
\end{eqnarray}
\end{corollaire}

We {{}}{then complement} the {{}}{above} property of monotonicity {{}}{with} a requirement concerning{{}}{ }random mixing.

Endly, we will observe the following property.

\begin{theoreme}\label{Axiommeasurementmap3}
The operation of measurement respects the induction process of a limit state from any sharpening family. In other words, for any  yes/no test{{}}{,} ${ \mathfrak{t}}$, the measurement map $({\centerdot} { \mathfrak{t}})$ is a Scott-continuous partial map.
\begin{eqnarray}
\forall { \mathfrak{t}}\in { \mathfrak{T}},\forall { \mathfrak{C}}\subseteq_{Chain} \lceil { \mathfrak{Q}}_{ \mathfrak{t}}\rceil,&& \bigsqcup{}_{\sigma\in { \mathfrak{C}}} (\sigma {\centerdot} { \mathfrak{t}})= (\bigsqcup{}_{{}_{ \mathfrak{S}}} { \mathfrak{C}}) {\centerdot} { \mathfrak{t}}.\label{Axiommeasurementmap3eq}
\end{eqnarray}
\end{theoreme}
\begin{proof}
Let us consider a sharpening family of preparation processes ${ \mathfrak{Q}}$. The existence of the limit state $(\bigsqcup_{{}_{ \mathfrak{S}}}  \lceil { \mathfrak{Q}} \rceil )$ is ensured by the collection of properties $\bigvee\!\!{}_{{}_{\sigma \in  \lceil { \mathfrak{Q}} \rceil }} {\widetilde{ \mathfrak{e}} }_{ \mathfrak{u}} (\sigma) =  {\widetilde{ \mathfrak{e}} }_{ \mathfrak{u}} (\bigsqcup_{{}_{ \mathfrak{S}}}  \lceil { \mathfrak{Q}} \rceil )$ considered for every yes/no test ${ \mathfrak{u}}\in { \mathfrak{T}}$. If we consider, in particular, the subset of statements associated {{}}{with} any yes/no test{{}}{, ${ \mathfrak{t}}_2$, made} by the observer concerning the states outcoming from the measurement associated {{}}{with} a yes/no test{{}}{,} ${ \mathfrak{t}}_1$, we deduce 
$\bigvee_{{}_{\sigma \in \lceil { \mathfrak{Q}} \rceil }} {\widetilde{ \mathfrak{e}} }_{{ \mathfrak{t}}_2} (\sigma {\centerdot} { \mathfrak{t}}_1) =  {\widetilde{ \mathfrak{e}} }_{{ \mathfrak{t}}_2}( (\bigsqcup_{{}_{ \mathfrak{S}}}  \lceil { \mathfrak{Q}} \rceil ) {\centerdot} { \mathfrak{t}}_1)$. And {{}}{thus}, for any ${ \mathfrak{t}}_1, { \mathfrak{t}}_2\in { \mathfrak{T}}$, we obtain ${\widetilde{ \mathfrak{e}} }_{{ \mathfrak{t}}_2} (\bigsqcup_{{}_{\sigma \in \lceil { \mathfrak{Q}} \rceil }}(\sigma {\centerdot} { \mathfrak{t}}_1)) =  {\widetilde{ \mathfrak{e}} }_{{ \mathfrak{t}}_2}( (\bigsqcup_{{}_{ \mathfrak{S}}}  \lceil { \mathfrak{Q}} \rceil ) {\centerdot} { \mathfrak{t}}_1)$. 
\end{proof}

The set of partial maps defined from ${ \mathfrak{S}}_1$ (the domain of the partial map has to be a Scott-closed subset of ${ \mathfrak{S}}_1$) to ${ \mathfrak{S}}_2$, which are order{{}}{ }preserving and Scott{{}}{ }continuous will be denoted $\left[ { \mathfrak{S}}_1 \dashrightarrow { \mathfrak{S}}_2\right]^\barwedge$.  \\

To summarize Theorem \ref{Axiommeasurementmap1} and Theorem \ref{Axiommeasurementmap3} , we will have :
\begin{eqnarray}
\forall { \mathfrak{t}}\in { \mathfrak{T}}, &&{\centerdot} { \mathfrak{t}} \in \left[ { \mathfrak{S}} \dashrightarrow { \mathfrak{S}}\right]^\barwedge \;\vert\;Dom^{ \mathfrak{S}}_{{\centerdot} { \mathfrak{t}}} =\lceil { \mathfrak{Q}}_{ \mathfrak{t}}\rceil.
\end{eqnarray}

\subsection{Minimally disturbing measurements}
\label{subsectionminimallydisturbingmeasurements}

As analyzed {{}}{above}, the 'certainty' of the observer{{}}{ }about the occurrence (or not) of a given 'property' for a given state, is formulated as a {{}}{counterfactual} statement ('actuality' or 'impossibility') about the tests that 'could' be realized on any sample corresponding to this state (and this certainty has been produced after having tested this property on similarly prepared samples). {\it Stricto sensu} this statement is then formulated without disturbing in any way the considered new sample, according to the definition of the 'elements of reality' for the system given in the celebrated paper of A. Einstein, B. Podolsky, and N. Rosen :
\begin{quote}
{\it 'If, without in any way disturbing a system, we can predict with certainty (i.e., with probability equal to unity) the value of a physical quantity, then there exists an element of physical reality corresponding to this physical quantity.'} \cite{Einstein}.
\end{quote}
Nevertheless, in order to establish an interpretation of 'properties' as elements of reality, the observer must be able to establish and confirm 'conjointly' {{}}{ }statements about the different properties that {{}}{are considered} 'simultaneously' actual for a given collection of similarly prepared samples. It is {{}}{thus} necessary to restrict the measurement operations{{}}{ }that will be used by the observer, {{}}{with regard to} the disturbance they {{}}{cause in} the measured sample. 
These measurement operations should {{}}{guarantee} that the state of the system after measurement be characterized by the properties established as actual beforehand, through 'successive' measurement operations. The possibility (and necessity) to characterize minimally disturbing measurements exists in the classical and in the quantum situation as well. However, although the existence of such measurements does not pose any conceptual problem in the classical situation, things are more complex in the quantum situation as soon as the measurement process irreducibly 'alters' the state of the measured system.\\
Despite the indeterministic character of quantum measurements, we note that the realization of a 'careful' yes/no test {{}}{does} allow the observer to {{}}{make} some statements about the state 'after' the measurement, although it appears {{}}{risky} to extend these conclusions to the state of the system 'before' the measurement (due to the irreducible alteration of the state during the measurement operation). At least, a 'careful' measurement of a given property on a given sample\footnote{Here we mean that the test is effectuated on the considered sample, according to the procedure defined by a yes/no test associated {{}}{with} this property, and the 'answer' received by the observer is 'positive'.} may {{}}{guarantee} the actuality of this property just after the measurement. As a consequence, the immediate repetition of this test should produce with certainty the same 'answer'. {{}}{These} sort of careful measurements do exist in the classical {{}}{and} in the quantum situation as well, {{}}{and} have been called {\em first{{}}{-}kind measurements} by W. Pauli :
\begin{quote}
{\it 'The method of measurement [...] has the property that a repetition of measurement gives the same value for the quantity measured as in the first measurement. In other words, if the result of using the measuring apparatus is not known, but only the fact of its use is known [...], the probability that the quantity measured has a certain value is the same, both before and after the measurement. We shall call such measurements the measurements of the first kind.'} \cite[p.75]{1958Pauli}\footnote{Note the distinction made by W.Pauli {{}}{between} measurements of the {{}}{first and} second kind : {\it 'On the other hand it can also happen that the system is changed but in a controllable fashion by the measurement — even when, in the state before the measurement, the quantity measured had with certainty a definite value. In this method, the result of a repeated measurement is not the same as that of the first measurement. But still it may be that from the result of this measurement, an un-ambiguous conclusion can be drawn regarding the quantity being measured for the concerned system before the measurement. Such measurements, we call the measurements of the second kind.'} \cite[p.75]{1958Pauli} To be concrete: (i) the determination of the position of a particle by a test of the presence of the particle in a given box appears to be  a measurement of the first kind, (ii) the determination of the momentum of a particle by the evaluation of the 'impact' of this particle on a given detector appears to be a measurement of the second kind. }
\end{quote}
We will adopt the following formal definition for this type of {{}}{measurement}.
\begin{MyDef} 
A yes/no test ${ \mathfrak{t}}\in { \mathfrak{T}}$ is said to lead to a {\em first{{}}{-}kind measurement} associated {{}}{with} the corresponding testable property $\lfloor { \mathfrak{t}} \rfloor$ iff (i) a positive result {{}}{of} the test ${ \mathfrak{t}}$ realized on any input sample is necessarily confirmed by an immediate repetition of this test realized on the samples {{}}{outcoming} from the first test, and (ii) the observer cannot establish if this new 'check' has been {{}}{performed} or not on the basis of the new tests {{}}{that could be performed} on the {{}}{outcoming} samples of the experiment.  In other words, the testable property $\lfloor { \mathfrak{t}} \rfloor$ can be considered as 'actual' after the measurement by ${ \mathfrak{t}}$, and the actuality of the property $\lfloor { \mathfrak{t}} \rfloor$ can be 'confirmed' through any repetition of the measurement by ${ \mathfrak{t}}$, because this repeated measurement {{}}{leaves} the state of the system {{}}{unchanged}.\\  
The subset of yes/no tests leading to first-kind measurements will be denoted ${ \mathfrak{T}}_{\!{}_{FKM}}$. In other words,
\begin{eqnarray}
\forall { \mathfrak{t}}\in { \mathfrak{T}},&&{ \mathfrak{t}}\in { \mathfrak{T}}_{\!{}_{FKM}}\;\; :\Leftrightarrow\;\;   %(\, 
\left\{
\begin{array}{ll}
\textit{\rm (i)} &\forall { \mathfrak{p}}\in { \mathfrak{Q}}_{ \mathfrak{t}},\;\; ({ \mathfrak{p}} {\centerdot} { \mathfrak{t}})\in { \mathfrak{A}}_{ \mathfrak{t}}\\
\textit{\rm (ii)} &\forall { \mathfrak{p}}\in { \mathfrak{A}}_{ \mathfrak{t}},\;\; ({ \mathfrak{p}} {\centerdot} { \mathfrak{t}}) \sim_{{}_{{ \mathfrak{P}}}} { \mathfrak{p}}
%\,)
\end{array}
\right.
\label{firstkind}
\end{eqnarray} 
\end{MyDef} 
\begin{lemme}
In terms of the action on the space of states, we then have (using {\bf Notion \ref{actionmeasurementsonstates}}) :
\begin{eqnarray}
\forall { \mathfrak{t}}\in { \mathfrak{T}}_{\!{}_{FKM}}, && %(\, 
\left\{
\begin{array}{ll}
\textit{\rm (i)} &\forall \sigma \in \lceil { \mathfrak{Q}}_{ \mathfrak{t}}\rceil,\;\; (\sigma {\centerdot} { \mathfrak{t}})\in \lceil { \mathfrak{A}}_{ \mathfrak{t}}\rceil \\
\textit{\rm (ii)} &\forall \sigma \in \lceil { \mathfrak{A}}_{ \mathfrak{t}}\rceil ,\;\; (\sigma {\centerdot} { \mathfrak{t}}) = \sigma
%\,)
\end{array}
\right.
\label{firstkindstates}
\end{eqnarray} 
\end{lemme}

As a remark, we also note the following trivial lemmas.
\begin{lemme}{}
\begin{eqnarray}
\forall { \mathfrak{t}}\in { \mathfrak{T}}_{\!{}_{FKM}},&& ({ \mathfrak{p}}_{{ \bot}} {\centerdot} { \mathfrak{t}}) \in { \mathfrak{A}}_{ \mathfrak{t}} \not= \varnothing\label{tfkmatnonempty}\\
\forall { \mathfrak{t}}\in { \mathfrak{T}}_{\!{}_{FKM}}, \forall { \mathfrak{p}}\in { \mathfrak{Q}}_{ \mathfrak{t}},&& (({ \mathfrak{p}}{\centerdot} { \mathfrak{t}}){\centerdot} { \mathfrak{t}}) \sim_{{}_{{ \mathfrak{P}}}} ({ \mathfrak{p}}{\centerdot} { \mathfrak{t}}) \label{tfkmidempotent}\\
\forall { \mathfrak{t}}\in { \mathfrak{T}}_{\!{}_{FKM}}, \forall { \mathfrak{p}}\in { \mathfrak{Q}}_{ \mathfrak{t}},&& (\, ({ \mathfrak{p}} {\centerdot} { \mathfrak{t}}) \sim_{{}_{{ \mathfrak{P}}}} { \mathfrak{p}}\,) \;\;\Rightarrow \;\; { \mathfrak{p}}\in { \mathfrak{A}}_{ \mathfrak{t}}\label{tfkmidempotent2}\\
\forall { \mathfrak{t}}\in { \mathfrak{T}}_{\!{}_{FKM}},&&  \lceil { \mathfrak{A}}_{ \mathfrak{t}}\rceil =Im^{ \mathfrak{S}}_{{\centerdot} { \mathfrak{t}}} =\{\, \sigma \in { \mathfrak{S}} \;\vert\;  \sigma=(\sigma {\centerdot} { \mathfrak{t}}) \,\}.\label{tfkmAtImFix}
\end{eqnarray} 
\end{lemme}
%\begin{proof} The equations (\ref{tfkmidempotent})(\ref{tfkmidempotent2})are obtained by combining straightforwardly the properties (\ref{firstkind} \textit{\rm (i)}) and (\ref{firstkind} \textit{\rm (ii)}).\\
%(\ref{firstkind} \textit{\rm (i)}) can be rewritten $Im^{ \mathfrak{P}}_{{\centerdot} { \mathfrak{t}}} \subseteq { \mathfrak{A}}_{ \mathfrak{t}}$ and (\ref{firstkind} \textit{\rm (ii)}) can be rewritten ${ \mathfrak{A}}_{ \mathfrak{t}} \subseteq \{\, { \mathfrak{p}}\in  { \mathfrak{P}}\;\vert\; ({ \mathfrak{p}} {\centerdot} { \mathfrak{t}}) \sim_{{}_{{ \mathfrak{P}}}} { \mathfrak{p}}\,\}$. Moreover, from (\ref{tfkmidempotent}) we know that $\forall { \mathfrak{p}}\in { \mathfrak{Q}}_{ \mathfrak{t}}$, $({ \mathfrak{p}} {\centerdot} { \mathfrak{t}}) \sim_{{}_{{ \mathfrak{P}}}} { \mathfrak{p}}$ implies $(\lceil { \mathfrak{p}}\rceil  {\centerdot} { \mathfrak{t}}) = \lceil { \mathfrak{p}}\rceil $, i.e. $ \lceil { \mathfrak{p}}\rceil \in Im^{ \mathfrak{S}}_{{\centerdot} { \mathfrak{t}}}$. We then obtain the equality (\ref{tfkmAtImFix}).
 %\end{proof}

 \begin{MyDef}
The subset of testable properties that can be tested through first-kind measurements will be denoted ${ \mathcal{L}}_{{}_{FKM}}$. The definition of this subset of ${ \mathcal{L}}$ is then summarized as follows
\begin{eqnarray}
{ \mathfrak{l}}\in { \mathcal{L}}_{{}_{FKM}} &:\Leftrightarrow &
\exists ({\centerdot} { \mathfrak{t}}) \in \left[ { \mathfrak{S}} \dashrightarrow { \mathfrak{S}}\right]^\barwedge\;\;\vert\;\;
\left\{
\begin{array}{l}
Dom^{ \mathfrak{S}}_{{\centerdot} { \mathfrak{t}}}=\lceil { \mathfrak{Q}}_{{ \mathfrak{l}}}\rceil\\
Im^{ \mathfrak{S}}_{{\centerdot} { \mathfrak{t}}}=\lceil { \mathfrak{A}}_{{ \mathfrak{l}}}\rceil= (\uparrow^{{}^{ \mathfrak{S}}}\!\!{\Sigma}_{{ \mathfrak{l}}})\\
\forall \sigma \in Dom^{ \mathfrak{S}}_{{\centerdot} { \mathfrak{t}}},\;\; (\sigma {\centerdot} { \mathfrak{t}}){\centerdot} { \mathfrak{t}}=\sigma {\centerdot} { \mathfrak{t}}
\end{array}
\right. \label{axiommeasurementmapfkm}
\end{eqnarray}
\end{MyDef}

In order to pronounce synthetic statements concerning the actuality of a 'collection' of properties for a given state, it appears necessary to clarify how 'successive' measurements of different properties can be used to pronounce the actuality of these properties 'conjointly' for a given sample. Indeed, an {{}}{inconvenient}
aspect of quantum measurements emerges when different properties are tested successively on a given sample: it generically happens that the effective measurements associated {{}}{with} different properties 'interfere', {{}}{prohibiting} the actuality {{}}{of} two such 'incompatible' properties to be affirmed simultaneously for a given preparation process (in this context, the definition of first-kind measurements appears to be insufficient for our purpose, {{}}{as} this subset of measurement maps is not closed under the succession of distinct measurement operations).\\
This limitation of effective measurement leads to no-go theorems in the pathway to the construction of a classical logic for the description of the properties of the system, as {{}}{noted} originally by E.Specker :
\begin{quote}
{\it 'Is it possible to extend the description of a quantum mechanical system through the introduction of supplementary — fictitious — propositions in such a way that in the extended domain the classical propositional logic holds (whereby, of course, for simultaneously decidable proposition negation, conjunction and disjunction must retain their meaning)? The answer to this question is negative, except in the case of Hilbert spaces of dimension 1 and 2.'}\cite{Specker}
\end{quote}
The origins of this puzzling fact can be presented in terms of concrete measurements {{}}{performed} on the system. Let us consider any preparation procedure ${ \mathfrak{p}}\in { \mathfrak{P}}$ {{}}{guaranteeing} the actuality of a given property ${ \mathfrak{l}}\in { \mathcal{L}}$ for its outcoming samples (i.e., ${ \mathfrak{p}}\in { \mathfrak{A}}_{{ \mathfrak{l}}}$). The actuality of this property may then be checked by applying any yes/no test ${ \mathfrak{t}}\in { \mathfrak{T}}$ associated {{}}{with} this property (i.e., ${ \mathfrak{l}}=\lfloor { \mathfrak{t}}\rfloor$) on these prepared samples before any other experiment as {{}}{long} as this yes/no test leads to a first-kind measurement, i.e., $({ \mathfrak{p}}{\centerdot} { \mathfrak{t}})\in { \mathfrak{A}}_{{ \mathfrak{l}}}$. However, {{}}{there} generically exists another property ${ \mathfrak{l}}'\in { \mathcal{L}}$ (and a yes/no test ${ \mathfrak{t}}'\in { \mathfrak{T}}$ associated {{}}{with} it, i.e., ${ \mathfrak{l}}'=\lfloor { \mathfrak{t}}'\rfloor$) such that:
(i) the result of any yes/no test associated {{}}{with} this second property on these prepared samples is fundamentally indeterminate (i.e., ${ \mathfrak{e}}({ \mathfrak{p}}, { \mathfrak{t}}')={ \bot}$), (ii) if we select among the {{}}{outcomes} of these yes/no tests the samples exhibiting this second property (i.e., the samples produced {{}}{through} the preparation procedure $({ \mathfrak{p}}{\centerdot} { \mathfrak{t}}')\in { \mathfrak{A}}_{{ \mathfrak{l}}'}$), then{{}}{ }any {{}}{subsequent}
measurement relative to the first property on these selected samples will exhibit an indeterminacy (i.e., ${ \mathfrak{e}}(({ \mathfrak{p}}{\centerdot} { \mathfrak{t}}'), { \mathfrak{t}})={ \bot}$). In other words, the actuality of the first property can {{}}{no longer} be affirmed{{}}{ }(i.e., $({ \mathfrak{p}}{\centerdot} { \mathfrak{t}}')\notin { \mathfrak{A}}_{{ \mathfrak{l}}}$) after the actuality of the second property has been effectively established by a measurement, even if the actuality of the first property had been established beforehand on these same prepared samples! {{}}{To summarize}, contrary to the context of classical measurements, for ${ \mathfrak{p}}\in { \mathfrak{P}}$ and ${ \mathfrak{t}}\in { \mathfrak{T}}$, we can not affirm that ${ \mathfrak{p}}\sqsubseteq_{{}_{ \mathfrak{P}}}({ \mathfrak{p}}{\centerdot} { \mathfrak{t}})$.\\
Despite this severe limitation on the determination process of the actual properties of a quantum system, it is however possible to singularize some measurements, chosen for their ability to '{{}}{minimally perturb} the system'. C.Piron summarizes the proposal for these ideal measurements as follows : 
\begin{quote}
{\it 'In general if we test a property ${\mathfrak{a}}$ by performing $\alpha$, one of the corresponding questions, we disturb completely the given physical system even if ${\mathfrak{a}}$ is actually true. We will say that a question $\alpha$ is an ideal measurement if, when we perform it, we can assert that (i) If the answer is "yes", then the corresponding proposition ${\mathfrak{a}}$ is true afterwards, and (ii) If the answer is "yes" and if a property ${\mathfrak{b}}$ is true and compatible with ${\mathfrak{a}}$, then ${\mathfrak{b}}$ is still true afterwards.'}\cite{Piron1981} 
\end{quote}
%Shortly, these measurements will be chosen such that: any property that is stated as actual before the measurement, and which is "compatible" with the property fixed by the measurement, will remain actual after this measurement. \\
Nevertheless, the definitions of 'compatible properties' and{{}}{ }'ideal first-kind measurements' seem to be trapped in a vicious circle: two properties are {\em compatible} as soon as they can be 'simultaneously' stated as actual using successive ideal first-kind measurements, and measurements are defined to be ideal as soon as they respect the actuality of the properties that are compatible with the measured property!\footnote{A basic solution to this problem has been formalized by C. Piron \cite{Piron1981}. This construction relies on an orthomodular lattice structure introduced on the space of properties (Note: we do not expect any such a construction in our perspective). In Piron's vocabulary, 
\begin{quote}
two  properties are compatible as soon as they form a boolean sub-algebra in the orthomodular lattice of properties (this requirement about the sub- boolean structure is a remnant of the particular structure on the space of properties in the classical situation).\cite[p.295]{Piron1972}
\end{quote}
}
In order to establish a consistent description, it appears necessary to clarify these notions in our vocabulary. 

\begin{MyDef} 
A family of testable properties ${ \mathfrak{L}}=(\,{ \mathfrak{l}}_i)_{ i\in I}\subseteq { \mathcal{L}}$  will be said to be {\em a compatible family of properties} (this fact will be denoted $\overbrace{\,{ \mathfrak{L}}\;\,}$), iff %(i) any collection of samples, similarly prepared according to a given preparation process ${ \mathfrak{p}} \in { \mathfrak{P}}$, and exhibiting one of these properties as 'actual' necessarily exhibits the other properties of this family as 'potential', and (ii) 
{{}}{there} exists at least one preparation process ${ \mathfrak{p}} \in { \mathfrak{P}}$ producing collections of samples exhibiting all of these properties as 'actual' (the statements about the occurrence of the properties $(\,{ \mathfrak{l}}_i)_{ i\in I}$ for the samples prepared {{}}{through} ${ \mathfrak{p}}$ will all be simultaneously 'positive with certainty'). In other words,
\begin{eqnarray}
% \forall { \mathfrak{L}}\subseteq_{fin}^{\not= \varnothing} { \mathcal{L}},\;\;\;\;\;  
 \overbrace{\,{ \mathfrak{L}}\;\,} 
 & :\Leftrightarrow & \bigcap_{{ \mathfrak{l}}\in { \mathfrak{L}}} { \mathfrak{A}}_{ \mathfrak{l}} \not= \varnothing %\;\;\textit{\rm and}\;\; \bigcup_{{ \mathfrak{l}}\in { \mathfrak{L}}} { \mathfrak{A}}_{ \mathfrak{l}} \subseteq \bigcap_{{ \mathfrak{l}}\in { \mathfrak{L}}} { \mathfrak{Q}}_{ \mathfrak{l}}.
 \label{defconsistentfamilyproperty}
\end{eqnarray}
In particular, a property ${ \mathfrak{l}}_1$ is said to be compatible with a property ${ \mathfrak{l}}_2$ iff $\overbrace{{ \mathfrak{l}}_1 { \mathfrak{l}}_2}$. This defines a binary relation called {{}}{the} {\em compatibility relation} on ${ \mathcal{L}}$.\\ The compatibility relation is a reflexive and symmetric relation.  Moreover, ${ \mathfrak{t}}_1 \sqsubseteq_{{}_{ \mathfrak{P}}} { \mathfrak{t}}_2$ implies $\overbrace{\lfloor { \mathfrak{t}}_1\rfloor \lfloor { \mathfrak{t}}_2\rfloor }$.
\end{MyDef} 

The {\em ideal measurements} will be characterized as follows. 

\begin{MyDef} \label{defidealmeasurement}
A yes/no test ${ \mathfrak{t}}_1\in { \mathfrak{T}}$ is said to {\em lead to an ideal measurement} of the corresponding testable property $\lfloor { \mathfrak{t}}_1 \rfloor$ (this fact is denoted ${ \mathfrak{t}}_1\in { \mathfrak{T}}_{{}_{Ideal}}$) iff, for any property $\lfloor { \mathfrak{t}}_2 \rfloor$ compatible with $\lfloor { \mathfrak{t}}_1 \rfloor$, the statement pronounced by the observer beforehand concerning the 'actuality' of the property $\lfloor { \mathfrak{t}}_2 \rfloor$ is conserved after the measurement operation associated {{}}{with} ${ \mathfrak{t}}_1$ has been realized, i.e.,
\begin{eqnarray}
\forall { \mathfrak{t}}_1\in { \mathfrak{T}},\;\;\;{ \mathfrak{t}}_1\in { \mathfrak{T}}_{{}_{Ideal}} & :\Leftrightarrow & 
(\,\forall { \mathfrak{t}}_2 \in { \mathfrak{T}}\;\vert\;\overbrace{\lfloor { \mathfrak{t}}_1 \rfloor \lfloor { \mathfrak{t}}_2 \rfloor }, \forall  { \mathfrak{p}}\in { \mathfrak{Q}}_{{ \mathfrak{t}}_1},\; { \mathfrak{e}}({ \mathfrak{p}},{ \mathfrak{t}}_2)=\textit{\bf Y}\;\Rightarrow\;  { \mathfrak{e}}({ \mathfrak{p}}{\centerdot} { \mathfrak{t}}_1,{ \mathfrak{t}}_2)=\textit{\bf Y}\,)\;\;\;\;\;\;\;\;\;\;\;\;\\
&\Leftrightarrow & (\, \forall { \mathfrak{t}}_2 \in { \mathfrak{T}},\;\;\overbrace{\lfloor { \mathfrak{t}}_1 \rfloor \lfloor { \mathfrak{t}}_2 \rfloor }\;\Rightarrow\;  ({ \mathfrak{Q}}_{{ \mathfrak{t}}_1} \cap { \mathfrak{A}}_{{ \mathfrak{t}}_2} ) {\centerdot} { \mathfrak{t}}_1 \subseteq  { \mathfrak{A}}_{{ \mathfrak{t}}_2} \,)   \label{axiomidealmeasurement}
\end{eqnarray}
We will adopt the following notation
\begin{eqnarray}
{ \mathfrak{T}}_{{}_{IFKM}}  &:=& { \mathfrak{T}}_{{}_{Ideal}} \cap { \mathfrak{T}}_{{}_{FKM}}.
\end{eqnarray}
\end{MyDef} 
\begin{remark}
When an ideal measurement operation associated {{}}{with} a yes/no test ${ \mathfrak{t}}_1$ is {{}}{performed} on a given collection of samples{{}}{,} similarly prepared in such a way that {{}}{ }property $\lfloor { \mathfrak{t}}_2 \rfloor$ was actual before this measurement operation, the outcoming samples are such that the observer {{}}{cannot} distinguish if {{}}{ } property $\lfloor { \mathfrak{t}}_2 \rfloor$ {{}}{was tested (or not)} after this measurement{{}}{. In effect}, 
\begin{eqnarray}
\forall { \mathfrak{t}}_1, { \mathfrak{t}}_2 \in { \mathfrak{T}}_{{}_{IFKM}} \;\vert\;  \overbrace{\lfloor { \mathfrak{t}}_1 \rfloor \lfloor { \mathfrak{t}}_2\rfloor} && \forall { \mathfrak{p}}\in { \mathfrak{A}}_{{ \mathfrak{t}}_2}\cap { \mathfrak{Q}}_{{ \mathfrak{t}}_1},\;\;\;\;\;\; { \mathfrak{p}}{\centerdot} { \mathfrak{t}}_1{\centerdot} { \mathfrak{t}}_2 \sim_{{}_{{ \mathfrak{P}}}} { \mathfrak{p}}{\centerdot} { \mathfrak{t}}_1.
\end{eqnarray}
\end{remark}

\begin{MyDef}
The subset of testable properties that can be tested through ideal first-kind measurements will be denoted ${ \mathcal{L}}_{{}_{IFKM}}$. The definition of this subset of ${ \mathcal{L}}$ is then summarized as follows
\begin{eqnarray}
{ \mathfrak{l}}\in { \mathcal{L}}_{{}_{IFKM}} &:\Leftrightarrow &
\exists ({\centerdot} { \mathfrak{t}}) \in \left[ { \mathfrak{S}} \dashrightarrow { \mathfrak{S}}\right]^\barwedge\;\;\vert\;\;
\left\{
\begin{array}{l}
Dom^{ \mathfrak{S}}_{{\centerdot} { \mathfrak{t}}}=\lceil { \mathfrak{Q}}_{{ \mathfrak{l}}}\rceil\\
Im^{ \mathfrak{S}}_{{\centerdot} { \mathfrak{t}}}=\lceil { \mathfrak{A}}_{{ \mathfrak{l}}}\rceil= (\uparrow^{{}^{ \mathfrak{S}}}\!\!{\Sigma}_{{ \mathfrak{l}}})\\
\forall \sigma \in Dom^{ \mathfrak{S}}_{{\centerdot} { \mathfrak{t}}},\;\; (\sigma {\centerdot} { \mathfrak{t}}){\centerdot} { \mathfrak{t}}=\sigma {\centerdot} { \mathfrak{t}}\\
\forall \sigma \in { \mathfrak{S}}\;\vert\;\widehat{\;\sigma \Sigma_{{ \mathfrak{l}}}}{}^{{}^{ \mathfrak{S}}},\; ( Dom^{ \mathfrak{S}}_{{\centerdot} { \mathfrak{t}}} \cap (\uparrow^{{}^{ \mathfrak{S}}}\!\!\!\! \sigma)) {\centerdot} { \mathfrak{t}} \subseteq  ( (\uparrow^{{}^{ \mathfrak{S}}}\!\!\!\! \sigma) \cap \lceil { \mathfrak{A}}_{{ \mathfrak{l}}}\rceil)
\end{array}
\right. \label{axiommeasurementmapifkm}
\end{eqnarray}
\end{MyDef}

The motivation to introduce 'ideal first-kind measurements' is very clear from an operational point of view.\footnote{These measurements {{}}{played} a fundamental role in {{}}{Mackey's traditional} axiomatic approach to quantum theory \cite{Cassinelli1975}{{}}{.}} 
If such measurement operations exist for a basic set of compatible properties, they can be used to build preparation processes {{}}{designed} to produce {{}}{collections} of samples for which these properties {{}}{will be found to be} 
'conjointly actual'. It is also completely clear that ideal first-kind measurements exist in concrete quantum mechanical experiments \cite{PhysRevLett.124.080401} and it is then natural to impose their existence at the center of a quantum axiomatics. 
Nevertheless, nothing guaranties that such a measurement operation exists for any given property ${ \mathfrak{l}}$ (or, equivalently, that a measurement map satisfying property (\ref{axiommeasurementmapifkm}) exists for the property ${ \mathfrak{l}}$). 
Our aim, {{}}{by} the end {{}}{of} the present subsection, will be to prove that an ideal first-kind measurement map exists{{}}{if} 
the corresponding 'property' satisfies a quasi-classicality criterion.

\begin{lemme}{}\label{t1dott2IFKM}
For any ${ \mathfrak{t}}_1,{ \mathfrak{t}}_2\in { \mathfrak{T}}_{{}_{IFKM}}$ two yes/no tests leading to ideal first-kind measurements such that ${ \mathfrak{t}}_1$ is compatible with ${ \mathfrak{t}}_2$, 
 the yes/no test $({ \mathfrak{t}}_1 {\centerdot} { \mathfrak{t}}_2)$ leads to ideal first-kind measurements associated {with} the conjunction of the properties $\lfloor { \mathfrak{t}}_1 \rfloor$ and $\lfloor { \mathfrak{t}}_2 \rfloor$. In other words,
\begin{eqnarray}
\forall { \mathfrak{t}}_1,{ \mathfrak{t}}_2 \in { \mathfrak{T}}_{{}_{IFKM}}\;\vert\;\overbrace{\lfloor { \mathfrak{t}}_1 \rfloor \lfloor { \mathfrak{t}}_2 \rfloor }, &&({ \mathfrak{t}}_1 {\centerdot} { \mathfrak{t}}_2) \in  { \mathfrak{T}}_{{}_{IFKM}}\\
&&{ \mathfrak{A}}_{{ \mathfrak{t}}_1 {\centerdot} { \mathfrak{t}}_2} = ({ \mathfrak{A}}_{{ \mathfrak{t}}_1}\cap { \mathfrak{A}}_{{ \mathfrak{t}}_2})\label{At1dott2IFKM},\\
&&Dom^{ \mathfrak{S}}_{{\centerdot} ({{ \mathfrak{t}}_1}{\centerdot} { \mathfrak{t}}_2)}={ \mathfrak{Q}}_{{{ \mathfrak{t}}_1}{\centerdot} { \mathfrak{t}}_2} 
={ \mathfrak{Q}}_{{ \mathfrak{t}}_1} \cap ({\centerdot} { \mathfrak{t}}_1)^{-1} ({ \mathfrak{Q}}_{{ \mathfrak{t}}_2} \cap { \mathfrak{A}}_{{ \mathfrak{t}}_1}),\label{Qt1dott2IFKM}\\
&&{ \mathfrak{A}}_{\overline{{{ \mathfrak{t}}_1}{\centerdot} { \mathfrak{t}}_2}} = { \mathfrak{A}}_{\overline{{ \mathfrak{t}}_1}} \cup ({\centerdot} { \mathfrak{t}}_1)^{-1} ({ \mathfrak{A}}_{\overline{{ \mathfrak{t}}_2}} \cap { \mathfrak{A}}_{{ \mathfrak{t}}_1})= { \mathfrak{S}}\smallsetminus { \mathfrak{Q}}_{{{{ \mathfrak{t}}_1}{\centerdot} { \mathfrak{t}}_2}}\label{Abart1dott2IFKM}
\end{eqnarray}
\end{lemme} 
\begin{proof}
Using equations (\ref{At1dott2}) and (\ref{firstkind} (ii)), we deduce (\ref{At1dott2IFKM}). Let us now prove that $({ \mathfrak{t}}_1 {\centerdot} { \mathfrak{t}}_2)\in  { \mathfrak{T}}_{{}_{IFKM}}$.
For any ${ \mathfrak{p}} \in { \mathfrak{Q}}_{({ \mathfrak{t}}_1 {\centerdot} { \mathfrak{t}}_2)}$, we know from equation (\ref{Qt1dott2}) that $({ \mathfrak{p}} {\centerdot} { \mathfrak{t}}_1)\in { \mathfrak{A}}_{{ \mathfrak{t}}_1} \cap { \mathfrak{Q}}_{{ \mathfrak{t}}_2}$. Using \underline{ }equations  (\ref{firstkind} (i)) and (\ref{axiomidealmeasurement}), we then deduce that $(({ \mathfrak{p}} {\centerdot} { \mathfrak{t}}_1){\centerdot} { \mathfrak{t}}_2)\in { \mathfrak{A}}_{{ \mathfrak{t}}_1} \cap { \mathfrak{A}}_{{ \mathfrak{t}}_2}={ \mathfrak{A}}_{({ \mathfrak{t}}_1 {\centerdot} { \mathfrak{t}}_2)}$ (using (\ref{At1dott2IFKM})), i.e., equation (\ref{firstkind} (i)) applied to the yes/no test $({ \mathfrak{t}}_1 {\centerdot} { \mathfrak{t}}_2)$. For any ${ \mathfrak{p}} \in { \mathfrak{A}}_{({ \mathfrak{t}}_1 {\centerdot} { \mathfrak{t}}_2)}=({ \mathfrak{A}}_{{ \mathfrak{t}}_1}\cap { \mathfrak{A}}_{{ \mathfrak{t}}_2})$  (using (\ref{At1dott2IFKM})), we deduce from {equations} (\ref{firstkind} (ii)) and (\ref{pt1dott2}) that $({ \mathfrak{p}} {\centerdot} ({ \mathfrak{t}}_1 {\centerdot} { \mathfrak{t}}_2)) = (({ \mathfrak{p}} {\centerdot} { \mathfrak{t}}_1) {\centerdot} { \mathfrak{t}}_2) \sim_{{}_{{ \mathfrak{P}}}} ({ \mathfrak{p}} {\centerdot} { \mathfrak{t}}_2) \sim_{{}_{{ \mathfrak{P}}}} { \mathfrak{p}}$, i.e., equation (\ref{firstkind} (ii)) applied to the yes/no test $({ \mathfrak{t}}_1 {\centerdot} { \mathfrak{t}}_2)$. Let us now consider a yes/no test ${ \mathfrak{t}} \in { \mathfrak{T}}$ compatible with the yes/no test $({ \mathfrak{t}}_1 {\centerdot} { \mathfrak{t}}_2)$. In other words, we have ${ \mathfrak{A}}_{{ \mathfrak{t}}} \cap { \mathfrak{A}}_{({ \mathfrak{t}}_1 {\centerdot} { \mathfrak{t}}_2)}\not= \varnothing$. As a result, we have then ${ \mathfrak{A}}_{{ \mathfrak{t}}} \cap { \mathfrak{A}}_{{ \mathfrak{t}}_1}\not= \varnothing$ and ${ \mathfrak{A}}_{{ \mathfrak{t}}} \cap { \mathfrak{A}}_{{ \mathfrak{t}}_2}\not= \varnothing$, i.e., $\overbrace{\lfloor{ \mathfrak{t}}\rfloor\lfloor{ \mathfrak{t}}_1\rfloor}$ and $\overbrace{\lfloor{ \mathfrak{t}}\rfloor\lfloor{ \mathfrak{t}}_2\rfloor}$. Using successively (\ref{pt1dott2}), (\ref{Qt1dott2}), {and} the compatibility relations $\overbrace{\lfloor{ \mathfrak{t}}\rfloor\lfloor{ \mathfrak{t}}_1\rfloor}$ and $\overbrace{\lfloor{ \mathfrak{t}}\rfloor\lfloor{ \mathfrak{t}}_2\rfloor}$ coupled with\underline{ }property (\ref{axiomidealmeasurement}), we obtain 
$({ \mathfrak{Q}}_{({ \mathfrak{t}}_1 {\centerdot} { \mathfrak{t}}_2)} \cap { \mathfrak{A}}_{{ \mathfrak{t}}} ) {\centerdot}({{ \mathfrak{t}}_1 {\centerdot} { \mathfrak{t}}_2}) \subseteq   (({ \mathfrak{Q}}_{{ \mathfrak{t}}_2} \cap { \mathfrak{A}}_{{ \mathfrak{t}}} ) {\centerdot} { \mathfrak{t}}_2)
\subseteq  { \mathfrak{A}}_{{ \mathfrak{t}}}$, i.e. property (\ref{axiomidealmeasurement}) for the compatibility property $\overbrace{\;\lfloor{ \mathfrak{t}}\rfloor\;\lfloor({{ \mathfrak{t}}_1 {\centerdot} { \mathfrak{t}}_2})\rfloor}$. \\
Properties (\ref{Qt1dott2IFKM}) and (\ref{Abart1dott2IFKM}) are explicit rewriting of properties (\ref{Qt1dott2}) and (\ref{Abart1dott2}).
\end{proof}   

\begin{MyDef}
A preparation process ${ \mathfrak{p}}\in { \mathfrak{P}}$ is said to be {\em consistent with the actuality of a given testable property ${ \mathfrak{l}}\in { \mathcal{L}}$} iff {{}}{there} exists a preparation process ${ \mathfrak{p}}'\in { \mathfrak{P}}$,  sharper than ${ \mathfrak{p}}$, and for which the property ${ \mathfrak{l}}$ is actual. We denote by ${ \mathfrak{K}}_{ \mathfrak{l}}$ the set of preparation processes consistent with the actuality of the testable property ${ \mathfrak{l}}$, i.e., 
\begin{eqnarray}
\forall { \mathfrak{l}}\in { \mathcal{L}},&& { \mathfrak{K}}_{ \mathfrak{l}}:=\{\,{ \mathfrak{p}}\in { \mathfrak{P}} \;\vert\; \exists { \mathfrak{p}}'\in { \mathfrak{A}}_{ \mathfrak{l}},\; { \mathfrak{p}} \sqsubseteq_{{}_{ \mathfrak{P}}} { \mathfrak{p}}' \,\}.
\end{eqnarray}
$\lceil { \mathfrak{K}}_{ \mathfrak{l}}\rceil$ will be called  {\em the consistency{{}}{ }domain of the property ${ \mathfrak{l}}$}. We have 
\begin{eqnarray}
\forall { \mathfrak{l}}\in { \mathcal{L}},&& \lceil { \mathfrak{K}}_{ \mathfrak{l}}\rceil =\downarrow_{{}_{ \mathfrak{S}}}\!\!\!\lceil { \mathfrak{A}}_{ \mathfrak{l}}  \rceil =\{\,\sigma'\in { \mathfrak{S}} \;\vert\; \widehat{{\Sigma}_{ \mathfrak{l}} \sigma'}^{{}_{ \mathfrak{S}}}\,\}.
\end{eqnarray}
\end{MyDef}

\begin{lemme}{}\label{Klscottclosed}  
For any testable property ${ \mathfrak{l}}\in { \mathcal{L}}$,  the consistency domain $ \lceil { \mathfrak{K}}_{ \mathfrak{l}}\rceil$ is a Scott-closed subset of $\lceil { \mathfrak{Q}}_{ \mathfrak{l}}\rceil$.
\end{lemme}
\begin{proof} {{}}{First}, for any $\sigma'\in  \lceil { \mathfrak{K}}_{ \mathfrak{l}}\rceil$ the property $\widehat{{\Sigma}_{ \mathfrak{l}} \sigma'}^{{}_{ \mathfrak{S}}}$ implies $\exists \sigma''\in \lceil { \mathfrak{A}}_{ \mathfrak{l}} \rceil$ with $\sigma'\sqsubseteq_{{}_{ \mathfrak{S}}} \sigma''$, and {{}}{therefore} $\widetilde{ \mathfrak{e}}_{ \mathfrak{t}}(\sigma')\leq \widetilde{ \mathfrak{e}}_{ \mathfrak{t}}(\sigma'') =\textit{\bf Y}$ for any ${ \mathfrak{t}}$ such that ${ \mathfrak{l}}=\lfloor { \mathfrak{t}}\rfloor$. As a consequence, $ \lceil { \mathfrak{K}}_{ \mathfrak{l}}\rceil\subseteq \lceil { \mathfrak{Q}}_{ \mathfrak{l}}\rceil $. Moreover, $ \lceil { \mathfrak{K}}_{ \mathfrak{l}}\rceil$ is obviously a downset. {{}}{Lastly}, let us consider ${ \mathfrak{C}} \subseteq_{Chain} { \mathfrak{S}}$ such that $(\, \forall { \mathfrak{c}}\in { \mathfrak{C}},\;  \widehat{{\Sigma}_{ \mathfrak{l}}{ \mathfrak{c}}}^{{}_{ \mathfrak{S}}}\,)$. For any ${ \mathfrak{c}}\in { \mathfrak{C}}$, we can define the element $({\Sigma}_{ \mathfrak{l}} \sqcup_{{}_{ \mathfrak{S}}}{ \mathfrak{c}})$. The chain ${ \mathfrak{C}}':=\{\, {\Sigma}_{ \mathfrak{l}} \sqcup_{{}_{ \mathfrak{S}}}{ \mathfrak{c}}\;\vert\; { \mathfrak{c}}\in { \mathfrak{C}}\,\}$ admits a supremum in ${ \mathfrak{S}}$ satisfying (i) $\forall { \mathfrak{c}}'\in { \mathfrak{C}}',\; {\Sigma}_{ \mathfrak{l}} \sqsubseteq_{{}_{ \mathfrak{S}}} { \mathfrak{c}}'$ and {{}}{thus} ${\Sigma}_{ \mathfrak{l}} \sqsubseteq_{{}_{ \mathfrak{S}}} \bigsqcup_{{}_{ \mathfrak{S}}}{ \mathfrak{C}}'$, and (ii) $\forall { \mathfrak{c}}\in { \mathfrak{C}},\; \exists { \mathfrak{c}}'\in { \mathfrak{C}}'\;\vert\; { \mathfrak{c}} \sqsubseteq_{{}_{ \mathfrak{S}}} { \mathfrak{c}}'$ and {{}}{thus} ${ \mathfrak{c}} \sqsubseteq_{{}_{ \mathfrak{S}}} \bigsqcup_{{}_{ \mathfrak{S}}}{ \mathfrak{C}}'$ and {{}}{therefore} $\bigsqcup_{{}_{ \mathfrak{S}}}{ \mathfrak{C}}\sqsubseteq_{{}_{ \mathfrak{S}}} \bigsqcup_{{}_{ \mathfrak{S}}}{ \mathfrak{C}}'$. As a result, we have obtained $(\, \widehat{{\Sigma}_{ \mathfrak{l}} \bigsqcup{}_{{}_{ \mathfrak{S}}}\!{ \mathfrak{C}}}^{{}_{ \mathfrak{S}}} \,)$ and {{}}{thus} $(\bigsqcup_{{}_{ \mathfrak{S}}}\!{ \mathfrak{C}})\in  \lceil { \mathfrak{K}}_{ \mathfrak{l}}\rceil$. This chain-completeness property implies the directed{{}}{ }completeness of $ \lceil { \mathfrak{K}}_{ \mathfrak{l}}\rceil$.
\end{proof}

\begin{lemme}{}\label{lemmathetal}
For any testable property ${ \mathfrak{l}}\in { \mathcal{L}}$, the map 
\begin{eqnarray}
&&
\begin{array}{rcrcl}
{\theta}_{ \mathfrak{l}}& : & \lceil { \mathfrak{K}}_{ \mathfrak{l}}\rceil & \rightarrow &\lceil { \mathfrak{A}}_{ \mathfrak{l}}\rceil\\
& & \sigma & \mapsto &  {\theta}_{ \mathfrak{l}}(\sigma) :=  {\Sigma}_{ \mathfrak{l}} \sqcup_{{}_{ \mathfrak{S}}}\sigma.
\end{array}
\end{eqnarray}
is an idempotent, order-preserving, Scott-continuous partial map.  It preserves also filtered{{}}{ }infima and existing suprema. 
\end{lemme}
\begin{proof}
Firstly, from the basic properties of $\sqcup_{{}_{ \mathfrak{S}}}$, we know that ${\theta}_{ \mathfrak{l}}$ is idempotent and order{{}}{ }preserving. Secondly, if we denote {{}}{by} $\iota$ the inclusion of $\lceil { \mathfrak{A}}_{ \mathfrak{l}}\rceil$ in $\lceil { \mathfrak{K}}_{ \mathfrak{l}}\rceil$, we have $id \sqsubseteq \iota \circ {\theta}_{ \mathfrak{l}}$ and ${\theta}_{ \mathfrak{l}}\circ \iota =id$. As a result, the right{{}}{ }adjoint ${\theta}_{ \mathfrak{l}}$ of this Galois connection preserves existing suprema. In particular, ${\theta}_{ \mathfrak{l}}$ is Scott{{}}{ }continuous.\\
Thirdly, from the join-continuity property satisfied in ${ \mathfrak{S}}$ (Conjecture \ref{conjecturejoincontinuity}), we deduce that ${\theta}_{ \mathfrak{l}}$ preserves filtered{{}}{ }infima.\\
The idempotency is a direct consequence of the properties of $\sqcup_{{}_{ \mathfrak{S}}}$.
\end{proof}

\begin{MyDef}
A yes/no test ${ \mathfrak{t}}\in { \mathfrak{T}}$ is said to {\em lead to a minimally disturbing measurement} of the corresponding testable property $\lfloor { \mathfrak{t}} \rfloor$ (this fact is denoted ${ \mathfrak{t}}\in { \mathfrak{T}}_{{}_{min}}$) iff (i) this measurement is first-kind, and (ii) for any preparation ${ \mathfrak{p}}$ 'consistent with the actuality of $\lfloor { \mathfrak{t}} \rfloor$', the observer is {{}}{able} to pronounce statements about the measured state $\lceil { \mathfrak{p}} {\centerdot} { \mathfrak{t}}\rceil$ {{}}{that} are the minimal statements simultaneously finer than the statements pronounced separately about $\lceil { \mathfrak{p}} \rceil$ and about ${\Sigma}_{\lfloor { \mathfrak{t}} \rfloor}$ before the measurement. In other words,
\begin{eqnarray}
\forall { \mathfrak{t}}\in { \mathfrak{T}},&& { \mathfrak{t}}\in { \mathfrak{T}}_{{}_{min}} \; :\Leftrightarrow \;
\left\{\begin{array}{l}
Dom^{ \mathfrak{P}}_{{\centerdot} { \mathfrak{t}}}={ \mathfrak{Q}}_{\lfloor {\mathfrak{t}}\rfloor}\\
{ \mathfrak{t}} \in { \mathfrak{T}}_{{}_{FKM}}\\
\forall { \mathfrak{p}} \in { \mathfrak{K}}_{\lfloor {\mathfrak{t}}\rfloor},\;\;\lceil { \mathfrak{p}} {\centerdot} { \mathfrak{t}}\rceil = \lceil { \mathfrak{p}} \rceil  \sqcup_{{}_{ \mathfrak{S}}} {\Sigma}_{\lfloor {\mathfrak{t}}\rfloor}.
\end{array}\right.
\label{Tmin}
\end{eqnarray}
\end{MyDef}    

\begin{MyDef} 
The subset of properties that can be tested through minimally disturbing measurements will be denoted ${ \mathcal{L}}_{{}_{min}}$. In other words, 
\begin{eqnarray}
\forall { \mathfrak{l}}\in { \mathcal{L}},&& { \mathfrak{l}}\in { \mathcal{L}}_{{}_{min}} \; :\Leftrightarrow \;
\exists ({\centerdot} { \mathfrak{t}}) \in \left[ { \mathfrak{S}} \dashrightarrow { \mathfrak{S}}\right]^\barwedge\;\;\vert\;\;
\left\{\begin{array}{l}
Dom^{ \mathfrak{S}}_{{\centerdot} { \mathfrak{t}}}=\lceil { \mathfrak{Q}}_{{ \mathfrak{l}}}\rceil \\
Im^{ \mathfrak{S}}_{{\centerdot} { \mathfrak{t}}}=\lceil { \mathfrak{A}}_{{ \mathfrak{l}}}\rceil= (\uparrow^{{}^{ \mathfrak{S}}}\!\!{\Sigma}_{{ \mathfrak{l}}})\\
\forall \sigma \in Dom^{ \mathfrak{S}}_{{\centerdot} { \mathfrak{t}}},\;\; (\sigma {\centerdot} { \mathfrak{t}}){\centerdot} { \mathfrak{t}}=\sigma {\centerdot} { \mathfrak{t}}\\
\forall \sigma \in \lceil { \mathfrak{K}}_{{ \mathfrak{l}}}\rceil,\;\;\sigma {\centerdot} { \mathfrak{t}} = \sigma \sqcup_{{}_{ \mathfrak{S}}} {\Sigma}_{{ \mathfrak{l}}}.
\end{array}\right.\;\;\;\;\;\;\;\;\;\; \label{Lmin}
\end{eqnarray}
\end{MyDef}     
\begin{remark}
We note that $\sigma \in \lceil { \mathfrak{K}}_{{ \mathfrak{l}}}\rceil$ means $\widehat{\,\sigma  {\Sigma}_{{ \mathfrak{l}}}\,}^{{}_{ \mathfrak{S}}}$ and then the supremum $(\sigma \sqcup_{{}_{ \mathfrak{S}}} {\Sigma}_{{ \mathfrak{l}}})$ exists, due to the consistent{{}}{ }completeness of ${ \mathfrak{S}}$. 
\end{remark}

\begin{remark}\label{prooffkmi}
Note the following implicit property of minimally disturbing measurement maps :
\begin{eqnarray}
  { \mathfrak{t}}\in { \mathfrak{T}}_{{}_{min}}
 & \Rightarrow & Im^{ \mathfrak{P}}_{{\centerdot} { \mathfrak{t}}}={ \mathfrak{A}}_{\lfloor {\mathfrak{t}}\rfloor}
\end{eqnarray}
Let us{{}}{ }distinguish two sub-cases.\\
As a first sub-case, let us suppose that ${ \mathfrak{p}}\in (\downarrow_{{}_{ \mathfrak{P}}}{ \mathfrak{A}}_{ \mathfrak{t}})$.  Applying property (\ref{Tmin}), we deduce {{}}{immediately} that $\lceil { \mathfrak{p}}{\centerdot} { \mathfrak{t}}\rceil \sqsupseteq_{{}_{ \mathfrak{S}}} {\Sigma}_{ \mathfrak{t}}$, i.e. $\lceil { \mathfrak{p}}{\centerdot} { \mathfrak{t}}\rceil \in (\uparrow^{{}_{ \mathfrak{S}}}\!\!{\Sigma}_{ \mathfrak{t}}) = \lceil { \mathfrak{A}}_{{ \mathfrak{t}}}\rceil$, and {{}}{therefore} $({ \mathfrak{p}}{\centerdot} { \mathfrak{t}}) \in { \mathfrak{A}}_{{ \mathfrak{t}}}$.\\
As a second sub-case, we now suppose ${ \mathfrak{p}}\in { \mathfrak{Q}}_{ \mathfrak{t}}\!\smallsetminus \!(\downarrow_{{}_{ \mathfrak{P}}}{ \mathfrak{A}}_{ \mathfrak{t}})$. %From $\lfloor { \mathfrak{t}} \rfloor\in { \mathcal{L}}_{{}_{min}}={ \mathcal{L}}_{{}_{q-cl}}$, we know that the supremum $\bigsqcup_{{}_{ \mathfrak{S}}} (\, (\downarrow_{{}_{ \mathfrak{S}}} \lceil { \mathfrak{p}}\rceil)\cap (\downarrow_{{}_{ \mathfrak{S}}}\lceil { \mathfrak{A}}_{ \mathfrak{t}}\rceil)\,)$ exists and is an element of $(\downarrow_{{}_{ \mathfrak{P}}}\lceil { \mathfrak{A}}_{ \mathfrak{t}}\rceil ) ={ \mathfrak{K}}_{\lfloor { \mathfrak{t}}\rfloor}$ (Lemma \ref{propquasiclassical}). Let us denote by ${ \mathfrak{p}}''$ an element of ${ \mathfrak{P}}$ which state corresponds to this supremum. 
Let us denote by ${ \mathfrak{p}}''$ an element of $(\downarrow_{{}_{ \mathfrak{P}}} { \mathfrak{p}}) \cap { \mathfrak{K}}_{\lfloor { \mathfrak{t}}\rfloor}$ which is non-empty (it contains $\bot_{{}_{ \mathfrak{P}}}$). Firstly, we have ${ \mathfrak{p}}''\sqsubseteq_{{}_{ \mathfrak{P}}} { \mathfrak{p}}$. We note in particular that ${ \mathfrak{p}}''$ is then an element of ${ \mathfrak{Q}}_{ \mathfrak{t}}$, because ${ \mathfrak{Q}}_{ \mathfrak{t}}$ is a downset. Moreover, the monotonicity of the measurement map associated {{}}{with} ${ \mathfrak{t}}$ implies $({ \mathfrak{p}}'' {\centerdot} { \mathfrak{t}}) \sqsubseteq_{{}_{ \mathfrak{P}}} ({ \mathfrak{p}} {\centerdot} { \mathfrak{t}})$. Secondly, we have ${ \mathfrak{p}}''\in (\downarrow_{{}_{ \mathfrak{P}}} { \mathfrak{A}}_{ \mathfrak{t}})={ \mathfrak{K}}_{ \mathfrak{t}}$, and {{}}{thus} $({ \mathfrak{p}}'' {\centerdot} { \mathfrak{t}}) \in { \mathfrak{A}}_{ \mathfrak{t}}$ as {{}}{proved} in the first sub-case. We now use the fact that ${ \mathfrak{A}}_{ \mathfrak{t}}$ is an upper-set 
to conclude that $({ \mathfrak{p}} {\centerdot} { \mathfrak{t}})\in { \mathfrak{A}}_{ \mathfrak{t}}$. 
\end{remark}

\begin{lemme}{}\label{lemmetIFKMimpliestmin}
We will assume the following preservation property
\begin{eqnarray}
\forall { \mathfrak{F}}\subseteq_{Fil} \lfloor { \mathfrak{Q}}_{ \mathfrak{t}}\rfloor, && (\bigsqcap{}^\veebar_{{}_{ \mathfrak{S}}}{ \mathfrak{F}}){\centerdot} { \mathfrak{t}}=\bigsqcap{}^\veebar_{{}_{\sigma\in { \mathfrak{F}}}}(\sigma {\centerdot} { \mathfrak{t}}).\label{filterdott}
\end{eqnarray}
We then have
\begin{eqnarray}
{ \mathfrak{t}}\in { \mathfrak{T}}_{{}_{IFKM}}
& \Rightarrow & { \mathfrak{t}}\in { \mathfrak{T}}_{{}_{min}}.
\end{eqnarray}
\end{lemme}
\begin{proof}
Let us consider ${ \mathfrak{t}}\in { \mathfrak{T}}_{{}_{IFKM}}$ and $\sigma \in \lceil { \mathfrak{K}}_{\lfloor { \mathfrak{t}}\rfloor }\rceil$. Using the fact that the measurement map $({\centerdot} { \mathfrak{t}})$ preserves filtered-infima, we deduce
\begin{eqnarray*}
(\sigma {\centerdot} { \mathfrak{t}})&=&(\bigsqcap{}_{{}_{ \mathfrak{S}}}  (\uparrow^{{}^{ \mathfrak{S}}}\!\!\!\! \sigma)) {\centerdot} { \mathfrak{t}}\\
&=& \bigsqcap{}_{{}_{ \mathfrak{S}}} ( ( Dom^{ \mathfrak{S}}_{{\centerdot} { \mathfrak{t}}} \cap (\uparrow^{{}^{ \mathfrak{S}}}\!\!\!\! \sigma)) {\centerdot} { \mathfrak{t}})\\
&\sqsupseteq_{{}_{ \mathfrak{S}}} & \bigsqcap{}_{{}_{ \mathfrak{S}}}  ((\uparrow^{{}^{ \mathfrak{S}}}\!\!\!\! \sigma) \cap \lceil { \mathfrak{A}}_{\lfloor { \mathfrak{t}}\rfloor }\rceil)=(\sigma \sqcup_{{}_{ \mathfrak{S}}} \Sigma_{\lfloor { \mathfrak{t}}\rfloor })
\end{eqnarray*}
Using $\sigma \sqsubseteq_{{}_{ \mathfrak{S}}} (\sigma \sqcup_{{}_{ \mathfrak{S}}} \Sigma_{\lfloor { \mathfrak{t}}\rfloor })$ and the monotonicity of the measurement map, {{}}{along with} property (\ref{firstkind} (ii)), we also obtain, for any $\sigma \in \lceil { \mathfrak{K}}_{\lfloor { \mathfrak{t}}\rfloor }\rceil$, the property $\sigma {\centerdot} { \mathfrak{t}}  \sqsubseteq_{{}_{ \mathfrak{S}}} (\sigma \sqcup_{{}_{ \mathfrak{S}}} \Sigma_{\lfloor { \mathfrak{t}}\rfloor }){\centerdot} { \mathfrak{t}}= \sigma \sqcup_{{}_{ \mathfrak{S}}} \Sigma_{\lfloor { \mathfrak{t}}\rfloor }$. As a result, 
\begin{eqnarray}
\forall \sigma \in \lceil { \mathfrak{K}}_{\lfloor { \mathfrak{t}}\rfloor }\rceil, &&  (\sigma {\centerdot} { \mathfrak{t}}) = (\sigma \sqcup_{{}_{ \mathfrak{S}}} \Sigma_{\lfloor { \mathfrak{t}}\rfloor }).
\end{eqnarray}
\end{proof}

\begin{lemme}{}\label{lemmeminimpliesifkm}
Let ${ \mathfrak{l}}$ be a testable property and let ${ \mathfrak{t}}$ be a yes/no test leading to a minimally disturbing measurement of the property ${ \mathfrak{l}}=\lfloor { \mathfrak{t}} \rfloor$ (i.e., ${ \mathfrak{l}}=\lfloor { \mathfrak{t}} \rfloor$ and ${ \mathfrak{t}}\in { \mathfrak{T}}_{{}_{min}}$), then necessarily ${ \mathfrak{t}}$ leads to ideal first-kind measurements of the property ${ \mathfrak{l}}$. As a conclusion,
\begin{eqnarray}
 { \mathfrak{t}}\in { \mathfrak{T}}_{{}_{min}}
 & \Rightarrow & { \mathfrak{t}}\in { \mathfrak{T}}_{{}_{IFKM}}.
\end{eqnarray}
\end{lemme}
\begin{proof}
{{}}{First}, ${ \mathfrak{t}}$ being in ${ \mathfrak{T}}_{{}_{min}}$, we know that $\lfloor { \mathfrak{t}} \rfloor$ is in ${ \mathcal{L}}_{{}_{min}}$ and {{}}{therefore} in ${ \mathcal{L}}_{{}_{q-cl}}$ using Lemma \ref{lminimplieslqcl}. We now {{}}{intend} to prove that the yes/no tests ${ \mathfrak{t}}$ satisfying the properties given in (\ref{Tmin}) {{}}{also satisfy} the properties (\ref{firstkind} (i)), (\ref{firstkind} (ii)) and (\ref{axiomidealmeasurement}), as {{}}{long} as the property $\lfloor { \mathfrak{t}} \rfloor$ is quasi-classical.\\
We have proved{{}}{ }property (\ref{firstkind} (i)) in Remark \ref{prooffkmi}.\\
Using property (\ref{Tmin}) for the particular case ${ \mathfrak{p}}\in { \mathfrak{A}}_{ \mathfrak{t}}$, we obtain $\lceil { \mathfrak{p}}{\centerdot} { \mathfrak{t}}\rceil = \lceil { \mathfrak{p}}\rceil$, because by definition ${\Sigma}_{ \mathfrak{t}}=\bigsqcap{}_{{}_{ \mathfrak{S}}} { \mathfrak{A}}_{ \mathfrak{t}}$. We have then proved {{}}{ }property (\ref{firstkind} (ii)).\\
Let us now consider a second yes/no test ${ \mathfrak{t}}'\in { \mathfrak{T}}$ such that $\overbrace{\lfloor{ \mathfrak{t}}\rfloor\lfloor{ \mathfrak{t}}'\rfloor}$, i.e., ${ \mathfrak{A}}_{ \mathfrak{t}} \cap { \mathfrak{A}}_{{ \mathfrak{t}}'}\not= \varnothing$. We note in particular that this compatibility property implies $\widetilde{ \mathfrak{e}}_{{ \mathfrak{t}}'}({\Sigma}_{ \mathfrak{t}})\leq {\bf Y}$. Let us also consider ${ \mathfrak{p}}\in { \mathfrak{Q}}_{ \mathfrak{t}}$ such that ${ \mathfrak{e}}({ \mathfrak{p}},{ \mathfrak{t}}')={\bf Y}$, i.e., ${ \mathfrak{p}}\in { \mathfrak{A}}_{{ \mathfrak{t}}'}$. We will distinguish two sub-cases as before.\\
As a first sub-case, let us suppose that ${ \mathfrak{p}}\in (\downarrow_{{}_{ \mathfrak{P}}}{ \mathfrak{A}}_{ \mathfrak{t}})$. Applying property (\ref{Tmin}), we have ${ \mathfrak{e}}({ \mathfrak{p}}{\centerdot} { \mathfrak{t}},{ \mathfrak{t}}')\geq { \mathfrak{e}}({ \mathfrak{p}},{{ \mathfrak{t}}'})={\bf Y}$ using the hypotheses.\\
As a second sub-case, we now suppose ${ \mathfrak{p}}\in { \mathfrak{Q}}_{ \mathfrak{t}}\!\smallsetminus \!(\downarrow_{{}_{ \mathfrak{P}}}{ \mathfrak{A}}_{ \mathfrak{t}})$. Let us consider once again any preparation process 
${ \mathfrak{p}}''\in { \mathfrak{P}}$ {{}}{whose} state 
corresponds to the supremum $\bigsqcup_{{}_{ \mathfrak{S}}} (\, (\downarrow_{{}_{ \mathfrak{S}}} \lceil { \mathfrak{p}}\rceil)\cap (\downarrow_{{}_{ \mathfrak{S}}}\lceil { \mathfrak{A}}_{ \mathfrak{t}}\rceil)\,)$.  We conclude as before that ${ \mathfrak{p}}''\in { \mathfrak{Q}}_{ \mathfrak{t}}$ and $({ \mathfrak{p}}'' {\centerdot} { \mathfrak{t}}) \sqsubseteq_{{}_{ \mathfrak{P}}} ({ \mathfrak{p}} {\centerdot} { \mathfrak{t}})$. In particular ${ \mathfrak{e}}({ \mathfrak{p}}'' {\centerdot} { \mathfrak{t}},{ \mathfrak{t}}') \leq { \mathfrak{e}}({ \mathfrak{p}} {\centerdot} { \mathfrak{t}},{ \mathfrak{t}}')$.\\
We know also that ${ \mathfrak{p}}''\in (\downarrow_{{}_{ \mathfrak{P}}} { \mathfrak{A}}_{ \mathfrak{t}})$ and {{}}{therefore}, using property (\ref{Tmin}), ${ \mathfrak{e}}({ \mathfrak{p}}''{\centerdot} { \mathfrak{t}},{ \mathfrak{t}}')\geq { \mathfrak{e}}({ \mathfrak{p}}'',{{ \mathfrak{t}}'})$.\\  
{{}}{Lastly}, we know that ${\Sigma}_{{ \mathfrak{t}}'} \sqsubseteq_{{}_{ \mathfrak{S}}} \lceil { \mathfrak{p}}''\rceil = \bigsqcup_{{}_{ \mathfrak{S}}} (\, (\downarrow_{{}_{ \mathfrak{S}}} \lceil { \mathfrak{p}}\rceil)\cap (\downarrow_{{}_{ \mathfrak{S}}}\lceil { \mathfrak{A}}_{ \mathfrak{t}}\rceil)\,)$ because  (i) ${ \mathfrak{p}}\in { \mathfrak{A}}_{{ \mathfrak{t}}'}$ implies ${\Sigma}_{{ \mathfrak{t}}'} \sqsubseteq_{{}_{ \mathfrak{S}}} \lceil { \mathfrak{p}}\rceil${{}}{, which} implies ${\Sigma}_{{ \mathfrak{t}}'}\in (\downarrow_{{}_{ \mathfrak{S}}} \lceil { \mathfrak{p}}\rceil)$, and (ii) ${ \mathfrak{A}}_{ \mathfrak{t}} \cap { \mathfrak{A}}_{{ \mathfrak{t}}'}\not= \varnothing$ implies ${\Sigma}_{{ \mathfrak{t}}'} \in (\downarrow_{{}_{ \mathfrak{S}}}\lceil { \mathfrak{A}}_{ \mathfrak{t}}\rceil)$. But ${\Sigma}_{{ \mathfrak{t}}'} \sqsubseteq_{{}_{ \mathfrak{S}}} \lceil { \mathfrak{p}}''\rceil$ is equivalently rewritten as ${ \mathfrak{e}}({ \mathfrak{p}}'',{ \mathfrak{t}}')={\bf Y}$. We {{}}{therefore} deduce {{}}{that} ${ \mathfrak{e}}({ \mathfrak{p}}''{\centerdot} { \mathfrak{t}},{ \mathfrak{t}}')={\bf Y}$.\\
Using the two intermediary results, we conclude this analysis of the second sub-case by ${ \mathfrak{e}}({ \mathfrak{p}}{\centerdot} { \mathfrak{t}},{ \mathfrak{t}}')={\bf Y}$.\\
As a conclusion, we have for any ${ \mathfrak{p}}\in { \mathfrak{Q}}_{ \mathfrak{t}}$, ${ \mathfrak{e}}({ \mathfrak{p}}{\centerdot} { \mathfrak{t}},{ \mathfrak{t}}')={\bf Y}$. We have {{}}{thus demonstrated} property (\ref{axiomidealmeasurement}).
\end{proof}

\begin{MyDef}
A testable property ${ \mathfrak{l}}\in { \mathcal{L}}$ is said to be {\em quasi-classical} (this fact is denoted ${ \mathfrak{l}}\in { \mathcal{L}}_{{}_{q-cl}}$) iff the consistency-domain $\lceil { \mathfrak{K}}_{{ \mathfrak{l}}}\rceil$ is a continuous retract of the domain $\lceil { \mathfrak{Q}}_{ \mathfrak{l}}\rceil $, i.e.,
\begin{eqnarray}
\forall { \mathfrak{l}}\in { \mathcal{L}},\;\;\;{ \mathfrak{l}}\in { \mathcal{L}}_{{}_{q-cl}} \;\; :\Leftrightarrow 
\;\; \exists \;\;  {\pi}_{ \mathfrak{l}} : \lceil { \mathfrak{Q}}_{ \mathfrak{l}}\rceil  \longrightarrow  \lceil { \mathfrak{K}}_{ \mathfrak{l}}\rceil\;\;\vert && {\pi}_{ \mathfrak{l}} \;\textit{\rm is Scott{{}}{ }continuous} \label{prop1pil}\\
& &\forall \sigma\in \lceil {{ \mathfrak{K}}_{ \mathfrak{l}}}\rceil,\;\;{\pi}_{ \mathfrak{l}}(\sigma)=\sigma \label{prop2pil}\\
& &\forall \sigma'\in {\lceil { \mathfrak{Q}}_{ \mathfrak{l}}\rceil},\;\;{\pi}_{ \mathfrak{l}}(\sigma')\sqsubseteq_{{}_{ \mathfrak{S}}} \sigma'. \label{prop3pil}
\end{eqnarray}
\end{MyDef}

Let ${{\mathfrak{S}}'}$ be a Scott-closed subset of ${\mathfrak{S}}$.  ${\mathfrak{S}}'$ is said to be {\it a Scott ideal} of ${\mathfrak{S}}$ iff 
\begin{eqnarray}
\forall S\subseteq_{fin} {\mathfrak{S}}',&& \widehat{S}\,{}^{{}^{\mathfrak{S}}}\;\Rightarrow\; \widehat{S}\,{}^{{}^{\mathfrak{S}'}}.\label{defidealized}
\end{eqnarray}

\begin{lemme}
\label{alternativepropertyidealized}
A Scott-closed subset ${\mathfrak{S}}'$ of a domain ${\mathfrak{S}}$ is a Scott-ideal iff it satisfies the property:
\begin{eqnarray}
\forall s\in {\mathfrak{S}},&&\;\; \bigsqcup{}_{{}_{\mathfrak{S}}}(\,(\downarrow_{{}_{{\mathfrak{S}}}}\,s)\sqcap_{{}_{\mathfrak{S}}} {\mathfrak{S}}'\,)\;\;\in {\mathfrak{S}}'.\label{alternativedefidealized}
\end{eqnarray}
\end{lemme}
\begin{proof}
Let us consider $S\subseteq_{fin} {\mathfrak{S}}'$ such that it exists $s\in {\mathfrak{S}}$ which is a common upper-bound of elements of $S$. We have then $S\subseteq (\,(\downarrow_{{}_{{\mathfrak{S}}}}\,s)\sqcap_{{}_{\mathfrak{S}}} {\mathfrak{S}}'\,)$. Using property (\ref{alternativedefidealized}), the supremum $\bigsqcup{}_{{}_{\mathfrak{S}}}(\,(\downarrow_{{}_{{\mathfrak{S}}}}\,s)\sqcap_{{}_{\mathfrak{S}}} {\mathfrak{S}}'\,)$ exists as an element of ${\mathfrak{S}}'$. It is then a common upper-bound of elements of $S$ that belongs to ${\mathfrak{S}}'$. Hence, property (\ref{alternativedefidealized}) implies that ${\mathfrak{S}}'$ is a Scott-ideal of ${\mathfrak{S}}$.\\
Conversely, let us suppose ${\mathfrak{S}}'$ is a Scott-ideal of ${\mathfrak{S}}$ and let us fix any $s\in {\mathfrak{S}}$. The set $(\,(\downarrow_{{}_{{\mathfrak{S}}}}\,s)\sqcap_{{}_{\mathfrak{S}}} {\mathfrak{S}}'\,)$ is upper-bounded by $s$. Using property (\ref{defidealized}), for any $F\subseteq_{fin} (\,(\downarrow_{{}_{{\mathfrak{S}}}}\,s)\sqcap_{{}_{\mathfrak{S}}} {\mathfrak{S}}'\,)$, it exists $M_F\in {\mathfrak{S}}'$ upper-bound of $F$. Hence, $\bigsqcup{}_{{\mathfrak{S}}'}F$ exists and is in ${\mathfrak{S}}'$, and $\bigsqcup{}_{{}_{\mathfrak{S}}}F=\bigsqcup{}_{{}_{{\mathfrak{S}}'}}F$. As an example, for $F\subseteq_{fin}{\mathbb{A}}(s)\sqcap_{{}_{\mathfrak{S}}} {\mathfrak{S}}'$ 
We can build the following subset: $\{\,\bigsqcup{}_{{}_{\mathfrak{S}}}F\;\vert\; F\subseteq_{fin}{\mathbb{A}}(s)\sqcap_{{}_{\mathfrak{S}}} {\mathfrak{S}}'\,\}\;\subseteq\,(\,(\downarrow_{{}_{{\mathfrak{S}}}}\,s)\sqcap_{{}_{\mathfrak{S}}} {\mathfrak{S}}'\,)$. It is a directed subset, by construction. It then admits a directed-supremum in ${{\mathfrak{S}}}$ which is in ${\mathfrak{S}}'$, because ${\mathfrak{S}}'$ is Scott-closed : $\bigsqcup{}^\barwedge_{{}_{\mathfrak{S}}}\,\{\,\bigsqcup{}_{{}_{\mathfrak{S}}}F\;\vert\; F\subseteq_{fin}{\mathbb{A}}(s)\sqcap_{{}_{\mathfrak{S}}} {\mathfrak{S}}'\,\}\;\in {\mathfrak{S}}'$. Due to algebraicity of ${\mathfrak{S}}$ (Theorem \ref{theoremalgebraic}), we then obtain $(\,\bigsqcup{}_{{}_{\mathfrak{S}}}\,\{\, t\in {\mathfrak{S}}'\;\vert\; t\sqsubseteq_{{}_{\mathfrak{S}}} s\,\}\,)\;\in {\mathfrak{S}}'$. This  achieves the proof of property (\ref{alternativedefidealized}).
\end{proof}

\begin{lemme}{}\label{propquasiclassical}
Let ${ \mathfrak{l}}$ be a quasi-classical property.  Then, ${ \mathfrak{K}}_{ \mathfrak{l}}$ is a Scott-ideal of $\lceil { \mathfrak{Q}}_{ \mathfrak{l}}\rceil$. Conversely,  if $\lceil { \mathfrak{K}}_{ \mathfrak{l}}\rceil$ is a Scott-ideal of $\lceil { \mathfrak{Q}}_{ \mathfrak{l}}\rceil$, then ${ \mathfrak{l}}$ is a quasi-classical property. 
Moreover, the retraction ${\pi}_{ \mathfrak{l}}$ is given by
\begin{eqnarray}
&& \forall \sigma \in \lceil { \mathfrak{Q}}_{ \mathfrak{l}}\rceil,\;\;  {\pi}_{ \mathfrak{l}}(\sigma)=\bigsqcup{}_{{}_{ \mathfrak{S}}} ( \lceil { \mathfrak{K}}_{ \mathfrak{l}}\rceil \cap (\downarrow_{{}_{ \mathfrak{S}}} \!\!\sigma )).\label{expressionpil}
\end{eqnarray}
The retraction ${\pi}_{ \mathfrak{l}}$ is an idempotent, order-preserving and Scott-continuous map, which preserves infima.
\end{lemme}
\begin{proof}
Let us consider $\sigma_1,\sigma_2 \in  \lceil { \mathfrak{K}}_{ \mathfrak{l}}\rceil$ and let us suppose that $\exists \sigma \in \lceil { \mathfrak{Q}}_{ \mathfrak{l}}\rceil \;\vert\; \sigma_1,\sigma_2\sqsubseteq_{{}_{ \mathfrak{S}}}\sigma$. The monotonicity of ${\pi}_{ \mathfrak{l}}$ implies ${\pi}_{ \mathfrak{l}}(\sigma_1)\sqsubseteq_{{}_{ \mathfrak{S}}} {\pi}_{ \mathfrak{l}}(\sigma)$ and  ${\pi}_{ \mathfrak{l}}(\sigma_2)\sqsubseteq_{{}_{ \mathfrak{S}}}{\pi}_{ \mathfrak{l}}(\sigma)$. Secondly, property  (\ref{prop2pil}) implies $\sigma_1={\pi}_{ \mathfrak{l}}(\sigma_1)$ and $\sigma_2={\pi}_{ \mathfrak{l}}(\sigma_2)$. {{}}{Thirdly}, property (\ref{prop3pil}) implies ${\pi}_{ \mathfrak{l}}(\sigma)\sqsubseteq_{{}_{ \mathfrak{S}}}\sigma$. As a conclusion, $\exists \sigma'={\pi}_{ \mathfrak{l}}(\sigma)\in  \lceil { \mathfrak{K}}_{ \mathfrak{l}}\rceil \;\vert\;  \sigma_1,\sigma_2\sqsubseteq_{{}_{ \mathfrak{S}}}\sigma'$.  Furthermore, from Lemma \ref{Klscottclosed}, we know already that $ \lceil { \mathfrak{K}}_{ \mathfrak{l}}\rceil$ is Scott-closed. As a result, we then conclude that $ \lceil { \mathfrak{K}}_{ \mathfrak{l}}\rceil$ is a Scott{{}}{ }ideal in $\lceil { \mathfrak{Q}}_{ \mathfrak{l}}\rceil$.\\
Conversely, if $\lceil { \mathfrak{K}}_{ \mathfrak{l}}\rceil$ is a Scott{{}}{ }ideal in $\lceil { \mathfrak{Q}}_{ \mathfrak{l}}\rceil$,  we can use{{}}{ }Lemma \ref{alternativepropertyidealized} to conclude that  
\begin{eqnarray}
&& \forall \sigma \in \lceil { \mathfrak{Q}}_{ \mathfrak{l}}\rceil,\;\;  {\pi}(\sigma):=\bigsqcup{}_{{}_{ \mathfrak{S}}} ( \lceil { \mathfrak{K}}_{ \mathfrak{l}}\rceil \cap (\downarrow_{{}_{ \mathfrak{S}}} \!\!\sigma ))\;\;\in  \lceil { \mathfrak{K}}_{ \mathfrak{l}}\rceil.
\end{eqnarray}
${\pi}$ is a map defined from $\lceil { \mathfrak{Q}}_{ \mathfrak{l}}\rceil$ to $\lceil { \mathfrak{K}}_{ \mathfrak{l}}\rceil${{}}{,} which {{}}{trivially satisfies} properties (\ref{prop1pil}), (\ref{prop2pil}) and(\ref{prop3pil}).  ${ \mathfrak{l}}$ is then a quasi-classical property and the expression {{}}{for} the retraction ${\pi}_{ \mathfrak{l}}$ is given by (\ref{expressionpil}).\\
${\pi}_{ \mathfrak{l}}$ is also the Galois right{{}}{ adjoint} 
of the inclusion map from $\lceil { \mathfrak{K}}_{ \mathfrak{l}}\rceil$ to  $\lceil { \mathfrak{Q}}_{ \mathfrak{l}}\rceil$. As an immediate consequence, ${\pi}_{ \mathfrak{l}}$ is a surjective, order-preserving map, which preserves infima.
 \end{proof}

\begin{lemme}{}\label{lqclimplieslmin}
\begin{eqnarray}
 { \mathcal{L}}_{{}_{q-cl}}&\subseteq & { \mathcal{L}}_{{}_{min}}
\end{eqnarray}
More precisely, if ${ \mathfrak{l}}\in { \mathcal{L}}_{{}_{q-cl}}$, the map ${\centerdot} {\Theta}_{ \mathfrak{l}}$ defined by
\begin{eqnarray}
&&
\begin{array}{rcrcl}
{\centerdot} {\Theta}_{ \mathfrak{l}}& : & { \mathfrak{S}} & \dashrightarrow &{ \mathfrak{S}}\\
&\lceil { \mathfrak{Q}}_{ \mathfrak{l}}\rceil \;\ni & \sigma & \mapsto & \sigma {\centerdot} {\Theta}_{ \mathfrak{l}} :=  {\Sigma}_{ \mathfrak{l}} \sqcup_{{}_{ \mathfrak{S}}}{\pi}_{ \mathfrak{l}}(\sigma).
\end{array}\label{defThetal}
\end{eqnarray}
is an idempotent, order-preserving and Scott-continuous partial map from $\lceil { \mathfrak{Q}}_{ \mathfrak{l}}\rceil$ to $ \lceil { \mathfrak{K}}_{ \mathfrak{l}}\rceil$, which preserves filtered-infima, and satisfies $\forall \sigma \in \lceil { \mathfrak{K}}_{ \mathfrak{l}}\rceil$, $\sigma {\centerdot}  {\Theta}_{ \mathfrak{l}} = \sigma \sqcup_{{}_{ \mathfrak{S}}} {\Sigma}_{{\mathfrak{l}}}.$
\end{lemme} 
\begin{proof}
Let us consider ${ \mathfrak{l}}\in { \mathcal{L}}_{{}_{q-cl}}$. 
${\pi}_{ \mathfrak{l}}$ is an idempotent, order-preserving and Scott-continuous partial map from $\lceil { \mathfrak{Q}}_{ \mathfrak{l}}\rceil$ to $ \lceil { \mathfrak{K}}_{ \mathfrak{l}}\rceil$, which preserves arbitrary infima, as shown in Lemma \ref{propquasiclassical}.\\
As proved in Lemma \ref{lemmathetal}, the map 
\begin{eqnarray}
&&
\begin{array}{rcrcl}
& &  \lceil { \mathfrak{K}}_{ \mathfrak{l}}\rceil & \longrightarrow & \lceil { \mathfrak{A}}_{ \mathfrak{l}}\rceil \\
& & \sigma & \mapsto & {\Sigma}_{ \mathfrak{l}} \sqcup_{{}_{ \mathfrak{S}}} \sigma
\end{array}
\end{eqnarray}
is an idempotent, order-preserving and Scott-continuous partial map, which preserves filtered-infima and existing suprema. As a result, ${\centerdot} {\Theta}_{ \mathfrak{l}}$ is an order-preserving, Scott-continuous partial map, which preserves filtered-infima. \\
The idempotency of ${\centerdot} {\Theta}_{ \mathfrak{l}}$ is verified as follows.  Firstly, ${\pi}_{ \mathfrak{l}}({\Sigma}_{ \mathfrak{l}}\sqcup_{{}_{ \mathfrak{S}}} {\pi}_{ \mathfrak{l}}(\sigma))={\Sigma}_{ \mathfrak{l}}\sqcup_{{}_{ \mathfrak{S}}} {\pi}_{ \mathfrak{l}}(\sigma)$ because ${\Sigma}_{ \mathfrak{l}} \sqcup_{{}_{ \mathfrak{S}}} {\pi}_{ \mathfrak{l}}(\sigma)\in \lceil { \mathfrak{K}}_{ \mathfrak{l}}\rceil$. \\
Endly, we have ${\Sigma}_{ \mathfrak{l}} \sqcup_{{}_{ \mathfrak{S}}}{\Sigma}_{ \mathfrak{l}}= {\Sigma}_{ \mathfrak{l}}$.\\
Moreover, for any $\sigma \in  \lceil { \mathfrak{K}}_{ \mathfrak{l}}\rceil= (\downarrow_{{}_{ \mathfrak{S}}}\!\!\lceil { \mathfrak{A}}_{ \mathfrak{l}}\rceil)$, we have ${\pi}_{ \mathfrak{l}}(\sigma)=\sigma$. Hence, we obtain $\forall \sigma \in (\downarrow_{{}_{ \mathfrak{S}}}\!\!\lceil { \mathfrak{A}}_{ \mathfrak{l}}\rceil),\;\;\sigma {\centerdot}  {\Theta}_{ \mathfrak{l}} = \sigma \sqcup_{{}_{ \mathfrak{S}}} {\Sigma}_{{\mathfrak{l}}}.$\\
As a result, ${\centerdot} {\Theta}_{ \mathfrak{l}}$ satisfies all properties mentioned in (\ref{Lmin}), i.e., the properties required to conclude that ${ \mathfrak{l}}\in { \mathcal{L}}_{{}_{min}}$. As a conclusion, ${ \mathcal{L}}_{{}_{q-cl}}\subseteq  { \mathcal{L}}_{{}_{min}}$.
\end{proof}

\begin{lemme}{}\label{lminimplieslqcl}
\begin{eqnarray}
 { \mathcal{L}}_{{}_{q-cl}}&\supseteq & { \mathcal{L}}_{{}_{min}}
\end{eqnarray}
More precisely, if ${ \mathfrak{l}}\in { \mathcal{L}}_{{}_{min}}$ and ${\centerdot} { \mathfrak{t}}$ is a measurement map defined to satisfy the minimality requirement (\ref{Lmin}), then the partial map $\rho_{ \mathfrak{t}}$ defined on $\lceil { \mathfrak{Q}}_{ \mathfrak{l}}\rceil$ by 
\begin{eqnarray}
&&
\begin{array}{rcrcl}
{\rho}_{ \mathfrak{t}}& : & { \mathfrak{S}} & \dashrightarrow &{ \mathfrak{S}}\\
&\lceil { \mathfrak{Q}}_{ \mathfrak{l}}\rceil \;\ni & \sigma & \mapsto & {\rho}_{ \mathfrak{t}}(\sigma) :=  \sigma \sqcap_{{}_{ \mathfrak{S}}} (\sigma {\centerdot} { \mathfrak{t}}).\label{proppi=imcapdott}
\end{array}
\end{eqnarray}
is a Scott-continuous retraction from $\lceil { \mathfrak{Q}}_{ \mathfrak{l}}\rceil$ to $\lceil {{ \mathfrak{K}}_{ \mathfrak{l}}}\rceil$.
\end{lemme} 
\begin{proof}
We {{}}{first} note the obvious property $\forall \sigma'\in {\lceil { \mathfrak{Q}}_{ \mathfrak{l}}\rceil},\;\;{\rho}_{ \mathfrak{t}}(\sigma')\sqsubseteq_{{}_{ \mathfrak{S}}} \sigma'$. \\
Secondly, we note that, due to {{}}{the} monotonicity requirement on the measurement map $({\centerdot} { \mathfrak{t}})$, $(\sigma {\centerdot} { \mathfrak{t}})$ is in $\lceil { \mathfrak{A}}_{ \mathfrak{l}}\rceil$. We then deduce that the range of ${\rho}_{ \mathfrak{t}}$ is included in $(\downarrow_{{}_{ \mathfrak{S}}}\lceil { \mathfrak{A}}_{ \mathfrak{l}}\rceil)$, i.e., included in $\lceil { \mathfrak{K}}_{ \mathfrak{l}}\rceil$.\\
Thirdly, due to{{}}{ }property (\ref{Lmin}) {{}}{being} satisfied by ${\centerdot} { \mathfrak{t}}$, we know that $\forall \sigma\in \lceil {{ \mathfrak{K}}_{ \mathfrak{l}}}\rceil,\;\;{\rho}_{ \mathfrak{t}}(\sigma)=\sigma \sqcap_{{}_{ \mathfrak{S}}} (\sigma \sqcup_{{}_{ \mathfrak{S}}} {\Sigma}_{ \mathfrak{l}})=\sigma$.\\
The Scott{{}}{ }continuity of ${\rho}_{ \mathfrak{t}}$ is derived from: the Scott{{}}{ }continuity of the measurement map $({\centerdot} { \mathfrak{t}})$, the meet-continuity property in ${ \mathfrak{S}}$ and the property given in \cite[Proposition 2.1.12]{Abramsky}:
\begin{eqnarray*}
\forall { \mathfrak{C}}\subseteq_{Chain} {\lceil { \mathfrak{Q}}_{ \mathfrak{l}}\rceil},\;\;\;\;\;\;\;\; {\rho}_{ \mathfrak{t}}(\bigsqcup{}_{{}_{ \mathfrak{S}}} { \mathfrak{C}})&=&(\bigsqcup{}_{{}_{ \mathfrak{S}}} { \mathfrak{C}}) \sqcap_{{}_{ \mathfrak{S}}} (  (\bigsqcup{}_{{}_{ \mathfrak{S}}} { \mathfrak{C}}) {\centerdot} { \mathfrak{t}})\\ 
&=& (\bigsqcup{}_{{}_{{ \mathfrak{c}} \in { \mathfrak{C}}}} { \mathfrak{c}}) \sqcap_{{}_{ \mathfrak{S}}} (  \bigsqcup{}_{{}_{{ \mathfrak{c}}' \in { \mathfrak{C}}}} ({ \mathfrak{c}}' {\centerdot} { \mathfrak{t}}))\\
&=&\bigsqcup{}_{{}_{{ \mathfrak{c}} \in { \mathfrak{C}}}} \bigsqcup{}_{{}_{{ \mathfrak{c}}' \in { \mathfrak{C}}}}  ({ \mathfrak{c}} \sqcap_{{}_{ \mathfrak{S}}} ({ \mathfrak{c}}' {\centerdot} { \mathfrak{t}}))\\
&=&\bigsqcup{}_{{}_{{ \mathfrak{c}} \in { \mathfrak{C}}}} ({ \mathfrak{c}} \sqcap_{{}_{ \mathfrak{S}}} ({ \mathfrak{c}} {\centerdot} { \mathfrak{t}}))=
\bigsqcup{}_{{}_{{ \mathfrak{c}} \in { \mathfrak{C}}}}{\rho}_{ \mathfrak{t}}({ \mathfrak{c}}).
\end{eqnarray*}
Finally, chain-continuity is equivalent to Scott{{}}{ }continuity. \\ 
We have then checked properties (\ref{prop1pil}), (\ref{prop2pil}) {{}}{and} (\ref{prop3pil}) for ${\rho}_{ \mathfrak{t}}$. As a result, ${ \mathfrak{l}}\in { \mathcal{L}}_{{}_{q-cl}}$ and {{}}{thus} ${ \mathcal{L}}_{{}_{q-cl}}\supseteq { \mathcal{L}}_{{}_{min}}$.
\end{proof}

\begin{theoreme}{\bf [Characterization of minimally disturbing measurements]}\label{theoremlmin=lIFKM}\\
A property ${ \mathfrak{l}}$ can be measured through minimally disturbing measurements iff ${ \mathfrak{l}}$ is a testable quasi-classical property (i.e.,  iff $\lceil { \mathfrak{K}}_{ \mathfrak{l}}\rceil$ is a Scott-ideal in $\lceil { \mathfrak{Q}}_{ \mathfrak{l}}\rceil$).  In other words,
\begin{eqnarray}
&& { \mathcal{L}}_{{}_{q-cl}}={ \mathcal{L}}_{{}_{min}}\subseteq { \mathcal{L}}_{{}_{IFKM}}.
\end{eqnarray}
For any quasi-classical property ${ \mathfrak{l}}\in { \mathcal{L}}_{{}_{q-cl}}$, the map given by
\begin{eqnarray}
&&
\begin{array}{rcrcl}
{\centerdot} {\Theta}_{ \mathfrak{l}}& : & { \mathfrak{S}} & \dashrightarrow &{ \mathfrak{S}}\\
&\lceil { \mathfrak{Q}}_{ \mathfrak{l}}\rceil \;\ni & \sigma & \mapsto & \sigma {\centerdot} {\Theta}_{ \mathfrak{l}} :=  {\Sigma}_{ \mathfrak{l}} \sqcup_{{}_{ \mathfrak{S}}}\bigsqcup{}_{{}_{ \mathfrak{S}}} ( \lceil { \mathfrak{K}}_{ \mathfrak{l}}\rceil \cap (\downarrow_{{}_{ \mathfrak{S}}} \!\!\sigma )).
\end{array}\label{explicitThetal}
\end{eqnarray}
defines an idempotent, order-preserving, Scott-continuous partial map from $\lceil { \mathfrak{Q}}_{ \mathfrak{l}}\rceil$ to $ \lceil { \mathfrak{K}}_{ \mathfrak{l}}\rceil$, preserving filtered-infima and satisfying $\forall \sigma \in \lceil { \mathfrak{K}}_{ \mathfrak{l}}\rceil$, $\sigma {\centerdot}  {\Theta}_{ \mathfrak{l}} = \sigma \sqcup_{{}_{ \mathfrak{S}}} {\Sigma}_{{\mathfrak{l}}}.$ This is the explicit form of the minimally disturbing measurement map (which is also an ideal first-kind measurement map) associated {{}}{with} the property ${ \mathfrak{l}}$.
\end{theoreme}
\begin{proof} 
Direct consequence of Lemmas \ref{propquasiclassical}, \ref{lqclimplieslmin},  \ref{lminimplieslqcl},  \ref{lemmeminimpliesifkm} and  \ref{lemmetIFKMimpliestmin}.
\end{proof}

\begin{theoreme}{\bf [The Chu space $({ \mathfrak{S}}, { \mathfrak{T}}_{{}_{min}}, \widetilde{ \mathfrak{e}})$ is {\em bi-extensional}]}\label{chuextensional}
\begin{eqnarray}
\begin{array}{rcrclc}
\mu & : & { \mathfrak{T}}_{{}_{min}} & \longrightarrow & \left[{ \mathfrak{S}}\rightarrow { \mathfrak{B}} \right]^\barwedge_{{}_\sqcap} & %\textit{\rm is an injective map}
\\
& & { \mathfrak{t}} & \mapsto & \widetilde{ \mathfrak{e}}_{ \mathfrak{t}} &
\end{array} \;\;\;\;\textit{\rm is injective}
\end{eqnarray}
As a result,  the Chu space $({ \mathfrak{S}}, { \mathfrak{T}}_{{}_{min}}, \widetilde{ \mathfrak{e}})$ is {\em extensional}, i.e.
\begin{eqnarray}
\forall { \mathfrak{t}}_1,{ \mathfrak{t}}_2\in { { \mathfrak{T}}}_{{}_{min}},&& 
(\,\forall \sigma \in { \mathfrak{S}},\;\; \widetilde{ \mathfrak{e}}_{{ \mathfrak{t}}_1}(\sigma)=\widetilde{ \mathfrak{e}}_{{ \mathfrak{t}}_2}(\sigma)\,)\;\;\Rightarrow\;\; (\,{ \mathfrak{t}}_1={ \mathfrak{t}}_2\,)
\end{eqnarray}
As a consequence of Lemma \ref{chuseparated},  this Chu space is {\em bi-extensional}.
\end{theoreme}
\begin{proof}
Once the evaluation map is given, we obtain unambiguously  (1) the Scott-closed subset $\lceil { \mathfrak{Q}}_{ \mathfrak{t}}\rceil$ of ${ \mathfrak{S}}$ as the reverse image of the subset $\{\bot,\textit{\bf Y}\}$ by the Scott-continuous map $\widetilde{ \mathfrak{e}}_{ \mathfrak{t}}$,  and (2) the element $\Sigma_{ \mathfrak{t}}$ as the {{}}{infimum} of the Scott-open filter determined as the non-empty reverse image of the subset $\{\textit{\bf Y}\}$ by the order-preserving and infima-preserving  map $\widetilde{ \mathfrak{e}}_{ \mathfrak{t}}$.  \\
Let us now consider two minimally disturbing tests ${ \mathfrak{t}}_1,{ \mathfrak{t}}_2\in { \mathfrak{T}}_{{}_{min}}$ such that $\Sigma_{{ \mathfrak{t}}_1}=\Sigma_{{ \mathfrak{t}}_2}$ and $\lceil { \mathfrak{Q}}_{{ \mathfrak{t}}_1}\rceil = \lceil { \mathfrak{Q}}_{{ \mathfrak{t}}_2}\rceil$.  We {{}}{then have,} firstly : 
\begin{eqnarray}
Dom^{ \mathfrak{S}}_{{\centerdot}{ \mathfrak{t}}_1}=\lceil { \mathfrak{Q}}_{{ \mathfrak{t}}_1}\rceil = \lceil { \mathfrak{Q}}_{{ \mathfrak{t}}_2}\rceil =Dom^{ \mathfrak{S}}_{{\centerdot}{ \mathfrak{t}}_2}.
\end{eqnarray}
Secondly, we have for any $\sigma$ in $Dom^{ \mathfrak{S}}_{{\centerdot}{ \mathfrak{t}}_1}=Dom^{ \mathfrak{S}}_{{\centerdot}{ \mathfrak{t}}_2}$
\begin{eqnarray}
&&\sigma {\centerdot} { \mathfrak{t}}_1 = \Sigma_{{ \mathfrak{t}}_1} \sqcup_{{}_{ \mathfrak{S}}} \bigsqcup{}_{{}_{ \mathfrak{S}}} \left( 
(\downarrow_{{}_{ \mathfrak{S}}} \uparrow^{{}^{ \mathfrak{S}}} \Sigma_{{ \mathfrak{t}}_1})
 \cap (\downarrow_{{}_{ \mathfrak{S}}} \!\!\sigma ) \right)=
\Sigma_{{ \mathfrak{t}}_2} \sqcup_{{}_{ \mathfrak{S}}} \bigsqcup{}_{{}_{ \mathfrak{S}}} \left(  
(\downarrow_{{}_{ \mathfrak{S}}} \uparrow^{{}^{ \mathfrak{S}}} \Sigma_{{ \mathfrak{t}}_2})
 \cap (\downarrow_{{}_{ \mathfrak{S}}} \!\!\sigma ) \right)=
\sigma {\centerdot} { \mathfrak{t}}_2 \;\;\;\;\;
\end{eqnarray} 
As a consequence, ${ \mathfrak{t}}_1={ \mathfrak{t}}_2$. As a result, the map $\mu$ is injective, and the Chu space $({ \mathfrak{S}}, { \mathfrak{T}}_{{}_{min}}, \widetilde{ \mathfrak{e}})$ is {{}}{therefore} extensional. Then, using Lemma \ref{chuseparated}, we deduce that the Chu space $({ \mathfrak{S}}, { \mathfrak{T}}_{{}_{min}}, \widetilde{ \mathfrak{e}})$ is bi-extensional.
\end{proof}  

\begin{MyDef}
We allow for a generalized definition of yes/no tests and of minimally disturbing yes/no tests. The corresponding set of {\em generalized yes/no tests} (resp. {\em generalized minimally disturbing yes/no tests}) will be denoted $\widetilde{ \mathfrak{T}}$ (resp. $\widetilde{ \mathfrak{T}}_{{}_{min}}$).  Let us consider $\Sigma,\Sigma' \in { \mathfrak{S}}$ (not necessarily compact) such that $\antiwidehat{\Sigma\Sigma'}{}^{{}^{ \mathfrak{S}}}$.  We denote a generic element of $\widetilde{ \mathfrak{T}}$ as ${ \mathfrak{t}}_{{}_{(\Sigma,\Sigma')}}$ and define it according to
\begin{eqnarray}
{ \mathfrak{A}}_{\,{ \mathfrak{t}}_{{}_{(\Sigma,\Sigma')}}} & := & \uparrow^{{}^{ \mathfrak{S}}}\!\!\Sigma  \\
{ \mathfrak{Q}}_{\,{ \mathfrak{t}}_{{}_{(\Sigma,\Sigma')}}} & := & ({ \mathfrak{S}}\smallsetminus \uparrow^{{}^{ \mathfrak{S}}}\!\!\Sigma') 
\end{eqnarray}
If ${ \mathfrak{t}}_{{}_{(\Sigma,\Sigma')}}\in \widetilde{ \mathfrak{T}}_{{}_{min}}$, i.e. if $\Sigma,\Sigma'$ are such that $(\downarrow_{{}_{ \mathfrak{S}}}\uparrow^{{}^{ \mathfrak{S}}}\!\!\Sigma)$ is an idealized sub- selection structure of ${ \mathfrak{Q}}:=({ \mathfrak{S}}\smallsetminus \uparrow^{{}^{ \mathfrak{S}}}\!\!\Sigma')$, we will obviously have 
\begin{eqnarray}
\forall \sigma\in { \mathfrak{Q}}_{\,{ \mathfrak{t}}_{{}_{(\Sigma,\Sigma')}}},\;\;\;\;\;\; \sigma {\centerdot} { \mathfrak{t}}_{{}_{(\Sigma,\Sigma')}} & := &\Sigma \sqcup_{{}_{ \mathfrak{S}}} \bigsqcup{}_{{}_{ \mathfrak{S}}} \left( 
(\downarrow_{{}_{ \mathfrak{S}}} \uparrow^{{}^{ \mathfrak{S}}} \Sigma)
 \cap (\downarrow_{{}_{ \mathfrak{S}}} \!\!\sigma ) \right) .
\end{eqnarray}
We define the corresponding {\em generalized property} ${ \mathfrak{l}}_{{}_{(\Sigma,\Sigma')}}$ straightforwardly and denote the set of generalized properties (resp. generalized minimally disturbing properties) by $\widetilde{ \mathcal{L}}$ (resp.  $\widetilde{ \mathcal{L}}_{{}_{min}}$). 
\end{MyDef}

\begin{MyDef} A generalized property ${ \mathfrak{l}}\in \widetilde{ \mathcal{L}}_{{}_{min}}$ will be said to be {\em a perfect property} (this fact will be denoted ${ \mathfrak{l}}\in \widetilde{ \mathcal{L}}_{{}_{perfect}}$) if $\overline{{ \mathfrak{l}}}$ is also a testable generalized minimally disturbing property. In other words, $\downarrow_{{}_{ \mathfrak{S}}}\uparrow^{{}^{ \mathfrak{S}}} \Sigma_{ \mathfrak{l}}$ is an idealized sub- selection structure of ${ \mathfrak{Q}}_{ \mathfrak{l}}=({ \mathfrak{S}}\smallsetminus \uparrow^{{}^{ \mathfrak{S}}}\Sigma_{\overline{ \mathfrak{l}}})$ and $\downarrow_{{}_{ \mathfrak{S}}}\uparrow^{{}^{ \mathfrak{S}}} \Sigma_{\overline{ \mathfrak{l}}}$ is an idealized sub- selection structure of ${ \mathfrak{Q}}_{\overline{ \mathfrak{l}}}=({ \mathfrak{S}}\smallsetminus \uparrow^{{}^{ \mathfrak{S}}}\Sigma_{{ \mathfrak{l}}})$. As a consequence, {{}}{there exist} two conjugate generalized minimally disturbing yes/no tests{{}}{,} ${ \mathfrak{t}}$ and $\overline{{ \mathfrak{t}}}${{}}{,} leading to ideal first-kind measurements of the respective generalized properties{{}}{,} ${ \mathfrak{l}}$ and $\overline{{ \mathfrak{l}}}$. The corresponding measurement $\centerdot { \mathfrak{t}}$ is said to be {\em a perfect measurement}. The yes/no test ${ \mathfrak{t}}$ will be said to be perfect as well (this fact will be denoted ${ \mathfrak{t}}\in \widetilde{ \mathfrak{T}}_{{}_{perfect}}$).  \end{MyDef}

\subsection{The space of 'Descriptions'}
\label{subsectionspacedescriptionsdomain}

\begin{MyDef}
We denote by ${ \mathcal{D}}$ the following subset of the powerset ${ \mathcal{P}}({ \mathcal{L}}_{{}_{IFKM}})$ : 
\begin{eqnarray}
{ \mathcal{D}}:= \{{ \mathfrak{L}}\subseteq { \mathcal{L}}_{{}_{IFKM}}\;\vert\; \overbrace{{ \mathfrak{L}}}\;\}
\end{eqnarray}
An element of this set corresponds to a family of properties that can be checked by 'conjoint minimally disturbing measurements' as being 'simultaneously actual' on a given sample.  Hence, ${ \mathcal{D}}$ will be called {\em the space of descriptions}.\\
An element $D\in { \mathcal{D}}$ will be eventually denoted $[\![ { \mathfrak{l}}_1\cdots { \mathfrak{l}}_n]\!]$ rather than $\{ { \mathfrak{l}}_1,\cdots ,{ \mathfrak{l}}_n\}$ to emphasize the compatibility between the properties constituting the description.
\end{MyDef}

\begin{lemme}{\bf ["Specker's principle"]} \label{speckerprinciple}
Let us consider a finite family of testable properties ${ \mathfrak{L}}=\{\,{ \mathfrak{l}}_1,\cdots,{ \mathfrak{l}}_N\}$ with ${ \mathfrak{l}}_k\in { \mathcal{L}}_{{}_{IFKM}}$ for any $ k\in \{1,\cdots,N\}$, such that the elements of this family are pairwise compatible, i.e., $\forall k, k'\in \{1,\cdots,N\},\;\overbrace{{ \mathfrak{l}}_k { \mathfrak{l}}_{k'}}$. The family ${ \mathfrak{L}}$ is then a compatible family. In other words,
\begin{eqnarray}
\forall { \mathfrak{L}}=\{\,{ \mathfrak{l}}_1,\cdots,{ \mathfrak{l}}_N\}\;\vert\; { \mathfrak{l}}_k\in { \mathcal{L}}_{{}_{IFKM}}, \;\;\;\;\;\;
\overbrace{ \mathfrak{L}} & \Leftrightarrow & (\,
\forall k, k'\in \{1,\cdots,N\},\;\overbrace{{ \mathfrak{l}}_k { \mathfrak{l}}_{k'}} \,).
\end{eqnarray}  
\end{lemme}
\begin{proof}
{{}}{Given the above} requirement, we consider $T=(\,{ \mathfrak{t}}_k)_{ k\in \{1,\cdots,N\}}$ with ${ \mathfrak{t}}_k\in {\mathfrak{T}}_{{}_{IFKM}}$ and ${ \mathfrak{l}}_k= \lfloor { \mathfrak{t}}_k\rfloor$ for any $ k=1,\cdots,N$.\\
{{}}{First}, using the compatibility relations $\overbrace{{ \mathfrak{l}}_{i} { \mathfrak{l}}_{k}}$ for $k=1,\cdots,i-1$ and Lemma 
\ref{t1dott2IFKM}, we prove (i) $\lfloor { \mathfrak{t}}_{1}{\centerdot} \cdots {\centerdot} { \mathfrak{t}}_{i-1}\rfloor \in { \mathcal{L}}_{{}_{IFKM}}$, (ii) the compatibility relation $\overbrace{{ \mathfrak{l}}_{i} \lfloor { \mathfrak{t}}_{1}{\centerdot} \cdots {\centerdot} { \mathfrak{t}}_{i-1}\rfloor}$ and (iii) ${ \mathfrak{A}}_{{ \mathfrak{t}}_{1}{\centerdot} \cdots {\centerdot} { \mathfrak{t}}_{i-1}}={ \mathfrak{A}}_{{ \mathfrak{t}}_{1}}\cap \cdots \cap { \mathfrak{A}}_{{ \mathfrak{t}}_{i-1}}$ for any $i=1,\cdots,N$.\\
Let us now consider ${ \mathfrak{p}} \in { \mathfrak{Q}}_{{ \mathfrak{t}}_1 {\centerdot} ({ \mathfrak{t}}_2 {\centerdot}(\cdots ({ \mathfrak{t}}_{N-1}{\centerdot} ({ \mathfrak{t}}_{N}))\cdots )}$ (we note that ${ \bot}_{{}_{ \mathfrak{S}}} \in { \mathfrak{Q}}_{{ \mathfrak{t}}_1 {\centerdot} ({ \mathfrak{t}}_2 {\centerdot}(\cdots ({ \mathfrak{t}}_{N-1}{\centerdot} ({ \mathfrak{t}}_{N}))\cdots )} \not= \varnothing $). 
We have then from equation (\ref{Qt1dott2}), $(\,{ \mathfrak{p}} \in  { \mathfrak{Q}}_{{ \mathfrak{t}}_1}\;\;\textit{\rm and}\;\; ({ \mathfrak{p}} {\centerdot}
{ \mathfrak{t}}_1) \in { \mathfrak{Q}}_{{ \mathfrak{t}}_2 {\centerdot}(\cdots ({ \mathfrak{t}}_{N-1}{\centerdot} ({ \mathfrak{t}}_{N}))\cdots )}\,)$.  Let us denote by $(P_i)$ the statement: $(\, ({ \mathfrak{p}}{\centerdot} { \mathfrak{t}}_1{\centerdot}\cdots {\centerdot} { \mathfrak{t}}_{i-1})\in { \mathfrak{A}}_{{ \mathfrak{t}}_1}\cap \cdots \cap { \mathfrak{A}}_{{ \mathfrak{t}}_{i-1}}\cap { \mathfrak{Q}}_{{ \mathfrak{t}}_i} \;\;\textit{\rm and}\;\; ({ \mathfrak{p}}{\centerdot} { \mathfrak{t}}_1{\centerdot}\cdots {\centerdot} { \mathfrak{t}}_{i})\in { \mathfrak{Q}}_{{ \mathfrak{t}}_{i+1} {\centerdot}(\cdots ({ \mathfrak{t}}_{N-1}{\centerdot} ({ \mathfrak{t}}_{N}))\cdots )}\,)$. Let us suppose that $(P_i)$ is satisfied. 
\\
Using the compatibility relation $\overbrace{{ \mathfrak{l}}_{i} \lfloor { \mathfrak{t}}_{1}{\centerdot} \cdots {\centerdot} { \mathfrak{t}}_{i-1}\rfloor}$ and the fact that ${ \mathfrak{t}}_{i}$ leads to an ideal measurement and {{}}{thus} satisfies property (\ref{axiomidealmeasurement}), we deduce that $({ \mathfrak{p}}{\centerdot} { \mathfrak{t}}_1{\centerdot}\cdots {\centerdot} { \mathfrak{t}}_{i})\in { \mathfrak{A}}_{{ \mathfrak{t}}_1}\cap \cdots \cap { \mathfrak{A}}_{{ \mathfrak{t}}_{i-1}}$. Using the fact that ${ \mathfrak{t}}_{i}$ leads to a first-kind measurement and {{}}{therefore} satisfies (\ref{firstkind} \textit{\rm (i)}), we {{}}{also have} $({ \mathfrak{p}}{\centerdot} { \mathfrak{t}}_1{\centerdot}\cdots {\centerdot} { \mathfrak{t}}_{i})\in { \mathfrak{A}}_{{ \mathfrak{t}}_i}$. {{}}{Now using} expression (\ref{Qt1dott2}), we conclude that{{}}{ } property $(P_{i+1})$ is satisfied. \\
As a result, we conclude by recursion that $({ \mathfrak{p}}{\centerdot} { \mathfrak{t}}_1{\centerdot}\cdots {\centerdot} { \mathfrak{t}}_{N})\in { \mathfrak{A}}_{{ \mathfrak{t}}_1}\cap \cdots \cap { \mathfrak{A}}_{{ \mathfrak{t}}_{N}}$. The set ${ \mathfrak{A}}_{{ \mathfrak{t}}_1}\cap \cdots \cap { \mathfrak{A}}_{{ \mathfrak{t}}_{N}}$ is then non-empty, and the family ${ \mathfrak{L}}$ is {{}}{therefore} a compatible family of properties.
\end{proof}
\begin{remark}
The history of this "principle" is recalled in \cite{cabello2012speckers} and some results are given in this paper that {\it "suggest that the principle that pairwise decidable propositions are jointly decidable, together with Boole’s condition that the sum of probabilities of jointly exclusive propositions cannot be higher than one [$\cdots$], which I will collectively call Specker’s principle, may explain quantum contextuality."} \footnote{See  \cite{Kochen} for the original results on non-contextuality in quantum mechanics{{}}{.}}
\end{remark}

\begin{theoreme}{}
The space of descriptions ${ \mathcal{D}}$ is a 'coherence domain' \cite{GirardProofsTypes} associated {{}}{with} the 'web' ${ \mathcal{L}}_{{}_{IFKM}}$ and {{}}{with} the coherence relation $\overbrace{\cdot \;\;\;\cdot}$. In other words, ${ \mathcal{D}}\subseteq { \mathcal{P}}({ \mathcal{L}}_{{}_{IFKM}})$ satisfies
\begin{eqnarray}
&&\forall { \mathfrak{L}}_1, { \mathfrak{L}}_2\in { \mathcal{P}}({ \mathcal{L}}_{{}_{IFKM}}),\;\;\; (\,{ \mathfrak{L}}_1 \subseteq { \mathfrak{L}}_2 \;\textit{\rm and}\; { \mathfrak{L}}_2 \in { \mathcal{D}}\,) \Rightarrow  (\,{ \mathfrak{L}}_1 \in { \mathcal{D}}\,)\\
&&\forall { \mathfrak{l}}\in { \mathcal{L}}_{{}_{IFKM}},\;\;\; \{{ \mathfrak{l}}\}\in  { \mathcal{D}}\\
&&\forall L\in { \mathcal{P}}({ \mathcal{D}})\;\vert\; (\,\forall { \mathfrak{L}}_1, { \mathfrak{L}}_2\in L,\; \overbrace{{ \mathfrak{L}}_1\cup { \mathfrak{L}}_{2}}\,),\;\;\; \bigcup L\;\in\; { \mathcal{D}}.
\end{eqnarray}
\end{theoreme}
\begin{proof}
The first property is a direct consequence of the definition of ${ \mathcal{D}}$. Indeed, $\forall { \mathfrak{L}}_1, { \mathfrak{L}}_2\in { \mathcal{P}}({ \mathcal{L}}_{{}_{IFKM}})$, {{}}{and} the properties ${ \mathfrak{L}}_1 \subseteq { \mathfrak{L}}_2$ and ${ \mathfrak{L}}_2 \in { \mathcal{D}}$ imply $\varnothing \not= \bigcap_{{ \mathfrak{l}}\in { \mathfrak{L}}_2} { \mathfrak{A}}_{ \mathfrak{l}} \subseteq \bigcap_{{ \mathfrak{l}}\in { \mathfrak{L}}_1} { \mathfrak{A}}_{ \mathfrak{l}}$.\\ 
The second property is a trivial consequence of the definition of consistency and the fact that we consider testable properties.\\
The property $(\forall M\in { \mathcal{P}}_{fin}({ \mathcal{D}})\;\vert\; (\,\forall { \mathfrak{L}}_1, { \mathfrak{L}}_2\in M,\; \overbrace{{ \mathfrak{L}}_1\cup { \mathfrak{L}}_{2}}\,),\;\;\; \bigcup M\;\in\; { \mathcal{D}})$ 
is a direct consequence of Lemma \ref{speckerprinciple}. Hence, for any $L\in { \mathcal{P}}({ \mathcal{D}})\;\vert\; (\,\forall { \mathfrak{L}}_1, { \mathfrak{L}}_2\in L,\; \overbrace{{ \mathfrak{L}}_1\cup { \mathfrak{L}}_{2}}\,)$, we deduce that $ N\;\in\; { \mathcal{D}}$ for any $N\subseteq_{fin}\bigcup L$, which allows{{}}{ }the third property {{}}{to be deduced}.
\end{proof}

\begin{MyDef}\label{Defsimultaneoustest}
For any description $D\in { \mathcal{D}}$, and as soon as $\Sigma_D := \bigsqcup{}_{{}_{{ \mathfrak{l}}\in D}}\Sigma_{ \mathfrak{l}}$ and $\Sigma'_D := \bigsqcap{}_{{}_{{ \mathfrak{l}}\in D}}\Sigma_{\overline{ \mathfrak{l}}}$ satisfy $\antiwidehat{\Sigma_D \Sigma_D'}{}^{{}^{ \mathfrak{S}}}$, we define a generalized yes/no test denoted ${ \mathfrak{t}}_{D}$ by ${ \mathfrak{t}}_{{}_D}:={ \mathfrak{t}}_{{}_{(\Sigma_D,\Sigma'_D)}}$. 
This generalized yes/no test characterizes the conjoint measurement of the compatible properties belonging to the description $D$.
\end{MyDef}

\subsection{From discriminating yes/no tests to the complete axiomatics on the space of states}

Let us introduce a notion that will reveal to be fundamental in the next subsection when we will have to consider the definition of an orthogonality relation on the space of states. 

\begin{MyDef}
A collection ${ \mathfrak{U}}$ of generalized yes/no tests is said to be {\em complete} iff
\begin{eqnarray}
&& \forall \Sigma \in { \mathfrak{S}}, \exists \Sigma'\in { \mathfrak{S}}\;\vert\; \;\;\;\;\; { \mathfrak{t}}_{{}_{(\Sigma,\Sigma')}} \in { \mathfrak{U}},\label{Defpropcompleteatom}
\end{eqnarray}
\end{MyDef}

\begin{MyDef}
A collection ${ \mathfrak{U}}$ of generalized yes/no tests is said to be {\em irredundant} iff
\begin{eqnarray}
\forall { \mathfrak{t}}_1, { \mathfrak{t}}_2\in { \mathfrak{U}} && \Sigma_{\lfloor { \mathfrak{t}}_1\rfloor} \sqsubseteq_{{}_{ \mathfrak{S}}} \Sigma_{\lfloor { \mathfrak{t}}_2\rfloor} \;\;\Leftrightarrow\;\;  \Sigma_{\overline{\lfloor { \mathfrak{t}}_2\rfloor}} \sqsubseteq_{{}_{ \mathfrak{S}}} \Sigma_{\overline{\lfloor { \mathfrak{t}}_1\rfloor}} \label{DefpropIrredundancy}
\end{eqnarray}
\end{MyDef}

\begin{MyDef}
A collection ${ \mathfrak{U}}$ of generalized yes/no tests is said to be {\em closed} iff
\begin{eqnarray}
\forall { \mathfrak{t}}\in  { \mathfrak{U}}, && \overline{ \mathfrak{t}} \in { \mathfrak{U}}\label{Defpropclose2}\\
%\forall { \mathfrak{C}},{ \mathfrak{C}}'\subseteq_{Chain} { \mathfrak{S}} \;\textit{\rm and}\; \Sigma:=\bigsqcup{}_{{}_{ \mathfrak{S}}} { \mathfrak{C}}, \Sigma':=\bigsqcap{}_{{}_{ \mathfrak{S}}} { \mathfrak{C}}' && { \mathfrak{t}}_{{}_{(\Sigma,\Sigma')}}\in { \mathfrak{U}} \label{Defproplimit}\\
\forall { \mathfrak{t}}_{{}_{(\Sigma_1,\Sigma'_1)}},{ \mathfrak{t}}_{{}_{(\Sigma_2,\Sigma'_2)}}\in  { \mathfrak{U}}\;\vert\; \widehat{\Sigma_1\Sigma_2}{}^{{}^{ \mathfrak{S}}}&& { \mathfrak{t}}_{{}_{(\Sigma_1\sqcup_{{}_{ \mathfrak{S}}}\Sigma_2,\Sigma'_1\sqcap_{{}_{ \mathfrak{S}}}\Sigma_2')}}\in { \mathfrak{U}}\label{Defpropfusion}
\end{eqnarray}
\end{MyDef}
\begin{remark}
The property (\ref{Defpropfusion}) implies more generally
\begin{eqnarray}
\forall D\in { \mathcal{D}}\;\vert \;\forall { \mathfrak{l}}\in D,\; { \mathfrak{l}} \in \lfloor { \mathfrak{U}}\rfloor, && { \mathfrak{t}}_{{}_{D}}\in { \mathfrak{U}} \label{Defpropcompletedemorgan}
\end{eqnarray}
\end{remark}

\begin{MyDef}
A {\em scheme of yes/no tests} is a complete, irredundant and closed collection of generalized yes/no tests.
\end{MyDef} 

Let us now introduce a class of generalized yes/no tests that will be fundamental to clarify the last axioms on the space of states.

\begin{MyDef}
A couple of states $(\sigma,\sigma')\in { \mathfrak{S}}^{\times 2}$ is said to be {\em quasi-consistent} (this fact is denoted $\sigma \bowtie_{{}_{ \mathfrak{S}}}\sigma'$) according to the following definition :
\begin{eqnarray}
\sigma \bowtie_{{}_{ \mathfrak{S}}}\sigma' & :\Leftrightarrow & (\,\forall \sigma'' \sqsubset_{{}_{ \mathfrak{S}}}\sigma', \widehat{\sigma\sigma''}{}^{{}^{ \mathfrak{S}}}\;\;\textit{\rm and}\;\; \forall \sigma'' \sqsubset_{{}_{ \mathfrak{S}}}\sigma, \widehat{\sigma'\sigma''}{}^{{}^{ \mathfrak{S}}}\,).\label{bowtie}
\end{eqnarray}
We have obviously $\widehat{\sigma\sigma'}{}^{{}^{ \mathfrak{S}}} \Rightarrow \sigma \bowtie_{{}_{ \mathfrak{S}}}\sigma'$.\\
We will denote $\sigma \antiwidehat{\bowtie}_{{}_{ \mathfrak{S}}}\sigma'  :\Leftrightarrow  (\sigma \bowtie_{{}_{ \mathfrak{S}}}\sigma'\;\textit{\rm and}\; \antiwidehat{\sigma\sigma'}{}^{{}^{ \mathfrak{S}}})$.
\end{MyDef}

\begin{MyDef}
We will say that a generalized yes/no test ${ \mathfrak{t}}_{{}_{(\Sigma,\Sigma')}}$ is a {\em discriminating yes/no test} iff $\Sigma \antiwidehat{\bowtie}_{{}_{ \mathfrak{S}}}\Sigma'$. In other words, 
\begin{eqnarray}
 \widetilde{ \mathfrak{T}}_{{}_{disc}} & := & \{\;
{ \mathfrak{t}}_{{}_{(\Sigma,\Sigma')}} \in \widetilde{ \mathfrak{T}}\;\;\vert\;\; \Sigma\; \antiwidehat{\bowtie}_{{}_{ \mathfrak{S}}}\Sigma'\;\}.\label{deftdisc}
\end{eqnarray}
The corresponding set of {\em discriminating properties} is denoted $\widetilde{ \mathcal{L}}_{{}_{disc}}$.
\end{MyDef}

We will constrain further the space of states by requiring the existence of schemes of discriminating yes/no tests.

\begin{MyReq}\label{axiomscheme}
The space of states is such that there exists a scheme of discriminating yes/no tests.
\end{MyReq}

Now, we can make a summary of the axioms postulated for the space of states.
\begin{theoreme}\label{quantumdomain}
The space of states is a projective domain,  with no complete meet-irreducible elements except maximal elements ({\bf Axiom \ref{Axiomnotype2}}),  such that a scheme of discriminating tests exists ({\bf Axiom \ref{axiomscheme}}).
\end{theoreme}

Remark \ref{exampleboolean} clarifies the precise consequences of {\bf Axiom \ref{axiomscheme}} for the space of states.

\begin{remark}\label{exampleboolean}
A natural question to address to our program is about the general existence of examples of domains solution of the constraints given in Theorem \ref{quantumdomain}. \\
In fact, it is easy to build a rather general class of examples of space of states.  Let us consider a projective lattice equipped with an orthocomplementation (i.e.  an involutive and order-reversing complementation). It will be denoted $({ \mathfrak{L}}, \sqcup_{{}_{ \mathfrak{L}}}, \sqcap_{{}_{ \mathfrak{L}}},\star)$.  We will require that the complete meet-irreducible elements be the co-atoms of ${ \mathfrak{L}}$.  Let us denote by $\bot_{{}_{ \mathfrak{L}}}$ (resp. $\top_{{}_{ \mathfrak{L}}}$) the bottom element (resp. top element) of ${ \mathfrak{L}}$.  We will assume that the join-irreducible elements of ${ \mathfrak{L}}$ are the atoms of ${ \mathfrak{L}}$.\\
Let us denote ${ \mathfrak{S}}:={ \mathfrak{L}}\smallsetminus \{\top_{{}_{ \mathfrak{L}}}\}$.  The laws $\sqcup_{{}_{ \mathfrak{S}}}$,  $\sqcap_{{}_{ \mathfrak{S}}}$ and $\star$ are induced from the laws on ${ \mathfrak{L}}$ and we have $\bot_{{}_{ \mathfrak{S}}}=\bot_{{}_{ \mathfrak{L}}}$. \\
${ \mathfrak{L}}$ being modular, ${ \mathfrak{S}}$ is then conditionally-modular. ${ \mathfrak{L}}$ being complete, ${ \mathfrak{S}}$ is then bounded-complete. ${ \mathfrak{L}}$ being atomistic, ${ \mathfrak{S}}$ is then atomistic.  ${ \mathfrak{L}}$ being meet-continuous, ${ \mathfrak{S}}$ is then meet-continuous. ${ \mathfrak{L}}$ being relatively-complemented, ${ \mathfrak{S}}$ is then relatively-complemented.  Endly, it is easy to check that ${ \mathfrak{S}}$ is directed-complete using the completeness of ${ \mathfrak{L}}$. As a result,  ${ \mathfrak{S}}$ is a projective domain.\\
Due to our requirement on the complete meet-irreducible elements of ${ \mathfrak{L}}$,  {\bf Axiom  \ref{Axiomnotype2}} is then immediately satisfied. \\
We now intent to consider the last axiom.  But before that, let us establish a simple result about the discriminating relation.\\
We firstly note that 
\begin{eqnarray}
\forall u,x\in { \mathfrak{S}}, && \antiwidehat{u\;x}{}^{{}^{ \mathfrak{S}}}  \Leftrightarrow  (u\sqcup_{{}_{ \mathfrak{L}}} x=\top_{{}_{ \mathfrak{L}}}).
\end{eqnarray}   
Let us consider $u,v, x\in { \mathfrak{S}}$ such that $u \sqcoversupset_{{}_{ \mathfrak{L}}} v$ and $u\parallel_{{}_{ \mathfrak{L}}} x$, we have the following case analysis : 
\begin{itemize}
\item $x \sqsubseteq_{{}_{ \mathfrak{L}}} v$ : impossible as it contradicts $x\parallel_{{}_{ \mathfrak{L}}} u$,\\
\item $x \sqsupset_{{}_{ \mathfrak{L}}} v$ : we have then $x \sqcup_{{}_{ \mathfrak{L}}} v=x$ and then $((x \sqcup_{{}_{ \mathfrak{L}}} u)  \sqsupset_{{}_{ \mathfrak{L}}} x  \sqsupset_{{}_{ \mathfrak{L}}} v,\;\; (x \sqcup_{{}_{ \mathfrak{L}}} u)  \sqsupset_{{}_{ \mathfrak{L}}} u  \sqcoversupset_{{}_{ \mathfrak{L}}} v)$ which implies $(x \sqcup_{{}_{ \mathfrak{L}}} u ) \sqcoversupset_{{}_{ \mathfrak{L}}} x = (x \sqcup_{{}_{ \mathfrak{L}}} v)$ using upper semi-modularity property in ${ \mathfrak{L}}$,\\
\item  $x \parallel_{{}_{ \mathfrak{L}}} v$ : we distinguish three sub-cases
\begin{itemize}
\item $(x \sqcup_{{}_{ \mathfrak{L}}} v)  \sqsupseteq_{{}_{ \mathfrak{L}}} u$ : impossible as it implies  $x  \sqsupseteq_{{}_{ \mathfrak{L}}} u$ which contradicts $u\parallel_{{}_{ \mathfrak{L}}} x$,\\
\item $(x \sqcup_{{}_{ \mathfrak{L}}} v)  \parallel_{{}_{ \mathfrak{L}}} u$ : we have then $((x \sqcup_{{}_{ \mathfrak{L}}} u)  \sqsupset_{{}_{ \mathfrak{L}}} (x\sqcup_{{}_{ \mathfrak{L}}} v)  \sqsupset_{{}_{ \mathfrak{L}}} v,\;\; (x \sqcup_{{}_{ \mathfrak{L}}} u)  \sqsupset_{{}_{ \mathfrak{L}}} u  \sqcoversupset_{{}_{ \mathfrak{L}}} v)$ which implies $(x \sqcup_{{}_{ \mathfrak{L}}} u ) \sqcoversupset_{{}_{ \mathfrak{L}}} x = (x \sqcup_{{}_{ \mathfrak{L}}} v)$ using upper semi-modularity property in ${ \mathfrak{L}}$,\\
\item $(x \sqcup_{{}_{ \mathfrak{L}}} v)  \sqsupset_{{}_{ \mathfrak{L}}} u$ : this sub-case implies immediately $(x \sqcup_{{}_{ \mathfrak{L}}} v)=(x \sqcup_{{}_{ \mathfrak{L}}} u)$, however this configuration $(x \sqcup_{{}_{ \mathfrak{L}}}u)=(x \sqcup_{{}_{ \mathfrak{L}}} v)$ implies $u=v$ (which contradicts $u \sqcoversupset_{{}_{ \mathfrak{L}}} v$) as soon as $(x \sqcap_{{}_{ \mathfrak{L}}}u)=(x \sqcap_{{}_{ \mathfrak{L}}} v)$, due to modularity property (\ref{propmodularity}).
\end{itemize}
\end{itemize}
As a conclusion,  using the defining property 
\begin{eqnarray}
(u \bowtie_{{}_{ \mathfrak{S}}} x) & \Leftrightarrow & (\forall v \sqcoversubset_{{}_{ \mathfrak{L}}} u,  (v\sqcup_{{}_{ \mathfrak{L}}} x)\sqsubset_{{}_{ \mathfrak{L}}} \top_{{}_{ \mathfrak{L}}}\;\;\textit{\rm and}\;\; \forall y \sqcoversubset_{{}_{ \mathfrak{L}}} x,  (u\sqcup_{{}_{ \mathfrak{L}}} y)\sqsubset_{{}_{ \mathfrak{L}}} \top_{{}_{ \mathfrak{L}}}),
\end{eqnarray} 
we finally obtain the simple equivalence :
\begin{eqnarray}
(u \;\antiwidehat{\bowtie}_{{}_{ \mathfrak{S}}} x ) & \Leftrightarrow & ((u\sqcup_{{}_{ \mathfrak{L}}} x)=\top_{{}_{ \mathfrak{L}}}
\;\;\textit{\rm and}\;\; 
(u\sqcap_{{}_{ \mathfrak{L}}}x)= \bot_{{}_{ \mathfrak{L}}}).
\end{eqnarray}   
Let us now come back to our last axiom.  Let us choose 
\begin{eqnarray}
{ \mathfrak{U}} &:=& \{\, { \mathfrak{t}}_{{}_{(\Sigma,\Sigma^\star)}}\;\vert\; \Sigma\in { \mathfrak{S}}\,\}.
\end{eqnarray}
The completeness condition for the scheme is satisfied as soon as
\begin{eqnarray}
&&\forall \sigma_1\in { \mathfrak{S}}^\ast,\exists \sigma'_1:=\sigma_1^\star\in { \mathfrak{S}} \;\vert (\sigma_1\antiwidehat{\bowtie}_{{}_{ \mathfrak{S}}}\sigma_1'),
\label{qdcomplete}
\end{eqnarray}
i.e., as soon as ${ \mathfrak{L}}$ is ortho-complemented.\\
Secondly, the irredundancy condition for the scheme is satisfied as soon as
\begin{eqnarray}
\forall \sigma_1,\sigma_2\in { \mathfrak{S}}^\ast, & &  (\sigma_1\sqsubseteq_{{}_{ \mathfrak{S}}}\sigma_2)\;\Rightarrow\; (\sigma_2^\star\sqsubseteq_{{}_{ \mathfrak{S}}}\sigma_1^\star),\;\;\;\;\;\;\;\;\;\;\;\;\label{qdirredundancy}
\end{eqnarray}
i.e., as soon as the star operation satisfies the order reversing property.\\
Endly, the closedness property for the scheme is satisfied as soon as the property
\begin{eqnarray}
\hspace{-0.5cm}\forall \sigma_1,\sigma_2\in { \mathfrak{S}}^\ast,\;\;\;\;\;(\widehat{\sigma_1\sigma_2}{}^{{}^{ \mathfrak{S}}}) &\Rightarrow & ((\sigma_1\sqcup_{{}_{ \mathfrak{S}}}\sigma_2)\antiwidehat{\bowtie}_{{}_{ \mathfrak{S}}}(\sigma_1^\star\sqcap_{{}_{ \mathfrak{S}}}\sigma_2^\star))\;\;\;\;\;\;\;\;\;\;\;\; \label{qdsimultaneous}
\end{eqnarray}
 expresses the DeMorgan's law satisfied by the star operation (this law is also a simple consequence of the order reversing property and the involutive property).\\
As a result, ${ \mathfrak{S}}$ is a well defined space of states.\\
The final conclusion of this remark is simple : {\bf Axiom \ref{axiomscheme}} expresses the necessity for ${ \mathfrak{S}}$ to be equipped with an ortho-complementation.
\end{remark}

Let us now formulate a conjecture relative to the set of discriminating yes/no tests. 

\begin{MyConj}
Any discriminating yes/no test is a perfect yes/no test. In other words,
\begin{eqnarray}
\widetilde{ \mathfrak{T}}_{{}_{disc}} & \subseteq & \widetilde{ \mathfrak{T}}_{{}_{perfect}}.
\end{eqnarray}
\end{MyConj}
\begin{remark}
This is not the purpose of this paper to clarify the proof of this property and we prefer to postpone this analysis to a forthcoming paper. Nevertheless, we want to note that this result is trivial in a basic class of spaces of states. Indeed, if $({ \mathfrak{L}}, \sqcup_{{}_{ \mathfrak{L}}}, \sqcap_{{}_{ \mathfrak{L}}},\star)$ is a finite boolean lattice and if we define ${ \mathfrak{S}}:=({ \mathfrak{L}}\smallsetminus \{\top_{{}_{ \mathfrak{L}}}\})$, we obtain immediately $({ \mathfrak{S}}\smallsetminus (\uparrow^{{}^{{ \mathfrak{S}}}} \Sigma^\star))=(\downarrow_{{}_{{ \mathfrak{S}}}}\!\uparrow^{{}^{{ \mathfrak{S}}}}\!\! \Sigma)$ for any $\Sigma\in { \mathfrak{S}}$, or, in other words, $\lceil { \mathfrak{Q}}_{ \mathfrak{l}}\rceil =\lceil { \mathfrak{K}}_{ \mathfrak{l}}\rceil$ for any discriminating property ${ \mathfrak{l}}$. Hence, any discriminating property is a perfect property.
\end{remark}

\subsection{Orthogonality relation on the space of states }

In the present subsection, ${ \mathfrak{U}}$ is a fixed scheme of generalized yes/no tests. When the choice ${ \mathfrak{U}}\subseteq \widetilde{ \mathfrak{T}}_{{}_{disc}}$ will have to be done, it will be mentioned explicitly.

\begin{MyDef} Two states $\sigma_1,\sigma_2 \in { \mathfrak{S}}^\ast$ are said to be {\em orthogonal} (this fact will be denoted $\sigma_1 \perp \sigma_2$, the dependence with respect to ${ \mathfrak{U}}$ is intentionally erased) iff 
they can be distinguished unambiguously by a statement associated {{}}{with} a perfect property. In other words,
 \begin{eqnarray} 
 \forall \sigma_1,\sigma_2 \in { \mathfrak{S}}^\ast,\;\; (\, \sigma_1 \perp \sigma_2\,)&\Leftrightarrow & (\,\exists { \mathfrak{l}}\in \lfloor{ \mathfrak{U}}\rfloor	 \;\vert\; \sigma_1\in \lceil { \mathfrak{A}}_{ \mathfrak{l}}\rceil \;\;\textit{\rm and}\;\; \sigma_2\in \lceil { \mathfrak{A}}_{\overline{ \mathfrak{l}}}\rceil \,).\label{def1ortho}\label{def2ortho}\end{eqnarray}
 This orthogonality relation is obviously symmetric and anti-reflexive.\\
We denote as usual
\begin{eqnarray}
\forall S \subseteq { \mathfrak{S}}^\ast, && S^{{\perp}} :=\{\,\sigma'\in { \mathfrak{S}}\;\vert\; \forall \sigma \in S,\; \sigma \perp \sigma'\,\}.
\end{eqnarray}
\end{MyDef}  
\begin{remark}
Moreover,  we have obviously
\begin{eqnarray} \forall \sigma_1,\sigma_2 \in { \mathfrak{S}}^\ast, && \sigma_1 \perp \sigma_2\;\;\Rightarrow \;\; \antiwidehat{\sigma_1\;\sigma_2}{}^{\!{}^{ \mathfrak{S}}}. \end{eqnarray}
Indeed, following{{}}{ }the definition of {{}}{the} consistency relation, we then note that, for any $\sigma_1,\sigma_2\in { \mathfrak{S}}^\ast$, the property $\widehat{\sigma_1\sigma_2}^{\!{}_{ \mathfrak{S}}}$ {{}}{immediately implies} $\forall { \mathfrak{t}}\in { \mathfrak{U}}, \widehat{{ \mathfrak{e}}(\sigma_1,{ \mathfrak{t}}){ \mathfrak{e}}(\sigma_2,{ \mathfrak{t}})}^{\!\!\!{}_{ \mathfrak{B}}}$. By negation, we obtain that $(\,\exists { \mathfrak{l}}\in \lfloor{ \mathfrak{U}}\rfloor \;\vert\; \sigma_1\in \lceil { \mathfrak{A}}_{ \mathfrak{l}}\rceil \;\;\textit{\rm and}\;\; \sigma_2\in \lceil { \mathfrak{A}}_{\overline{ \mathfrak{l}}}\rceil \,)$ implies $\antiwidehat{\sigma_1\;\sigma_2}{}^{\!{}^{ \mathfrak{S}}}$.
\end{remark}

\begin{lemme}
For any $\sigma\in { \mathfrak{S}}^\ast$, $\{\sigma\}^\perp$ is a non-empty filter. 
\end{lemme}
\begin{proof}
Firstly, let us consider $\sigma_1,\sigma_2\in { \mathfrak{S}}^\ast$ such that $\sigma_1 \sqsubseteq_{{}_{ \mathfrak{S}}}\sigma_2$ and let us assume that $\sigma_1 \perp \sigma$.  We have $(\,\exists { \mathfrak{l}}\in \lfloor{ \mathfrak{U}}\rfloor	 \;\vert\; \sigma\in \lceil { \mathfrak{A}}_{ \mathfrak{l}}\rceil \;\;\textit{\rm and}\;\; \sigma_1\in \lceil { \mathfrak{A}}_{\overline{ \mathfrak{l}}}\rceil \,)
\Rightarrow (\,\exists { \mathfrak{l}}\in \lfloor{ \mathfrak{U}}\rfloor	 \;\vert\; \sigma\in \lceil { \mathfrak{A}}_{ \mathfrak{l}}\rceil \;\;\textit{\rm and}\;\; \sigma_2\in \lceil { \mathfrak{A}}_{\overline{ \mathfrak{l}}}\rceil \,)
\Leftrightarrow \sigma\in \{\sigma_2\}^\perp.$\\ Secondly, let us consider $S\subseteq { \mathfrak{S}}$ such that $\forall \sigma'\in S, \sigma \perp \sigma'$. There exists a family of generalized properties $({ \mathfrak{l}}_{\sigma''})_{\sigma''\in S}\subseteq \lfloor{ \mathfrak{U}}\rfloor$ such that $\forall \sigma'\in S$,  $\Sigma_{{ \mathfrak{l}}_{\sigma'}}\sqsubseteq_{{}_{ \mathfrak{S}}} \sigma$ and $\Sigma_{\overline{{ \mathfrak{l}}_{\sigma'}}}\sqsubseteq_{{}_{ \mathfrak{S}}} \sigma'$. Using relation   (\ref{Defpropcompletedemorgan}), we deduce that there exists a generalized property ${ \mathfrak{l}}$ defined by $\Sigma_{ \mathfrak{l}}:= \bigsqcup{}_{{}_{\sigma'\in S}} \Sigma_{{ \mathfrak{l}}_{\sigma'}} \sqsubseteq_{{}_{ \mathfrak{S}}} \sigma$ and $\Sigma_{\overline{ \mathfrak{l}}} := \bigsqcap{}_{{}_{\sigma'\in S}} \Sigma_{\overline{{ \mathfrak{l}}_{\sigma'}}}$. And we have then $\Sigma_{ \mathfrak{l}}\sqsubseteq_{{}_{ \mathfrak{S}}} \sigma$ and $\Sigma_{\overline{ \mathfrak{l}}}\sqsubseteq_{{}_{ \mathfrak{S}}} \sigma',\forall \sigma'\in S$. In other words,  $\sigma \perp (\bigsqcap{}_{{}_{ \mathfrak{S}}} S)$.
\end{proof}

As a consequence, we will adopt the following definition
\begin{MyDef}
The space of states is equipped with a unary operation defined as follows
\begin{eqnarray}
\forall \sigma \in { \mathfrak{S}}^\ast,\; && \sigma^\star := \bigsqcap{}_{{}_{ \mathfrak{S}}} \{\sigma\}^\perp.\label{stardef}
\end{eqnarray}
\end{MyDef}

\begin{lemme}{}
\begin{eqnarray}
\forall \sigma \in { \mathfrak{S}}^\ast, && \{\sigma\}^\perp = \uparrow^{{}^{ \mathfrak{S}}} \!\!\!\sigma^\star, \label{starprop1}\\
\forall \sigma_1,\sigma_2 \in { \mathfrak{S}}^\ast, && ( \sigma_1 \perp \sigma_2 ) \Leftrightarrow ( \sigma_1^\star \sqsubseteq_{{}_{ \mathfrak{S}}} \sigma_2 ),\label{starprop2}\\
\forall \sigma_1,\sigma_2 \in { \mathfrak{S}}^\ast, && ( \sigma_1 \sqsubseteq_{{}_{ \mathfrak{S}}}\sigma_2 ) \Rightarrow ( \sigma_2^\star \sqsubseteq_{{}_{ \mathfrak{S}}} \sigma_1^\star ).\label{starorderreversing}\\
\forall \sigma \in { \mathfrak{S}}^\ast, \forall { \mathfrak{l}}\in { \mathcal{L}},&& \widetilde{ \mathfrak{e}}_{\overline{ \mathfrak{t}}}(\sigma)= \widetilde{ \mathfrak{e}}_{{ \mathfrak{t}}}(\sigma^\star).\label{esigmalbaresigmastarl}
\end{eqnarray}
\end{lemme}
\begin{proof}
The properties (\ref{starprop1}) and (\ref{starprop2}) are trivial consequences of the defining property (\ref{stardef}).\\
Let us consider $\sigma_1,\sigma_2\in { \mathfrak{S}}^\ast$ such that $\sigma_1 \sqsubseteq_{{}_{ \mathfrak{S}}}\sigma_2.$ We have $\sigma\in \{\sigma_1\}^\perp \Leftrightarrow (\,\exists { \mathfrak{l}}\in \lfloor{ \mathfrak{U}}\rfloor	 \;\vert\; \sigma_1\in \lceil { \mathfrak{A}}_{ \mathfrak{l}}\rceil \;\;\textit{\rm and}\;\; \sigma\in \lceil { \mathfrak{A}}_{\overline{ \mathfrak{l}}}\rceil \,)
\Rightarrow (\,\exists { \mathfrak{l}}\in \lfloor{ \mathfrak{U}}\rfloor \;\vert\; \sigma_2\in \lceil { \mathfrak{A}}_{ \mathfrak{l}}\rceil \;\;\textit{\rm and}\;\; \sigma\in \lceil { \mathfrak{A}}_{\overline{ \mathfrak{l}}}\rceil \,)
\Leftrightarrow \sigma\in \{\sigma_2\}^\perp.$
We then have
\begin{eqnarray}
\sigma_1 \sqsubseteq_{{}_{ \mathfrak{S}}}\sigma_2 &\Rightarrow & \{\sigma_1\}^\perp \subseteq \{\sigma_2\}^\perp\;\; \Rightarrow\;\; (\bigsqcap{}_{{}_{ \mathfrak{S}}} \{\sigma_1\}^\perp ) \sqsubseteq_{{}_{ \mathfrak{S}}} (\bigsqcap{}_{{}_{ \mathfrak{S}}}\{\sigma_2\}^\perp).
\end{eqnarray}
As a result, we obtain the order-reversing property (\ref{starorderreversing}) of the unary operation $\star$. 
\end{proof}

\begin{lemme}\label{lemmeconsequenceirredundancy}
\begin{eqnarray}
&& \forall \sigma \in { \mathfrak{S}},\; \existunique\; { \mathfrak{l}}_\sigma \in \lfloor{ \mathfrak{U}}\rfloor\;\vert\;  (\Sigma_{{ \mathfrak{l}}_\sigma}=\sigma,\;\;\;\; \Sigma_{\overline{{ \mathfrak{l}}_\sigma}}=\sigma^\star)
\end{eqnarray}
\end{lemme}
\begin{proof}
From $\sigma^\star \in \{\sigma\}^\perp$ and the defining property (\ref{def2ortho}), we deduce that there exists ${ \mathfrak{l}}_\sigma \in \lfloor{ \mathfrak{U}}\rfloor$ such that $\Sigma_{{ \mathfrak{l}}_\sigma}\sqsubseteq_{{}_{ \mathfrak{S}}}\sigma$ and $\Sigma_{\overline{{ \mathfrak{l}}_\sigma}}\sqsubseteq_{{}_{ \mathfrak{S}}}\sigma^\star$. From completeness property   (\ref{Defpropcompleteatom}),  we know that, for any $\sigma \in { \mathfrak{S}}$, there exists a $\Sigma'\in { \mathfrak{S}}$ such that ${ \mathfrak{l}}_{{}_{(\sigma,\Sigma')}}\in \lfloor{ \mathfrak{U}}\rfloor$.  We note by the way that $\Sigma'\in \{\sigma\}^\perp$ and then $\Sigma'\sqsupseteq_{{}_{ \mathfrak{S}}} \bigsqcap{}_{{}_{ \mathfrak{S}}}\{\sigma\}^\perp=\sigma^\star$. Using irredundancy property (\ref{DefpropIrredundancy}), we deduce that $\sigma^\star \sqsubseteq_{{}_{ \mathfrak{S}}}\Sigma' \sqsubseteq_{{}_{ \mathfrak{S}}} \Sigma_{\overline{{ \mathfrak{l}}_\sigma}}  \sqsubseteq_{{}_{ \mathfrak{S}}} \sigma^\star$, and then $\Sigma_{\overline{{ \mathfrak{l}}_\sigma}}  = \sigma^\star$. Using once again irredundancy property (\ref{DefpropIrredundancy}), we deduce that $\Sigma_{{{ \mathfrak{l}}_\sigma}}  = \sigma$.
\end{proof}

\begin{lemme}{}
\begin{eqnarray}
\forall \sigma \in { \mathfrak{S}}^\ast, && \sigma^{\star\star}=\sigma.\label{starinvolution}
\\
\forall S \subseteq_{fin}^{\not=\varnothing} { \mathfrak{S}}^\ast \;\vert\; \widehat{S} \,{}^{{}^{ \mathfrak{S}}}, && (\bigsqcup{}_{{}_{ \mathfrak{S}}} S)^\star = \bigsqcap{}_{{}_{\sigma \in S}}\;  \sigma^\star \label{stardemorgan}
\end{eqnarray}
\end{lemme}
\begin{proof}
Using Lemma \ref{lemmeconsequenceirredundancy} and property $(\;\forall { \mathfrak{l}}\in \lfloor{ \mathfrak{U}}\rfloor, \;\;{ \mathfrak{l}}=\overline{\overline{{ \mathfrak{l}}}}\;\;)$, we deduce property (\ref{starinvolution}).\\
As usual, property (\ref{stardemorgan}) is a direct consequence of properties (\ref{starinvolution}) and (\ref{starorderreversing}). To derive it directly, using Lemma \ref{lemmeconsequenceirredundancy}, we introduce, for any $\sigma\in S$,  the property ${ \mathfrak{l}}_\sigma$ such that $\Sigma_{{ \mathfrak{l}}_\sigma}=\sigma$ and $\Sigma_{\overline{{ \mathfrak{l}}_\sigma}}=\sigma^\star$. We now deduce, from property  (\ref{Defpropcompletedemorgan}), that there exists a generalized property ${ \mathfrak{l}}\in \lfloor{ \mathfrak{U}}\rfloor$ such that $\Sigma_{ \mathfrak{l}} = \bigsqcup{}_{{}_{\sigma\in S}} \Sigma_{{ \mathfrak{l}}_\sigma}=\bigsqcup{}_{{}_{ \mathfrak{S}}} S$ and $\Sigma'_{ \mathfrak{l}} = \bigsqcap {}_{{}_{\sigma\in S}} \Sigma_{\overline{{ \mathfrak{l}}_\sigma}}=\bigsqcap{}_{{}_{\sigma \in S}}\;  \sigma^\star$. Using Lemma \ref{lemmeconsequenceirredundancy}, we deduce $\Sigma'_{ \mathfrak{l}}=\Sigma_{ \mathfrak{l}}^\star$, i.e.  property (\ref{stardemorgan}).
\end{proof}

\begin{lemme}{}
The closure operator associated with the orthogonality relation $\perp$ satisfies
\begin{eqnarray}
\forall S \subseteq { \mathfrak{S}}^\ast, && S^{\perp \perp}=\uparrow^{{}^{ \mathfrak{S}}}\!\!\!(\bigsqcap{}_{{}_{ \mathfrak{S}}} S).\label{Sperpperp}
\end{eqnarray}
\end{lemme}
\begin{proof} First of all, we have $S^\perp  =  \bigcap{}_{{}_{\sigma\in S}} \{\sigma\}^\perp  =  \uparrow^{{}^{ \mathfrak{S}}} \bigsqcup{}_{{}_{\sigma\in S}} \sigma^\star$. As a consequence, we have $S^{\perp\perp}  = (\uparrow^{{}^{ \mathfrak{S}}} \bigsqcup{}_{{}_{\sigma\in S}} \sigma^\star)^\perp = \{ \bigsqcup{}_{{}_{\sigma\in S}} \{\sigma\}^\star\}^\perp = \uparrow^{{}^{ \mathfrak{S}}} (\bigsqcup{}_{{}_{\sigma\in S}} \sigma^\star)^\star$. Now, using De Morgan's law (\ref{stardemorgan}), we deduce $S^{\perp\perp}  = \uparrow^{{}^{ \mathfrak{S}}} (\bigsqcap{}_{{}_{\sigma\in S}} \sigma^{\star\star}).$ We now use the involutive property (\ref{starinvolution}) to deduce $S^{\perp\perp}= \uparrow^{{}^{ \mathfrak{S}}}\!\!\!(\bigsqcap{}_{{}_{ \mathfrak{S}}} S)$.
\end{proof}

\subsection{The space of ortho-closed subsets of pure states as a Hilbert lattice}

In this subsection, we will impose ${ \mathfrak{U}}\subseteq  \widetilde{ \mathfrak{T}}_{{}_{disc}}$.

\begin{MyDef}
The space of pure states ${ \mathfrak{S}}_{{}_{pure}}$ inherits an orthogonality relation (denoted $\underline{\perp}$) from the orthogonality relation $\perp$ defined on the whole space of states ${ \mathfrak{S}}$. 
\begin{eqnarray} \forall \sigma_1,\sigma_2 \in { \mathfrak{S}}_{{}_{pure}}, \;\;\; (\, \sigma_1 \underline{\perp} \sigma_2\,)& : \Leftrightarrow & (\, \sigma_1 {\perp} \sigma_2\,),\\
\forall S \subseteq { \mathfrak{S}}_{{}_{pure}},  \;\;\; S^{\underline{\perp}} & = & \{\,\sigma'\in { \mathfrak{S}}_{{}_{pure}}, \;\vert\; \forall \sigma \in S,\; \sigma \perp \sigma'\,\}=\underline{S^{\perp}}.
 \end{eqnarray}
\end{MyDef}

\begin{lemme}
\begin{eqnarray}
 \forall S \subseteq { \mathfrak{S}}_{{}_{pure}},  \;\;\; S^{\underline{\perp}\underline{\perp}} & = &\underline{ \{\bigsqcap{}_{{}_{ \mathfrak{S}}} S\}}.\label{Sperpbarperpbar}
  \end{eqnarray}
\end{lemme}

\begin{MyDef}
The set of ortho-closed subsets of the space of pure states equipped with the orthogonality relation $\underline{\perp}$ is denoted ${ \mathcal{C}}({ \mathfrak{S}}_{{}_{pure}})$.  The set ${ \mathcal{C}}({ \mathfrak{S}}_{{}_{pure}})$ is equipped with the following operations
\begin{eqnarray}
\forall { \mathfrak{c}}_1, { \mathfrak{c}}_2\in { \mathcal{C}}({ \mathfrak{S}}_{{}_{pure}}),\;\;\;\;\;\; { \mathfrak{c}}_1 \wedge { \mathfrak{c}}_2&:=&  { \mathfrak{c}}_1 \cap { \mathfrak{c}}_2\\
{ \mathfrak{c}}_1 \vee { \mathfrak{c}}_2&:=& (  { \mathfrak{c}}_1 \cup { \mathfrak{c}}_2 )^{\underline{\perp}\underline{\perp}}.
\end{eqnarray}
 and by the unary operation $\underline{\perp}$. 
\end{MyDef}

\begin{lemme}\label{lemmecomplete}
 ${ \mathcal{C}}({ \mathfrak{S}}_{{}_{pure}})$ is a complete ortho-lattice.
\end{lemme}

\begin{lemme}
${ \mathcal{C}}({ \mathfrak{S}}_{{}_{pure}})$ is atomic, i.e. 
\begin{eqnarray}
\forall \sigma\in { \mathfrak{S}}_{{}_{pure}}, \{\sigma\}^{\underline{\perp}\underline{\perp}}= \{\sigma\}.
\end{eqnarray}
\end{lemme}
\begin{proof}
Due to {\bf Axiom \ref{Axiomnotype2}} (i.e. the absence of type 2's pure states), we have immediately
\begin{eqnarray}
\forall \sigma\in { \mathfrak{S}}_{{}_{pure}}, \underline{\{\sigma\}}= \{\sigma\}.
\end{eqnarray}
\end{proof}

\begin{lemme}\label{lemmeatomistic}
The lattice ${ \mathcal{C}}({ \mathfrak{S}}_{{}_{pure}})$ is atomistic, i.e. 
\begin{eqnarray}
\forall { \mathfrak{c}}\in { \mathcal{C}}({ \mathfrak{S}}_{{}_{pure}}), && { \mathfrak{c}} = \bigvee{}_{\!\!{}_{\sigma\in { \mathfrak{c}}}} \;\{ \sigma\}^{\underline{\perp}\underline{\perp}}=\bigvee{}_{\!\!{}_{\sigma\in { \mathfrak{c}}}} \;\{ \sigma\}.
\end{eqnarray}
\end{lemme}

\begin{lemme} \label{propvetterlein}
We have the following property
\begin{eqnarray}
\forall S\subseteq_{fin}^{\not= \varnothing} { \mathfrak{S}},\forall \sigma\in { \mathfrak{S}}\;\vert\; \sigma \notin S^{\perp\perp},&& \exists \sigma'\in S^{\perp}\;\vert\; (S\cup \{\sigma\})^{\perp\perp} = (S\cup \{\sigma'\})^{\perp\perp}.\label{vetterlein1}
\end{eqnarray}
We have then immediately
\begin{eqnarray}
\forall S\subseteq_{fin}^{\not= \varnothing} { \mathfrak{S}}_{{}_{pure}},\forall \sigma\in { \mathfrak{S}}_{{}_{pure}} \;\vert\; \sigma \notin S^{\underline{\perp}\underline{\perp}},&& \exists \sigma'\in S^{\underline{\perp}}\;\vert\; (S\cup \{\sigma\})^{\underline{\perp}\underline{\perp}} = (S\cup \{\sigma'\})^{\underline{\perp}\underline{\perp}}.\label{vetterlein2}
\end{eqnarray}
\end{lemme}
\begin{proof}
First of all, from (\ref{Sperpperp}), we have $(S\cup \{\sigma\})^{\perp\perp}=\uparrow^{{}^{ \mathfrak{S}}}\!\!\!((\bigsqcap{}_{{}_{ \mathfrak{S}}} S) \sqcap_{{}_{ \mathfrak{S}}} \sigma)$. We have $((\bigsqcap{}_{{}_{ \mathfrak{S}}} S) \sqcap_{{}_{ \mathfrak{S}}} \sigma) \sqsubset_{{}_{ \mathfrak{S}}} (\bigsqcap{}_{{}_{ \mathfrak{S}}} S)$ because the condition $\sigma \notin S^{\perp\perp}$ means $\sigma \not\sqsupseteq_{{}_{ \mathfrak{S}}} (\bigsqcap{}_{{}_{ \mathfrak{S}}} S)$. Due to lemma \ref{lemmeconsequenceirredundancy}, there exists a unique discriminating property ${ \mathfrak{l}}$ such that $\Sigma_{ \mathfrak{l}}=(\bigsqcap{}_{{}_{ \mathfrak{S}}} S)$ and $\Sigma_{\overline{ \mathfrak{l}}}=(\bigsqcap{}_{{}_{ \mathfrak{S}}} S)^\star$. The discriminating character of the property ${ \mathfrak{l}}$ implies that $\forall \Sigma'' \sqsubset_{{}_{ \mathfrak{S}}}\Sigma_{ \mathfrak{l}},\; \widehat{\Sigma_{\overline{ \mathfrak{l}}}\Sigma''}{}^{{}^{ \mathfrak{S}}}$. In particular, choosing $\Sigma''=((\bigsqcap{}_{{}_{ \mathfrak{S}}} S) \sqcap_{{}_{ \mathfrak{S}}} \sigma)$, we obtain $\widehat{(\bigsqcap{}_{{}_{ \mathfrak{S}}} S)^\star ((\bigsqcap{}_{{}_{ \mathfrak{S}}} S) \sqcap_{{}_{ \mathfrak{S}}} \sigma)}$. Explicitely, there exists an element $\sigma' \in S^\perp=(\uparrow^{{}^{ \mathfrak{S}}}(\bigsqcap{}_{{}_{ \mathfrak{S}}} S)^\star)$ such that $\sigma' \sqsupseteq_{{}_{ \mathfrak{S}}}  (\sigma\sqcap_{{}_{ \mathfrak{S}}}(\bigsqcap{}_{{}_{ \mathfrak{S}}} S))$. In other words, there exists an element $\sigma' \in S^\perp$ such that $(S\cup \{\sigma'\})^{{\perp}{\perp}} = (S\cup \{\sigma\})^{{\perp}{\perp}}$. This proves the property (\ref{vetterlein1}) .\\
We deduce (\ref{vetterlein2}) immediately from (\ref{vetterlein1}) using (\ref{Sperpbarperpbar}).
\end{proof}

\begin{remark}\label{remarkseparation}
The property (\ref{vetterlein2}) implies the following simple property : 
\begin{eqnarray}
\forall \sigma_1,\sigma_2\in { \mathfrak{S}}_{{}_{pure}}\;\vert\; \sigma_1\not= \sigma_2,\;&& \exists \sigma_3\in { \mathfrak{S}}_{{}_{pure}}\;\vert\; (\, \sigma_1 \underline{\perp}\sigma_3\;\textit{\rm and}\; \sigma_2\not\!\!\underline{\perp}\sigma_3\,).\label{zhongseparation}
\end{eqnarray}
Indeed, we deduce from (\ref{vetterlein2}), for any $\sigma_1,\sigma_2\in { \mathfrak{S}}_{{}_{pure}}\;\vert\; \sigma_1\not= \sigma_2$, the fact that $\exists \sigma_3 \underline{\perp} \sigma_1$ such that $(\{\sigma_1\}\cup \{\sigma_2\})^{\underline{\perp}\underline{\perp}} = (\{\sigma_1\}\cup \{\sigma_3\})^{\underline{\perp}\underline{\perp}}$, and we cannot have $\sigma_2 \underline{\perp}\sigma_3$ because this would mean $\sigma_2 \in \{\sigma_1\}^{\underline{\perp}\underline{\perp}}=\{\sigma_1\}$ which contradicts the assumption.\\
The property (\ref{zhongseparation}) is called 'Separation' axiom in \cite[Definition 2.3]{Zhong2021}.  
\end{remark}

\begin{remark}\label{remarkrepresentation}
For each $S\subseteq { \mathfrak{S}}_{{}_{pure}}$, we can define the corresponding discriminating yes/no test ${ \mathfrak{t}}_{{}_{(\bigsqcap S, (\bigsqcap S)^\star)}}$ and for any $\sigma\notin S^{\underline{\perp}}$ define $\sigma\!\!\lrcorner_S:=\sigma {\centerdot} { \mathfrak{t}}_{{}_{(\bigsqcap S, (\bigsqcap S)^\star)}}$. We have, for any $\sigma'\in S$, the equivalence $(\sigma' \underline{\perp} \sigma) \Leftrightarrow (\sigma' \underline{\perp} \sigma\!\!\lrcorner_S)$.  This property is called 'Representation' axiom in \cite[Definition 2.3]{Zhong2021}.  Let us prove it in our context.  We will suppose that $S$ is not reduced to a single element,  this particular case being in fact separately and trivially checked. \\
First of all, it is straightforward to prove that $(\sigma' \underline{\perp} \sigma\!\!\lrcorner_S) \Rightarrow (\sigma' \underline{\perp} \sigma)$. Indeed,  $(\sigma' \underline{\perp} \sigma\!\!\lrcorner_S)$ can be rewritten  $\textit{\bf Y}=\widetilde{ \mathfrak{e}}_{{ \mathfrak{t}}_{(\sigma'^\star,\sigma')}}( \sigma\!\!\lrcorner_S)$, i.e.  $\textit{\bf Y}=\widetilde{ \mathfrak{e}}_{{ \mathfrak{t}}_{(\sigma'^\star,\sigma')}}( \sigma {\centerdot} { \mathfrak{t}}_{{}_{(\bigsqcap S, (\bigsqcap S)^\star)}})$. The minimally disturbing yes/no tests ${ \mathfrak{t}}_{(\sigma'^\star,\sigma')}$ and ${ \mathfrak{t}}_{{}_{(\bigsqcap S, (\bigsqcap S)^\star)}}$ are compatible. We then have 
$\widetilde{ \mathfrak{e}}_{{ \mathfrak{t}}_{(\sigma'^\star,\sigma')}}( \sigma {\centerdot} { \mathfrak{t}}_{{}_{(\bigsqcap S, (\bigsqcap S)^\star)}}) \leq \widetilde{ \mathfrak{e}}_{{ \mathfrak{t}}_{(\sigma'^\star,\sigma')}}( \sigma )$,  from (\ref{axiomidealmeasurement}). As a result, we obtain $\textit{\bf Y}=\widetilde{ \mathfrak{e}}_{{ \mathfrak{t}}_{(\sigma'^\star,\sigma')}}( \sigma )$, i.e. $(\sigma' \underline{\perp} \sigma)$\\
Let us now prove the implication $(\sigma' \underline{\perp} \sigma) \Rightarrow (\sigma' \underline{\perp} \sigma\!\!\lrcorner_S)$. Let us suppose that $\sigma\in \{\sigma'\}^{\underline{\perp}}=\uparrow^{{}^{ \mathfrak{S}}}\!\!\!\!\sigma'^\star$.  We note that $\sigma'^\star \sqsubset_{{}_{ \mathfrak{S}}} (\bigsqcap{}_{{}_{ \mathfrak{S}}} S)^\star$. Moreover, we have $(\bigsqcap{}_{{}_{ \mathfrak{S}}} S)^\star \bowtie_{{}_{ \mathfrak{S}}} \bigsqcap{}_{{}_{ \mathfrak{S}}} S$. As a consequence, we deduce that $(\uparrow^{{}^{ \mathfrak{S}}}\!\bigsqcap{}_{{}_{ \mathfrak{S}}} S)\cap (\uparrow^{{}^{ \mathfrak{S}}}\!\!\!\sigma'^\star)\not= \varnothing$, i.e.  $S\cap \{\sigma'\}^{\underline{\perp}}\not=\varnothing$. Let us then consider $\sigma'' \in S\cap \{\sigma'\}^{\underline{\perp}}$. We have obviously $(\sigma''\sqcap_{{}_{ \mathfrak{S}}} \sigma)\in (\downarrow_{{}_{ \mathfrak{S}}}\uparrow^{{}^{ \mathfrak{S}}}\!\bigsqcap{}_{{}_{ \mathfrak{S}}} S)\cap (\downarrow_{{}_{ \mathfrak{S}}}\sigma)$ and then $(\sigma''\sqcap_{{}_{ \mathfrak{S}}} \sigma) \sqsubseteq_{{}_{ \mathfrak{S}}} (\sigma\!\lrcorner_S\sqcap_{{}_{ \mathfrak{S}}} \sigma)$ because of the property (\ref{proppi=imcapdott}). Moreover,  as soon as (i) $(\downarrow_{{}_{ \mathfrak{S}}}\sigma)$ is a filter,  (ii) $\sigma''\in \{\sigma'\}^{\underline{\perp}}$ by construction, and (iii) $\sigma\in \{\sigma'\}^{\underline{\perp}}$ by assumption, we conclude that $(\sigma''\sqcap_{{}_{ \mathfrak{S}}} \sigma)\in \{\sigma'\}^{\underline{\perp}}$. As a result, we deduce that $(\sigma\!\lrcorner_S \sqcap_{{}_{ \mathfrak{S}}} \sigma)\in \{\sigma'\}^{\underline{\perp}}$, and in particular $\sigma\!\lrcorner_S  \in \{\sigma'\}^{\underline{\perp}}$. This concludes the proof.
\end{remark}

\begin{lemme}\label{lemmeorthomodular}
The ortho-lattice ${ \mathcal{C}}({ \mathfrak{S}}_{{}_{pure}})$ is ortho-modular.
\end{lemme}
\begin{proof}
Let us firstly prove that, for any maximal orthogonal subset $S$ of an ortho-closed set $A\in { \mathcal{C}}({ \mathfrak{S}}_{{}_{pure}})$, we have $A=S^{\underline{\perp}\underline{\perp}}$. Let us suppose it exists $\sigma\in A\smallsetminus S^{\underline{\perp}\underline{\perp}}$. Using previous lemma, we can identify $\sigma' \in S^{\underline{\perp}}$ such that $(S\cup \{\sigma\})^{\underline{\perp}\underline{\perp}}=(S\cup \{\sigma'\})^{\underline{\perp}\underline{\perp}}\subseteq A$. This result contradicts the maximality of the orthogonal subset $S$ in $A$. We have then necessarily $A=S^{\underline{\perp}\underline{\perp}}$. Using now \cite[Corollary 2 p.7]{Dacey} (see also \cite[Theorem 35]{WILCE2009443}), we then conclude that the ortho-lattice ${ \mathcal{C}}({ \mathfrak{S}}_{{}_{pure}})$ is orthomodular.
\end{proof}

\begin{lemme}\label{lemmecovering}
The ortho-lattice ${ \mathcal{C}}({ \mathfrak{S}}_{{}_{pure}})$ satisfies the {\em covering property}, i.e.
\begin{eqnarray}
\forall A\in { \mathcal{C}}({ \mathfrak{S}}_{{}_{pure}}), \forall \sigma \in { \mathfrak{S}}_{{}_{pure}}\; \vert\; \sigma \notin A,&& \{\sigma\} \vee A \;\;\textit{\rm covers}\;\; A.
\end{eqnarray}
\end{lemme}
\begin{proof}
Let $A\in { \mathcal{C}}({ \mathfrak{S}}_{{}_{pure}})$ and $\sigma \notin A$. From Lemma \ref{propvetterlein}, we know that there exists $\sigma' \in A^{\underline{\perp}}$ such that $A\vee \{\sigma\}=(A\cup \{\sigma\})^{\underline{\perp}\underline{\perp}} = (A\cup \{\sigma'\})^{\underline{\perp}\underline{\perp}}=A\vee \{\sigma'\}$. Since $\sigma'$ is an atom orthogonal to $A$, it follows from the orthomodularity of ${ \mathcal{C}}({ \mathfrak{S}}_{{}_{pure}})$ that $A\vee \{\sigma\}$ covers $A$.
\end{proof}

\begin{theoreme}\label{theorempropsys}
The ortho-lattice ${ \mathcal{C}}({ \mathfrak{S}}_{{}_{pure}})$ forms a {\em Piron's propositional system} (also called {\em Hilbert lattice}) (see \cite[Definition 5.9]{STUBBE2007477}).
\end{theoreme}
\begin{proof}
Direct consequence of Lemma \ref{lemmecomplete},
Lemma \ref{lemmeatomistic},
Lemma \ref{lemmeorthomodular},
Lemma \ref{lemmecovering},
\end{proof}

\begin{MyDef}
The orthogonality space ${ \mathfrak{S}}_{{}_{pure}}$ is said to be {\em reducible} iff ${ \mathfrak{S}}_{{}_{pure}}$ is the disjoint union of non-empty subsets ${ \mathfrak{S}}_1, { \mathfrak{S}}_2 \subseteq { \mathfrak{S}}_{{}_{pure}}$ such that $\sigma_1 \underline{\perp} \sigma_2$ for any $\sigma_1\in { \mathfrak{S}}_1$ and $\sigma_2\in { \mathfrak{S}}_2$. Otherwise, ${ \mathfrak{S}}_{{}_{pure}}$ is said to be {\em irreducible}.
\end{MyDef}

\begin{MyReq}\label{Axiomlinearity}
\begin{eqnarray}
\forall \sigma_1,\sigma_2\in { \mathfrak{S}}_{{}_{pure}}\;\vert\; \sigma_1\not= \sigma_2,
\;\;\;\;\; \exists \sigma_3\in { \mathfrak{S}}_{{}_{pure}} \;\vert\; (\;\sigma_3\not=\sigma_1,\;\;\;\; \sigma_3\not=\sigma_2, \;\;\textit{\rm and}\;\; \sigma_3\in \underline{\{\sigma_1 \sqcap_{{}_{ \mathfrak{S}}} \sigma_2\}} \;)\;\;\;\;\;&&
\end{eqnarray}
\end{MyReq}

\begin{remark}\label{remarksuperposition}
Previous axiom is nothing else than the 'Superposition' axiom in \cite[Definition 2.3]{Zhong2021}.  
\end{remark}
\begin{remark}
As a conclusion of remark \ref{remarkseparation}, remark \ref{remarkrepresentation}, and remark \ref{remarksuperposition}, we conclude that $({ \mathfrak{S}}_{{}_{pure}}, \underline{\perp})$ is a 'quantum Kripke frame' as defined in \cite[Definition 2.3]{Zhong2021}.
\end{remark}

\begin{lemme}
${ \mathfrak{S}}_{{}_{pure}}$ is irreducible.
\end{lemme}
\begin{proof}
Let us assume that ${ \mathfrak{S}}_{{}_{pure}}$ is reducible, then ${ \mathfrak{S}}_{{}_{pure}}$ is the disjoint union of  non-empty subsets ${ \mathfrak{S}}_1, { \mathfrak{S}}_2 \subseteq { \mathfrak{S}}_{{}_{pure}}$ such that $\sigma_1 \underline{\perp} \sigma_2$ for any $\sigma_1\in { \mathfrak{S}}_1$ and $\sigma_2\in { \mathfrak{S}}_2$. Let us consider $\sigma_1\in { \mathfrak{S}}_1$ and $\sigma_2\in { \mathfrak{S}}_2$ and $\sigma_3\in \{\sigma_1,\sigma_2\}^{\underline{\perp}\underline{\perp}}$. Necessarily $\sigma_3\in S_1$ or $\sigma_3\in S_2$, and then $\sigma_3\underline{\perp}\sigma_1$ or $\sigma_3\underline{\perp}\sigma_2$. If $\sigma_3\underline{\perp}\sigma_1$, we have $\{\sigma_3\}\subseteq \{\sigma_1,\sigma_2\}^{\underline{\perp}\underline{\perp}} \cap \{\sigma_1\}^{\underline{\perp}}=  (\{\sigma_1\}\vee \{\sigma_2\}) \cap \{\sigma_1\}^{\underline{\perp}}=\{\sigma_2\}$. 
In the same way, if $\sigma_3\underline{\perp}\sigma_2$, then $\{\sigma_3\}=\{\sigma_1\}$. As a result, we conclude that $\{\sigma_1,\sigma_2\}^{\underline{\perp}\underline{\perp}}=\{\sigma_1,\sigma_2\}$\\
This proof is given in the first part of the proof of \cite[Lemma 2.9]{Vetterlein2020}. 
From {\bf Axiom \ref{Axiomlinearity}} we then deduce that ${ \mathfrak{S}}_{{}_{pure}}$ is irreducible.
\end{proof}

\begin{corollaire}\label{lemmeirreducible}
${ \mathcal{C}}({ \mathfrak{S}}_{{}_{pure}})$ is an irreducible ortho-lattice.
\end{corollaire}
\begin{proof}
Let us suppose ${ \mathcal{C}}({ \mathfrak{S}}_{{}_{pure}})$ is reducible, then there exist a central element $A$  distinct from the bottom element $\varnothing$  and the top element ${ \mathfrak{S}}_{{}_{pure}}$. Then any atom is either below $A$ or below $A^{\underline{\perp}}$, and then any $\sigma\in { \mathfrak{S}}_{{}_{pure}}$ is either in $A$ or in $A^{\underline{\perp}}$. Hence ${ \mathfrak{S}}_{{}_{pure}}$ is reducible.
\end{proof} 

\begin{theoreme}
If the condition of the {\bf Axiom \ref{Axiomlinearity}} is satisfied, the ortho-lattice ${ \mathcal{C}}({ \mathfrak{S}}_{{}_{pure}})$ forms an irreducible Piron's propositional system.
\end{theoreme}
\begin{proof}
Direct consequence of Theorem \ref{theorempropsys} and Corollary \ref{lemmeirreducible}.
\end{proof}

\begin{remark}\label{examplebooleanreducible}
It is important to note that the examples of spaces of states built in Remark \ref{exampleboolean} are in fact interesting to reconstruct a consistent quantum space of states, because they generically lead to {\bf irreducible} ${ \mathcal{C}}({ \mathfrak{S}}_{{}_{pure}})$. This point can be easily verified. Indeed, if we suppose that it exists $\sigma_1$ and $\sigma_2$ in ${ \mathfrak{S}}_{{}_{pure}}$ such that $\exists \sigma_3\in { \mathfrak{S}}_{{}_{pure}} \;\vert\; (\;\sigma_3\not=\sigma_1,\;\;\;\; \sigma_3\not=\sigma_2, \;\;\textit{\rm and}\;\; \sigma_3\in \underline{\{\sigma_1 \sqcap_{{}_{ \mathfrak{S}}} \sigma_2\}}\,)$, then the sub-graph of ${ \mathfrak{L}}$ formed by the elements $\sigma_1, \sigma_2, \sigma_3,(\sigma_1 \sqcap_{{}_{ \mathfrak{S}}} \sigma_2), \top_{{}_{ \mathfrak{L}}}$ is a 'diamond sub-graph' of ${ \mathfrak{L}}$, which is totally compatible with the modular character of ${ \mathfrak{L}}$. 
\end{remark}

\section{Symmetries}
\label{sectionsymmetries}
Let us consider two observers{{}}{,} $O_1$ and $O_2${{}}{, who wish} to formalize 'transactions' concerning their experimental results about the system. 

\begin{MyDef}
{{}}{Observer} $O_1$ has chosen a preparation process{{}}{,} ${ \mathfrak{p}}_1\in { \mathfrak{P}}^{{}^{(O_1)}}${{}}{,} and {{}}{intends} to describe it to{{}}{ }observer $O_2$. {{}}{Observer} $O_2$ is able to interpret the macroscopic data defining ${ \mathfrak{p}}_1$ in terms of the elements of ${ \mathfrak{P}}^{{}^{(O_2)}}${{}}{ } using a map $f_{{}_{(12)}} : { \mathfrak{P}}^{{}^{(O_1)}} \rightarrow { \mathfrak{P}}^{{}^{(O_2)}}$ (i.e., $O_2$ knows how to identify a preparation procedure $f_{{}_{(12)}}({ \mathfrak{p}}_1)$ corresponding to any ${ \mathfrak{p}}_1$).\\
{{}}{Observer} $O_2$ has chosen a yes/no test ${ \mathfrak{t}}_2\in { \mathfrak{T}}^{{}^{(O_2)}}$ and {{}}{intends} to address the corresponding question to $O_1$. {{}}{Observer} $O_1$ is able to interpret the macroscopic data defining ${ \mathfrak{t}}_2$ in terms of the elements of ${ \mathfrak{T}}^{{}^{(O_1)}}${{}}{ } using a map $f^{{}^{(21)}} : { \mathfrak{T}}^{{}^{(O_2)}} \rightarrow { \mathfrak{T}}^{{}^{(O_1)}}$ (i.e., $O_1$ {{}}{ }knows how to fix a test $f^{{}^{(21)}}({ \mathfrak{t}}_2)$ corresponding to any ${ \mathfrak{t}}_2$).\\
The {{}}{pair} of maps $(f_{{}_{(12)}},f^{{}^{(21)}})$ where  $f_{{}_{(12)}} \in { \mathfrak{P}}^{{}^{(O_1)}} \rightarrow { \mathfrak{P}}^{{}^{(O_2)}}$ and $f^{{}^{(21)}} : { \mathfrak{T}}^{{}^{(O_2)}}\rightarrow { \mathfrak{T}}^{{}^{(O_1)}}$ defines {\em a dictionary} formalizing the transaction from $O_1$ to $O_2$.
\end{MyDef}
The first task these observers want to accomplish is to confront their knowledge, i.e., to compare their 'statements' about the system.
\begin{MyDef}
As soon as the transaction is formalized using a {{}}{dictionary}, {{}}{the two observers} can formulate {{}}{their} statements and {{}}{each} confront them {{}}{with} the statements of the other. \\
{{}}{First, observer} $O_1$ can interpret the macroscopic data defining ${ \mathfrak{t}}_2$ using the map $f^{{}^{(21)}}$. Then, he produces the statement ${ \mathfrak{e}}^{{}^{(O_1)}}({ \mathfrak{p}}_1, f^{{}^{(21)}}({ \mathfrak{t}}_2))$ concerning the results of this test on the chosen samples. \\ 
Secondly,{{}}{ }observer $O_2$ can interpret the macroscopic data defining ${ \mathfrak{p}}_1$ using the map $f_{{}_{(12)}}$. Then,{{}}{ }observer $O_2$ pronounces {{}}{her} statement ${ \mathfrak{e}}^{{}^{(O_2)}}(f_{{}_{(12)}}({ \mathfrak{p}}_1), { \mathfrak{t}}_2)$ concerning the results of test ${ \mathfrak{t}}_2$ on the correspondingly prepared samples. \\ 
The two observers{{}}{,} $O_1$ and $O_2${{}}{,} are said to {\em agree about their statements} iff 
\begin{eqnarray}
\forall { \mathfrak{p}}_1\in { \mathfrak{P}}^{{}^{(O_1)}}, \forall { \mathfrak{t}}_2 \in { \mathfrak{T}}^{{}^{(O_2)}},&&{ \mathfrak{e}}^{{}^{(O_2)}}(f_{{}_{(12)}}({ \mathfrak{p}}_1), { \mathfrak{t}}_2)={ \mathfrak{e}}^{{}^{(O_1)}}({ \mathfrak{p}}_1, f^{{}^{(21)}}({ \mathfrak{t}}_2)),\label{defchumorphism}
\end{eqnarray}
i.e., iff the adjoint pair $(f_{{}_{(12)}},f^{{}^{(21)}})$ defines a morphism of Chu spaces from $({ \mathfrak{P}}^{{}^{(O_1)}},{ \mathfrak{T}}^{{}^{(O_1)}},{ \mathfrak{e}}^{{}^{(O_1)}})$ to $({ \mathfrak{P}}^{{}^{(O_2)}},{ \mathfrak{T}}^{{}^{(O_2)}},{ \mathfrak{e}}^{{}^{(O_2)}})$ \cite{PRATT1999319}.
\end{MyDef}

\begin{lemme}{}
{{}}{If} the dictionary $(f_{{}_{(12)}},f^{{}^{(21)}})$ satisfies{{}}{ }property (\ref{defchumorphism}) (i.e., the adjoint pair defines a Chu morphism from $({ \mathfrak{P}}^{{}^{(O_1)}},{ \mathfrak{T}}^{{}^{(O_1)}},{ \mathfrak{e}}^{{}^{(O_1)}})$ to $({ \mathfrak{P}}^{{}^{(O_2)}},{ \mathfrak{T}}^{{}^{(O_2)}},{ \mathfrak{e}}^{{}^{(O_2)}})$) we have immediately
\begin{eqnarray}
\forall { \mathfrak{p}},{ \mathfrak{p}}'\in { \mathfrak{P}}^{{}^{(O_1)}},&& (\, { \mathfrak{p}}\sim_{{}_{ \mathfrak{P}^{{}^{(O_1)}}}}{ \mathfrak{p}}'\,)\;\Rightarrow\; (\, f_{{}_{(12)}}({ \mathfrak{p}})\sim_{{}_{ \mathfrak{P}^{{}^{(O_2)}}}} f_{{}_{(12)}}({ \mathfrak{p}}')\,)
\end{eqnarray}
\end{lemme}
\begin{proof}
${ \mathfrak{p}}\sim_{{}_{ \mathfrak{P}^{{}^{(O_1)}}}}{ \mathfrak{p}}'$ implies, in particular, ${ \mathfrak{e}}^{{}^{(O_1)}}({ \mathfrak{p}}, f^{{}^{(21)}}({ \mathfrak{t}}_2))={ \mathfrak{e}}^{{}^{(O_1)}}({ \mathfrak{p}}', f^{{}^{(21)}}({ \mathfrak{t}}_2))$ for any ${ \mathfrak{t}}_2\in { \mathfrak{T}}^{{}^{(O_2)}}$.  Using property (\ref{defchumorphism}), we obtain ${ \mathfrak{e}}^{{}^{(O_1)}}(f_{{}_{(12)}}({ \mathfrak{p}}), { \mathfrak{t}}_2)={ \mathfrak{e}}^{{}^{(O_1)}}(f_{{}_{(12)}}({ \mathfrak{p}}'), { \mathfrak{t}}_2)$ for any ${ \mathfrak{t}}_2\in { \mathfrak{T}}^{{}^{(O_2)}}$, i.e., $ f_{{}_{(12)}}({ \mathfrak{p}})\sim_{{}_{ \mathfrak{P}^{{}^{(O_2)}}}} f_{{}_{(12)}}({ \mathfrak{p}}')$.
\end{proof}

\begin{MyDef}
In order for{{}}{ }observers $O_1$ and $O_2$ to be in {\em complete agreement} about the system (once they do agree about their statements), it is necessary for them to be unable to distinguish the outcomes of the control tests, realized to confirm (or not) their statements. \\
Firstly, the measurement realized by {{}}{ }observer $O_2$ is given by $(f_{{}_{(12)}}({ \mathfrak{p}}_1) {\centerdot} { \mathfrak{t}}_2)$.\\
Secondly,{{}}{ }observer $O_2$ {{}}{interprets}, using $f_{{}_{(12)}}$, the measurement $({ \mathfrak{p}}_1 {\centerdot} f^{{}^{(21)}}({ \mathfrak{t}}_2))$ realized by {{}}{ }observer $O_1$. \\
In other words, it is necessary for the dictionary $(f_{{}_{(12)}},f^{{}^{(21)}})$ {{}}{also to satisfy} the following property
\begin{eqnarray}
\forall { \mathfrak{p}}_1\in { \mathfrak{P}}^{{}^{(O_1)}}, \forall { \mathfrak{t}}_2 \in { \mathfrak{T}}^{{}^{(O_2)}},&&f_{{}_{(12)}}({ \mathfrak{p}}_1) {\centerdot} { \mathfrak{t}}_2 \;\sim_{{}_{ \mathfrak{S}^{(O_2)}}}\;
f_{{}_{(12)}}({ \mathfrak{p}}_1 {\centerdot} f^{{}^{(21)}}({ \mathfrak{t}}_2)),\label{defchucenterdot1}
\end{eqnarray}
\end{MyDef}

\begin{MyDef}
If the dictionary $(f_{{}_{(12)}},f^{{}^{(21)}})$ satisfies {{}}{ }properties (\ref{defchumorphism}) {{}}{and} (\ref{defchucenterdot1}), {{}}{as well as} property (\ref{defchucenterdot2}) below 
\begin{eqnarray}
\forall { \mathfrak{t}},{ \mathfrak{t}}'\in { \mathfrak{T}}^{{}^{(O_2)}},&& f^{{}^{(21)}}({ \mathfrak{t}}{\centerdot}{ \mathfrak{t}}')=f^{{}^{(21)}}({ \mathfrak{t}}){\centerdot}f^{{}^{(21)}}({ \mathfrak{t}}'),\label{defchucenterdot2}
\end{eqnarray}
and {{}}{ }properties (\ref{defchuinjectivity}) {{}}{and} (\ref{defchusurjectivity}) below
\begin{eqnarray}
&&\forall { \mathfrak{p}},{ \mathfrak{p}}'\in { \mathfrak{P}}^{{}^{(O_1)}},\;\; (f_{{}_{(12)}}({ \mathfrak{p}})\sim_{{}_{ \mathfrak{P}^{(O_2)}}}f_{{}_{(12)}}({ \mathfrak{p}}'))\;\Rightarrow\; ({ \mathfrak{p}}\sim_{{}_{ \mathfrak{P}^{(O_1)}}}{ \mathfrak{p}}'),\label{defchuinjectivity}\\
&&\textit{\rm $f^{{}^{(21)}}$ surjective},\label{defchusurjectivity}
\end{eqnarray}
then this dictionary is said to {\em relate by a symmetry} $({ \mathfrak{P}}^{{}^{(O_1)}},{ \mathfrak{T}}^{{}^{(O_1)}},{ \mathfrak{e}}^{{}^{(O_1)}})$ to $({ \mathfrak{P}}^{{}^{(O_2)}},{ \mathfrak{T}}^{{}^{(O_2)}},{ \mathfrak{e}}^{{}^{(O_2)}})$. This fact will be denoted
\begin{eqnarray}
(f_{{}_{(12)}},f^{{}^{(21)}}) &\in & { \bf {Sym}}\left[ ({ \mathfrak{P}}^{{}^{(O_1)}},{ \mathfrak{T}}^{{}^{(O_1)}},{ \mathfrak{e}}^{{}^{(O_1)}})\rightarrow ({ \mathfrak{P}}^{{}^{(O_2)}},{ \mathfrak{T}}^{{}^{(O_2)}},{ \mathfrak{e}}^{{}^{(O_2)}})\right].
\end{eqnarray}
\end{MyDef}

\begin{remark}
We note that the axiom (\ref{defchucenterdot2}) has been designed to preserve the associativity property of the succession rule. Indeed,
for any ${ \mathfrak{p}}\in { \mathfrak{P}}^{{}^{(O_1)}},$ and any ${ \mathfrak{t}},{ \mathfrak{t}}' \in { \mathfrak{T}}^{{}^{(O_2)}}$, we have
$f_{{}_{(12)}}({ \mathfrak{p}}) {\centerdot} ({ \mathfrak{t}} {\centerdot} { \mathfrak{t}}')=(f_{{}_{(12)}}({ \mathfrak{p}}) {\centerdot} { \mathfrak{t}}) {\centerdot} { \mathfrak{t}}'= f_{{}_{(12)}}({ \mathfrak{p}}{\centerdot} f^{{}^{(21)}}({ \mathfrak{t}}))  {\centerdot} { \mathfrak{t}}'=f_{{}_{(12)}}(({ \mathfrak{p}}{\centerdot} f^{{}^{(21)}}({ \mathfrak{t}})) {\centerdot} f^{{}^{(21)}}({ \mathfrak{t}}'))=f_{{}_{(12)}}({ \mathfrak{p}}{\centerdot} (f^{{}^{(21)}}({ \mathfrak{t}}) {\centerdot} f^{{}^{(21)}}({ \mathfrak{t}}')))=f_{{}_{(12)}}({ \mathfrak{p}}{\centerdot} f^{{}^{(21)}}({ \mathfrak{t}} {\centerdot} { \mathfrak{t}}'))$.\\
\end{remark}

\begin{remark}
{{}}{Properties} (\ref{defchuinjectivity}) and (\ref{defchusurjectivity}) have been introduced in order to be able to derive Theorem \ref{symmetrypreserveminimal}.
\end{remark}

\begin{theoreme}{\bf [Composition of symmetries]} \label{chucomposition}
\begin{eqnarray}
&&\hspace{-2.5cm}\left. \begin{array}{r}(f_{{}_{(12)}},f^{{}^{(21)}}) \in  { \bf {Sym}}\left[ ({ \mathfrak{P}}^{{}^{(O_1)}},{ \mathfrak{T}}^{{}^{(O_1)}},{ \mathfrak{e}}^{{}^{(O_1)}})\rightarrow ({ \mathfrak{P}}^{{}^{(O_2)}},{ \mathfrak{T}}^{{}^{(O_2)}},{ \mathfrak{e}}^{{}^{(O_2)}})\right] \nonumber\\
(g_{{}_{(23)}},g^{{}^{(32)}}) \in  { \bf {Sym}}\left[ ({ \mathfrak{P}}^{{}^{(O_2)}},{ \mathfrak{T}}^{{}^{(O_2)}},{ \mathfrak{e}}^{{}^{(O_2)}})\rightarrow ({ \mathfrak{P}}^{{}^{(O_3)}},{ \mathfrak{T}}^{{}^{(O_3)}},{ \mathfrak{e}}^{{}^{(O_3)}})\right] \end{array}\right\} \\
& \Rightarrow & ( g_{{}_{(23)}}\circ f_{{}_{(12)}} ,f^{{}^{(21)}} \circ g^{{}^{(32)}}) \in  { \bf {Sym}}\left[ ({ \mathfrak{P}}^{{}^{(O_1)}},{ \mathfrak{T}}^{{}^{(O_1)}},{ \mathfrak{e}}^{{}^{(O_1)}})\rightarrow ({ \mathfrak{P}}^{{}^{(O_3)}},{ \mathfrak{T}}^{{}^{(O_3)}},{ \mathfrak{e}}^{{}^{(O_3)}})\right].
\end{eqnarray}
\end{theoreme}
\begin{proof}
We firstly note that 
\begin{eqnarray}
\forall { \mathfrak{p}}_1\in { \mathfrak{P}}^{{}^{(O_1)}}, \forall { \mathfrak{t}}_3 \in { \mathfrak{T}}^{{}^{(O_3)}},\;\;{ \mathfrak{e}}^{{}^{(O_3)}}(g_{{}_{(23)}}\circ f_{{}_{(12)}}({ \mathfrak{p}}_1), { \mathfrak{t}}_3)={ \mathfrak{e}}^{{}^{(O_2)}}(f_{{}_{(12)}}({ \mathfrak{p}}_1), g^{{}^{(32)}}({ \mathfrak{t}}_3))={ \mathfrak{e}}^{{}^{(O_1)}}({ \mathfrak{p}}_1,  f^{{}^{(21)}} \circ g^{{}^{(32)}}({ \mathfrak{t}}_3)).&&\;\;\;\;\;
\end{eqnarray}
Secondly, we have
\begin{eqnarray}
\forall { \mathfrak{p}}_1\in { \mathfrak{P}}^{{}^{(O_1)}}, \forall { \mathfrak{t}}_3 \in { \mathfrak{T}}^{{}^{(O_3)}},\;\; (g_{{}_{(23)}}\circ f_{{}_{(12)}})({ \mathfrak{p}}_1) {\centerdot} { \mathfrak{t}}_3 = g_{{}_{(23)}}(\,f_{{}_{(12)}}({ \mathfrak{p}}_1) {\centerdot} g^{{}^{(32)}}({ \mathfrak{t}}_3)\,)
=(g_{{}_{(23)}}\circ f_{{}_{(12)}})(\,({ \mathfrak{p}}_1) {\centerdot} (f^{{}^{(21)}} \circ  g^{{}^{(32)}})({ \mathfrak{t}}_3)\,).&&\;\;\;\;\;
\end{eqnarray}
The properties (\ref{defchucenterdot2}), (\ref{defchuinjectivity}), and (\ref{defchusurjectivity}) are trivially preserved by composition.
\end{proof}

\begin{lemme}{}\label{chumorphismlceil}
The dictionary $(h_{{}_{(12)}}, h^{{}^{(21)}})$ defined by 
\begin{eqnarray}
\begin{array}{rrclccrcl}
h_{{}_{(12)}}:& { \mathfrak{P}} &\rightarrow &{ \mathfrak{S}} &                  & h^{{}^{(21)}} :& { \mathfrak{T}} &\rightarrow &{ \mathfrak{T}} \\
  & { \mathfrak{p}} & \mapsto & \lceil { \mathfrak{p}} \rceil    &                   &                       & { \mathfrak{t}} & \mapsto & { \mathfrak{t}} 
\end{array}
\end{eqnarray}
satisfies {{}}{ }properties (\ref{defchumorphism}), (\ref{defchucenterdot1}), (\ref{defchucenterdot2}), (\ref{defchuinjectivity}) {{}}{and} (\ref{defchusurjectivity}). In other words, this dictionary relates by a symmetry $({ \mathfrak{P}},{ \mathfrak{T}}, { \mathfrak{e}})$ to $({ \mathfrak{S}},{ \mathfrak{T}}, \widetilde{ \mathfrak{e}})$.
\begin{eqnarray}
(h_{{}_{(12)}},h^{{}^{(21)}}) \in  { \bf {Sym}}\left[ ({ \mathfrak{P}},{ \mathfrak{T}},{ \mathfrak{e}})\rightarrow ({ \mathfrak{S}},{ \mathfrak{T}},\widetilde{ \mathfrak{e}})\right]
\end{eqnarray}
\end{lemme}
\begin{proof}
{{}}{Property} (\ref{defchumorphism}) is a direct consequence of the definition (\ref{defetilde}) of $\widetilde{ \mathfrak{e}}$. \\
{{}}{Property} (\ref{defchucenterdot1}) is a direct consequence of property (\ref{tsursimdef}).\\
{{}}{Property} (\ref{defchucenterdot2}) is tautologically verified.\\
{{}}{Property} (\ref{defchuinjectivity}) relies on the definition of the map $\lceil \cdot \rceil$.\\
{{}}{Property} (\ref{defchusurjectivity}) is trivial.
\end{proof}

As a consequence of Lemma \ref{chumorphismlceil} and Theorem \ref{chucomposition}, we can define the following notion.

\begin{MyDef}
{{}}{For} any dictionary $(f_{{}_{(12)}},f^{{}^{(21)}}) \in  { \bf {Sym}}\left[ ({ \mathfrak{P}}^{{}^{(O_1)}},{ \mathfrak{T}}^{{}^{(O_1)}},{ \mathfrak{e}}^{{}^{(O_1)}})\rightarrow ({ \mathfrak{P}}^{{}^{(O_2)}},{ \mathfrak{T}}^{{}^{(O_2)}},{ \mathfrak{e}}^{{}^{(O_2)}})\right]$,  we will associate the dictionary $(\widetilde{f}_{{}_{(12)}},f^{{}^{(21)}})$ defined by
\begin{eqnarray}
\forall { \mathfrak{p}}\in { \mathfrak{P}}^{{}^{(O_1)}},&&\widetilde{f}_{{}_{(12)}}(\lceil { \mathfrak{p}} \rceil_{{}_{1}}):=\lceil {f}_{{}_{(12)}}({ \mathfrak{p}} )\rceil_{{}_{2}}.
\end{eqnarray}
We have
\begin{eqnarray}
(\widetilde{f}_{{}_{(12)}},f^{{}^{(21)}})\in  { \bf {Sym}}\left[ ({ \mathfrak{S}}^{{}^{(O_1)}},{ \mathfrak{T}}^{{}^{(O_1)}},\widetilde{ \mathfrak{e}}^{{}^{(O_1)}})\rightarrow ({ \mathfrak{S}}^{{}^{(O_2)}},{ \mathfrak{T}}^{{}^{(O_2)}},\widetilde{ \mathfrak{e}}^{{}^{(O_2)}})\right].
\end{eqnarray}
{{}}{Explicitly}, $(\widetilde{f}_{{}_{(12)}},f^{{}^{(21)}})$ {{}}{has} to satisfy the following requirements :
\begin{eqnarray}
&&\forall \sigma_1\in { \mathfrak{S}}^{{}^{(O_1)}}, \forall { \mathfrak{t}}_2 \in { \mathfrak{T}}^{{}^{(O_2)}},\;\;\widetilde{ \mathfrak{e}}^{{}^{(O_2)}}_{{ \mathfrak{t}}_2}(\widetilde{f}_{{}_{(12)}}(\sigma_1))=\widetilde{ \mathfrak{e}}^{{}^{(O_1)}}_{f^{{}^{(21)}}({ \mathfrak{t}}_2)}(\sigma_1),\label{defchumorphismsigma}\\
&&\forall \sigma_1\in { \mathfrak{S}}^{{}^{(O_1)}}, \forall { \mathfrak{t}}_2 \in { \mathfrak{T}}^{{}^{(O_2)}},\;\; \widetilde{f}_{{}_{(12)}}(\sigma_1) {\centerdot} { \mathfrak{t}}_2 \;=\;
\widetilde{f}_{{}_{(12)}}(\sigma_1 {\centerdot} f^{{}^{(21)}}({ \mathfrak{t}}_2)),\label{defchucenterdot1sigma}\\
&&\forall { \mathfrak{t}},{ \mathfrak{t}}'\in { \mathfrak{T}}^{{}^{(O_2)}},\;\; f^{{}^{(21)}}({ \mathfrak{t}}\,{\centerdot}{ \mathfrak{t}}')=f^{{}^{(21)}}({ \mathfrak{t}}){\centerdot}f^{{}^{(21)}}({ \mathfrak{t}}'),\label{defchucenterdot2sigma}\\
&&\textit{\rm $\widetilde{f}_{{}_{(12)}}$ injective}\label{defchuinjectivitysigma}\\ 
&&\textit{\rm $f^{{}^{(21)}}$ surjective}\label{defchusurjectivitysigma}
\end{eqnarray}
\end{MyDef} 

\begin{lemme}{}\label{lemmaflawson} $\widetilde{f}_{{}_{(12)}}$ is as map satisfying 
\begin{eqnarray}
\widetilde{f}_{{}_{(12)}} \in \left[ { \mathfrak{S}}^{{}^{(O_1)}}\rightarrow { \mathfrak{S}}^{{}^{(O_2)}}\right]^\barwedge_{{}_{\sqcap}}\;\;\;\;\;\;\textit{\rm and}\;\;\;\;\;\; \widetilde{f}_{{}_{(12)}}(\bot_{{}_{{ \mathfrak{S}}^{{}^{(O_1)}}}})=\bot_{{}_{{ \mathfrak{S}}^{{}^{(O_2)}}}}.
\end{eqnarray}
In particular, $\widetilde{f}_{{}_{(12)}}$ is order-preserving.
\end{lemme}
\begin{proof}
\begin{eqnarray}
\forall { \mathfrak{t}}_2 \in { \mathfrak{T}}^{{}^{(O_2)}},\;\;\widetilde{ \mathfrak{e}}^{{}^{(O_2)}}_{{ \mathfrak{t}}_2}(\widetilde{f}_{{}_{(12)}}(\bot_{{}_{{ \mathfrak{S}}^{{}^{(O_1)}}}}))=\widetilde{ \mathfrak{e}}^{{}^{(O_1)}}_{f^{{}^{(21)}}({ \mathfrak{t}}_2)}(\bot_{{}_{{ \mathfrak{S}}^{{}^{(O_1)}}}})=\bot \;\;\;& \Rightarrow & \widetilde{f}_{{}_{(12)}}(\bot_{{}_{{ \mathfrak{S}}^{{}^{(O_1)}}}})=\bot_{{}_{{ \mathfrak{S}}^{{}^{(O_2)}}}}.
\end{eqnarray}
Secondly, from Theorem \ref{widetildeelawson}, we know that $\forall { \mathfrak{t}}\in { \mathfrak{T}}^{{}^{(O_2)}}$, $\widetilde{ \mathfrak{e}}^{{}^{(O_2)}}_{ \mathfrak{t}}$ is an order-preserving and Lawson-continuous map (i.e., $\widetilde{ \mathfrak{e}}^{{}^{(O_1)}}_{ \mathfrak{t}} \in \left[ { \mathfrak{S}}^{{}^{(O_1)}}\rightarrow { \mathfrak{B}}\right]^\barwedge_{{}_{\sqcap}}$).  We can then deduce that $\widetilde{f}_{{}_{(12)}}$ is an order-preserving and Lawson-continuous map.  Let us {{}}{first} check the Scott{{}}{ }continuity of $\widetilde{f}_{{}_{(12)}}$ :
\begin{eqnarray}
\forall { \mathfrak{t}}\in { \mathfrak{T}}^{{}^{(O_2)}}, \forall { \mathfrak{C}}\subseteq_{Chain} { \mathfrak{S}}^{{}^{(O_1)}},\;\;
\;\;\widetilde{ \mathfrak{e}}^{{}^{(O_2)}}_{ \mathfrak{t}}(\widetilde{f}_{{}_{(12)}}(\bigsqcup{}_{{}_{ \mathfrak{S}^{{}^{(O_1)}}}}{ \mathfrak{C}}))&=& \widetilde{ \mathfrak{e}}^{{}^{(O_1)}}_{f^{{}^{(21)}}({ \mathfrak{t}})}(\bigsqcup{}_{{}_{ \mathfrak{S}^{{}^{(O_1)}}}}{ \mathfrak{C}})\;\;\;\;\;\;\;\;\;\;\textit{\rm from eq. (\ref{defchumorphismsigma})}\nonumber\\
&=& \bigsqcup{}_{{}_{\sigma\in { \mathfrak{C}}}}\widetilde{ \mathfrak{e}}^{{}^{(O_1)}}_{f^{{}^{(21)}}({ \mathfrak{t}})}(\sigma)\;\;\;\;\;\;\;\;\;\;\textit{\rm from eq. (\ref{continuityrequirement2})}\nonumber\\
&=& \bigsqcup{}_{{}_{\sigma\in { \mathfrak{C}}}}\widetilde{ \mathfrak{e}}^{{}^{(O_2)}}_{{ \mathfrak{t}}}(f_{{}_{(12)}}(\sigma))\;\;\;\;\;\;\;\;\;\;\textit{\rm from eq. (\ref{defchumorphismsigma})}\nonumber\\
&=& \widetilde{ \mathfrak{e}}^{{}^{(O_2)}}_{{ \mathfrak{t}}}(\bigsqcup{}_{{}_{\sigma\in { \mathfrak{C}}}} f_{{}_{(12)}}(\sigma))\;\;\;\;\;\;\;\;\;\;\textit{\rm from eq. (\ref{continuityrequirement2})}.\nonumber\\
\end{eqnarray}
Then, using Lemma \ref{chuseparated}, we conclude that $\widetilde{f}_{{}_{(12)}}$ is Scott{{}}{ }continuous (and in particular order{{}}{ }preserving)
\begin{eqnarray}
\forall { \mathfrak{C}}\subseteq_{Chain} { \mathfrak{S}}^{{}^{(O_1)}},\;\; \widetilde{f}_{{}_{(12)}}(\bigsqcup{}_{{}_{ \mathfrak{S}^{{}^{(O_1)}}}}{ \mathfrak{C}}) &=& \bigsqcup{}_{{}_{\sigma\in { \mathfrak{C}}}} \widetilde{f}_{{}_{(12)}}(\sigma).\label{f12scott}
\end{eqnarray}
We can prove property (\ref{f12lawson}) below, using properties (\ref{infcontinuity}) and (\ref{defchumorphismsigma}), along the same line of proof : 
\begin{eqnarray}
\forall { \mathfrak{Q}}\subseteq { \mathfrak{S}}^{{}^{(O_1)}},\;\; \widetilde{f}_{{}_{(12)}}(\bigsqcap{}_{{}_{ \mathfrak{S}^{{}^{(O_1)}}}}{ \mathfrak{Q}}) &=& \bigsqcap{}_{{}_{\sigma\in { \mathfrak{Q}}}} \widetilde{f}_{{}_{(12)}}(\sigma).\label{f12lawson}
\end{eqnarray}
\end{proof}

\begin{theoreme}{\bf [Preservation of the class of minimally disturbing measurements by symmetry]} \label{symmetrypreserveminimal}
Let $(\widetilde{f}_{{}_{(12)}},f^{{}^{(21)}})$ be a dictionary relating by symmetry $({ \mathfrak{S}}^{{}^{(O_1)}},{ \mathfrak{T}}^{{}^{(O_1)}},\widetilde{ \mathfrak{e}}^{{}^{(O_1)}})$ to $({ \mathfrak{S}}^{{}^{(O_2)}},{ \mathfrak{T}}^{{}^{(O_2)}},\widetilde{ \mathfrak{e}}^{{}^{(O_2)}})$, and let ${ \mathfrak{t}}\in { \mathfrak{T}}^{{}^{(O_2)}}_{{}_{IFKM}}$ be a yes/no test leading to an ideal first-kind measurement, {{}}{therefore} $f^{{}^{(21)}}({ \mathfrak{t}})$ is a yes/no test {{}}{that leads} to an ideal first-kind {{}}{measurement}, i.e., $f^{{}^{(21)}}({ \mathfrak{t}})\in { \mathfrak{T}}^{{}^{(O_1)}}_{{}_{IFKM}}$. In other words,
\begin{eqnarray}
&&\forall (\widetilde{f}_{{}_{(12)}},f^{{}^{(21)}})\in  { \bf {Sym}}\left[ ({ \mathfrak{S}}^{{}^{(O_1)}},{ \mathfrak{T}}^{{}^{(O_1)}},\widetilde{ \mathfrak{e}}^{{}^{(O_1)}})\rightarrow ({ \mathfrak{S}}^{{}^{(O_2)}},{ \mathfrak{T}}^{{}^{(O_2)}},\widetilde{ \mathfrak{e}}^{{}^{(O_2)}})\right],\;\;\;\; { \mathfrak{t}}\in { \mathfrak{T}}^{{}^{(O_2)}}_{{}_{IFKM}} \;\;\;\Rightarrow \;\;\; f^{{}^{(21)}}({ \mathfrak{t}})\in { \mathfrak{T}}^{{}^{(O_1)}}_{{}_{IFKM}}.\;\;\;\;\;\;\;\;\;\;\;\;\\
\textit{\rm i.e.}&&(\widetilde{f}_{{}_{(12)}},f^{{}^{(21)}})\in  { \bf {Sym}}\left[ ({ \mathfrak{S}}^{{}^{(O_1)}},{ \mathfrak{T}}_{{}_{min}}^{{}^{(O_1)}},\widetilde{ \mathfrak{e}}^{{}^{(O_1)}})\rightarrow ({ \mathfrak{S}}^{{}^{(O_2)}},{ \mathfrak{T}}_{{}_{min}}^{{}^{(O_2)}},\widetilde{ \mathfrak{e}}^{{}^{(O_2)}})\right].\;\;\;\;\;\;\;\;\;\;\;\;
\end{eqnarray}
\end{theoreme}
\begin{proof} 
Let us {{}}{first} prove that the symmetry preserves the Scott-continuity property of measurement maps.  
\begin{eqnarray}
\forall { \mathfrak{C}}\subseteq_{Chain} { \mathfrak{Q}}^{{}^{(O_1)}}_{f^{{}^{(21)}}({ \mathfrak{t}})},\;\; \widetilde{f}_{{}_{(12)}} ( (\bigsqcup{}_{{}_{ \mathfrak{S}^{{}^{(O_1)}}}}{ \mathfrak{C}}) {\centerdot} f^{{}^{(21)}}({ \mathfrak{t}})) &=& 
  \widetilde{f}_{{}_{(12)}}(\bigsqcup{}_{{}_{ \mathfrak{S}^{{}^{(O_1)}}}}{ \mathfrak{C}}) {\centerdot} { \mathfrak{t}}\;\;\;\;\;\;\;\;\;\;\textit{\rm from eq.  (\ref{defchucenterdot1sigma})}\nonumber\\
  &=&( \bigsqcup{}_{{}_{\sigma\in { \mathfrak{C}}}} \widetilde{f}_{{}_{(12)}}(\sigma)){\centerdot} { \mathfrak{t}}\;\;\;\;\;\;\;\;\;\;\textit{\rm from eq. (\ref{f12scott})}\nonumber\\
  &=& \bigsqcup{}_{{}_{\sigma\in { \mathfrak{C}}}} (\widetilde{f}_{{}_{(12)}}(\sigma){\centerdot} { \mathfrak{t}})\;\;\;\;\;\;\;\;\;\;\textit{\rm from eq. (\ref{Axiommeasurementmap3eq})}\nonumber\\
  &=& \bigsqcup{}_{{}_{\sigma\in { \mathfrak{C}}}} \widetilde{f}_{{}_{(12)}} ( \sigma {\centerdot} f^{{}^{(21)}}({ \mathfrak{t}})) \;\;\;\;\;\;\;\;\;\;\textit{\rm from eq.  (\ref{defchucenterdot1sigma})}\nonumber\\
  &=&\widetilde{f}_{{}_{(12)}} (  \bigsqcup{}_{{}_{\sigma\in { \mathfrak{C}}}} ( \sigma {\centerdot} f^{{}^{(21)}}({ \mathfrak{t}}))) \;\;\;\;\;\;\;\;\;\;\textit{\rm from eq.  (\ref{f12scott})}\;\;\;\;\;\;\;\;\;\;\;\;\;\;\;\;\;\;
\end{eqnarray}
We now use the injectivity property (\ref{defchuinjectivitysigma}) to {{}}{confirm} the preservation of Scott{{}}{ }continuity of measurement{{}}{ }maps by symmetries: 
\begin{eqnarray}
\forall { \mathfrak{C}}\subseteq_{Chain} { \mathfrak{Q}}^{{}^{(O_1)}}_{f^{{}^{(21)}}({ \mathfrak{t}})},\;\;  (\bigsqcup{}_{{}_{ \mathfrak{S}^{{}^{(O_1)}}}}{ \mathfrak{C}}) {\centerdot} f^{{}^{(21)}}({ \mathfrak{t}}) &=&   \bigsqcup{}_{{}_{\sigma\in { \mathfrak{C}}}} ( \sigma {\centerdot} f^{{}^{(21)}}({ \mathfrak{t}}))
\end{eqnarray}
The second continuity{{}}{ }property is proved along the same line of proof, using properties (\ref{defchucenterdot1sigma}), (\ref{f12lawson}), (\ref{filterdott}), and (\ref{defchuinjectivitysigma})
\begin{eqnarray} \forall { \mathfrak{F}}\subseteq_{Fil} { \mathfrak{Q}}^{{}^{(O_1)}}_{f^{{}^{(21)}}({ \mathfrak{t}})},\;\;  (\bigsqcap{}^\veebar_{{}_{ \mathfrak{S}^{{}^{(O_1)}}}}{ \mathfrak{F}}) {\centerdot} f^{{}^{(21)}}({ \mathfrak{t}}) &=&   \bigsqcap{}^\veebar_{{}_{\sigma\in { \mathfrak{F}}}} ( \sigma {\centerdot} f^{{}^{(21)}}({ \mathfrak{t}})).\end{eqnarray}
Secondly, let us prove that $f^{{}^{(21)}}({ \mathfrak{t}})\in { \mathfrak{T}}^{{}^{(O_1)}}_{{}_{FKM}}$.  Let us consider any $\sigma\in { \mathfrak{S}}^{{}^{(O_1)}},$ and ${ \mathfrak{t}}\in { \mathfrak{T}}^{{}^{(O_2)}}_{{}_{FKM}}$. \\The preservation of {{}}{ }equation (\ref{firstkindstates} (i)) is proved as follows
\begin{eqnarray}
\widetilde{ \mathfrak{e}}^{{}^{(O_1)}}_{f^{{}^{(21)}}({ \mathfrak{t}})}(\sigma)\leq {\rm Y}&\Rightarrow  &\widetilde{ \mathfrak{e}}^{{}^{(O_1)}}_{{ \mathfrak{t}}}(\widetilde{f}_{{}_{(12)}}(\sigma))\leq {\rm Y}\;\;\;\;\;\;\;\;\;\;\textit{\rm from eq. (\ref{defchumorphismsigma})}\nonumber\\
&\Rightarrow & \widetilde{ \mathfrak{e}}^{{}^{(O_1)}}_{{ \mathfrak{t}}}(\widetilde{f}_{{}_{(12)}}(\sigma) {\centerdot} { \mathfrak{t}})= {\rm Y}\;\;\;\;\;\;\;\;\;\;\textit{\rm from ${ \mathfrak{t}}\in { \mathfrak{T}}^{{}^{(O_2)}}_{{}_{FKM}}$}\nonumber\\
&\Rightarrow & \widetilde{ \mathfrak{e}}^{{}^{(O_1)}}_{{ \mathfrak{t}}}(\widetilde{f}_{{}_{(12)}}(\sigma {\centerdot} f^{{}^{(21)}}({ \mathfrak{t}})))= {\rm Y}\;\;\;\;\;\;\;\;\;\;\textit{\rm from eq. (\ref{defchucenterdot1sigma})}\nonumber\\
&\Rightarrow & \widetilde{ \mathfrak{e}}^{{}^{(O_1)}}_{f^{{}^{(21)}}({ \mathfrak{t}})}(\sigma {\centerdot} f^{{}^{(21)}}({ \mathfrak{t}}))= {\rm Y}\;\;\;\;\;\;\;\;\;\;\textit{\rm from eq. (\ref{defchumorphismsigma})}\;\;\;\;\;\;\;\;\;\;\;\;\;\;\;\;\;\;\;\;
\end{eqnarray}
The preservation of {{}}{ }equation (\ref{firstkindstates} (ii)) is proved as follows
\begin{eqnarray}
\widetilde{ \mathfrak{e}}^{{}^{(O_1)}}_{f^{{}^{(21)}}({ \mathfrak{t}})}(\sigma)= {\rm Y}&\Rightarrow  &\widetilde{ \mathfrak{e}}^{{}^{(O_1)}}_{{ \mathfrak{t}}}(\widetilde{f}_{{}_{(12)}}(\sigma))= {\rm Y}\;\;\;\;\;\;\;\;\;\;\textit{\rm from eq. (\ref{defchumorphismsigma})}\nonumber\\
&\Rightarrow & 
\widetilde{f}_{{}_{(12)}}(\sigma){\centerdot} { \mathfrak{t}} = \widetilde{f}_{{}_{(12)}}(\sigma)
\;\;\;\;\;\;\;\;\;\;\textit{\rm from ${ \mathfrak{t}}\in { \mathfrak{T}}^{{}^{(O_2)}}_{{}_{FKM}}$}\nonumber\\
&\Rightarrow & \widetilde{f}_{{}_{(12)}}(\sigma {\centerdot} {f}^{{}^{(21)}}({ \mathfrak{t}})) = \widetilde{f}_{{}_{(12)}}(\sigma)\;\;\;\;\;\;\;\;\;\;\textit{\rm from eq. (\ref{defchucenterdot1sigma})}\nonumber\\
&\Rightarrow &\sigma {\centerdot} {f}^{{}^{(21)}}({ \mathfrak{t}}) = \sigma  \;\;\;\;\;\;\;\;\;\;\textit{\rm from eq. (\ref{defchuinjectivitysigma})}\;\;\;\;\;\;\;\;\;\;\;\;\;\;\;\;\;\;\;\;
\end{eqnarray}
Thirdly, it remains to {{}}{be shown} that $f^{{}^{(21)}}({ \mathfrak{t}})\in { \mathfrak{T}}^{{}^{(O_1)}}_{{}_{Ideal}}$. Let us consider ${ \mathfrak{u}}\in { \mathfrak{T}}^{{}^{(O_1)}}$ such that $\overbrace{\lfloor f^{{}^{(21)}}({ \mathfrak{t}})\rfloor \lfloor { \mathfrak{u}}\rfloor}$.  The surjectivity of $f^{{}^{(21)}}$ (equation (\ref{defchusurjectivitysigma})) implies that {{}}{there} exists ${ \mathfrak{t}}'\in { \mathfrak{T}}^{{}^{(O_2)}}$ such that ${ \mathfrak{u}}=f^{{}^{(21)}}({ \mathfrak{t}}')$. \\
The compatibility relation $\overbrace{\lfloor f^{{}^{(21)}}({ \mathfrak{t}})\rfloor \lfloor f^{{}^{(21)}}({ \mathfrak{t}}')\rfloor}$ implies the compatibility relation $\overbrace{\lfloor { \mathfrak{t}}\rfloor \lfloor { \mathfrak{t}}'\rfloor}$. Indeed
\begin{eqnarray}
\overbrace{\lfloor f^{{}^{(21)}}({ \mathfrak{t}})\rfloor \lfloor f^{{}^{(21)}}({ \mathfrak{t}}')\rfloor} & \Leftrightarrow & \exists \sigma \in {\mathfrak{S}}\;\vert\; \widetilde{ \mathfrak{e}}^{{}^{(O_1)}}_{f^{{}^{(21)}}({ \mathfrak{t}})}(\sigma)=\widetilde{ \mathfrak{e}}^{{}^{(O_1)}}_{f^{{}^{(21)}}({ \mathfrak{t}}')}(\sigma)= {\rm Y}\nonumber\\
& \Rightarrow & \exists \sigma' = \widetilde{f}_{{}_{(12)}}(\sigma) \in {\mathfrak{S}}\;\vert\; \widetilde{ \mathfrak{e}}^{{}^{(O_1)}}_{{ \mathfrak{t}}}(\sigma')=\widetilde{ \mathfrak{e}}^{{}^{(O_1)}}_{{ \mathfrak{t}}'}(\sigma')= {\rm Y}\;\;\;\;\;\;\;\;\;\;\textit{\rm from eq. (\ref{defchumorphismsigma})}\nonumber\\
& \Rightarrow & \overbrace{\lfloor { \mathfrak{t}}\rfloor \lfloor { \mathfrak{t}}'\rfloor}.
\end{eqnarray}
Let us now consider any $\sigma\in { \mathfrak{S}}$ and ${ \mathfrak{t}}'\in { \mathfrak{T}}^{{}^{(O_2)}}$ such that $\overbrace{\lfloor f^{{}^{(21)}}({ \mathfrak{t}})\rfloor \lfloor f^{{}^{(21)}}({ \mathfrak{t}}')\rfloor}$, we then have
\begin{eqnarray}
(\, \widetilde{ \mathfrak{e}}^{{}^{(O_1)}}_{f^{{}^{(21)}}({ \mathfrak{t}})}(\sigma)\leq {\rm Y} \;\;\;\textit{\rm and}\;\;\;  \widetilde{ \mathfrak{e}}^{{}^{(O_1)}}_{f^{{}^{(21)}}({ \mathfrak{t}}')}(\sigma)= {\rm Y}\,)& \Rightarrow & (\, \widetilde{ \mathfrak{e}}^{{}^{(O_1)}}_{{ \mathfrak{t}}}(\widetilde{f}_{{}_{(12)}}(\sigma))\leq {\rm Y} \;\;\;\textit{\rm and}\;\;\;  \widetilde{ \mathfrak{e}}^{{}^{(O_1)}}_{{ \mathfrak{t}}'}(\widetilde{f}_{{}_{(12)}}(\sigma))= {\rm Y}\,)\;\;\;\;\;\;\;\;\;\;\textit{\rm from eq. (\ref{defchumorphismsigma})}\nonumber\\
& \Rightarrow & (\,  \widetilde{ \mathfrak{e}}^{{}^{(O_1)}}_{{ \mathfrak{t}}'}(\widetilde{f}_{{}_{(12)}}(\sigma) {\centerdot} { \mathfrak{t}})= {\rm Y}\,)\;\;\;\;\;\;\;\;\;\;\textit{\rm from ${ \mathfrak{t}}\in { \mathfrak{T}}^{{}^{(O_2)}}_{{}_{Ideal}}$ and $\widehat{{ \mathfrak{t}}{ \mathfrak{t}}'}$}\nonumber\\
& \Rightarrow & (\,  \widetilde{ \mathfrak{e}}^{{}^{(O_1)}}_{{ \mathfrak{t}}'}(\widetilde{f}_{{}_{(12)}}(\sigma {\centerdot} f^{{}^{(21)}}({ \mathfrak{t}})))= {\rm Y}\,)\;\;\;\;\;\;\;\;\;\;\textit{\rm from eq.  (\ref{defchucenterdot1sigma})}\nonumber\\
& \Rightarrow & (\,  \widetilde{ \mathfrak{e}}^{{}^{(O_1)}}_{f^{{}^{(21)}}({ \mathfrak{t}}')}(\sigma {\centerdot} f^{{}^{(21)}}({ \mathfrak{t}}))= {\rm Y}\,)\;\;\;\;\;\;\;\;\;\;\textit{\rm from eq.  (\ref{defchumorphismsigma})}
\end{eqnarray}
%As a final result, we have proved $f^{{}^{(21)}}({ \mathfrak{t}})\in { \mathfrak{T}}^{{}^{(O_1)}}_{{}_{IFKM}}$.
\end{proof}
\begin{lemme}\label{f12surjective}
$\widetilde{f}_{{}_{(12)}}$ is surjective.
\end{lemme}
\begin{proof}
Let us introduce the following map on ${ \mathfrak{S}}$:
\begin{eqnarray}
\begin{array}{rcrcl}
{f}^{\downarrow}_{{}_{(21)}} & :& { \mathfrak{S}}^{{}^{(O_2)}} & \longrightarrow & { \mathfrak{S}}^{{}^{(O_1)}}\\
& & \Sigma & \mapsto & {f}^{\downarrow}_{{}_{(21)}}(\Sigma):=\bigsqcap{}_{{}_{{ \mathfrak{S}}^{{}^{(O_1)}}}} \{\, \sigma \;\vert\; \widetilde{ \mathfrak{e}}^{{}^{(O_1)}}_{f^{{}^{(21)}}({ \mathfrak{t}}_{{}_{(\Sigma,\Sigma^\star)}})}(\sigma)=\textit{\bf Y}\,\},
\end{array}
\label{definf21downarrow}
\end{eqnarray}
where ${ \mathfrak{t}}_{{}_{(\Sigma,\Sigma^\star)}}$ designates the unique discriminating yes/no test in ${ \mathfrak{U}}$ with $\Sigma_{\;{ \mathfrak{t}}_{{}_{(\Sigma,\Sigma^\star)}}}=\Sigma$.\\
We have 
\begin{eqnarray}
\forall \Sigma\in { \mathfrak{S}}^{{}^{(O_2)}}, \Sigma'\in { \mathfrak{S}}^{{}^{(O_1)}},\;\;\;\;
{f}^{\downarrow}_{{}_{(21)}}(\Sigma) \sqsubseteq_{{}_{{ \mathfrak{S}}^{{}^{(O_1)}}}} \Sigma' & \Leftrightarrow & \widetilde{ \mathfrak{e}}^{{}^{(O_1)}}_{f^{{}^{(21)}}({ \mathfrak{t}}_{{}_{(\Sigma,\Sigma^\star)}})}   (\Sigma')=\textit{\bf Y}\nonumber\\
& \Leftrightarrow & \widetilde{ \mathfrak{e}}^{{}^{(O_1)}}_{{ \mathfrak{t}}_{{}_{(\Sigma,\Sigma^\star)}}}(\widetilde{f}_{{}_{(12)}}(\Sigma'))=\textit{\bf Y}\nonumber\\
& \Leftrightarrow &  \Sigma \sqsubseteq_{{}_{{ \mathfrak{S}}^{{}^{(O_2)}}}} \widetilde{f}_{{}_{(12)}}(\Sigma') 
\end{eqnarray}
$\widetilde{f}_{{}_{(12)}}$ is then the right Galois adjunct of ${f}^{\downarrow}_{{}_{(21)}}$.  Then, $\widetilde{f}_{{}_{(12)}}$ is surjective and preserves infima (this last property was already proved as a part of Lemma \ref{lemmaflawson}).
\end{proof}

\begin{lemme}{}\label{chutbar}
For any $(\widetilde{f}_{{}_{(12)}},f^{{}^{(21)}})\in  { \bf {Sym}}\left[ ({ \mathfrak{S}}^{{}^{(O_1)}},{ \mathfrak{T}}^{{}^{(O_1)}},\widetilde{ \mathfrak{e}}^{{}^{(O_1)}})\rightarrow ({ \mathfrak{S}}^{{}^{(O_2)}},{ \mathfrak{T}}^{{}^{(O_2)}},\widetilde{ \mathfrak{e}}^{{}^{(O_2)}})\right]$ and ${ \mathfrak{t}}\in { \mathfrak{T}}^{{}^{(O_2)}}$
\begin{eqnarray}
&& f^{{}^{(21)}}(\overline{\, { \mathfrak{t}} \,})=\overline{ f^{{}^{(21)}}({ \mathfrak{t}}) }.\;\;\;\;\;\;\;\;\;\;\;\;\;\;\;\;\label{f21tbar}
\end{eqnarray}
As a consequence,
\begin{eqnarray}
&& \widetilde{f}_{{}_{(12)}}(\Sigma^\star)= (\widetilde{f}_{{}_{(12)}}(\Sigma))^\star.\;\;\;\;\;\;\;\;\;\;\;\;\;\;\;\;\label{f12sigmastar}%%%%SSSS
\end{eqnarray}
\end{lemme}
\begin{proof}
Using successively two times properties (\ref{defchumorphismsigma}) and (\ref{etbar}), we obtain : 
$\forall \sigma\in { \mathfrak{S}}^{{}^{(O_1)}},\forall { \mathfrak{t}}\in { \mathfrak{T}}^{{}^{(O_2)}},\;\; \widetilde{ \mathfrak{e}}^{{}^{(O_1)}}_{f^{{}^{(21)}}(\overline{\, { \mathfrak{t}} \,})}(\sigma) = \widetilde{ \mathfrak{e}}^{{}^{(O_2)}}_{\overline{\, { \mathfrak{t}} \,}}(\widetilde{f}_{{}_{(12)}}(\sigma)) =\overline{\;  \widetilde{ \mathfrak{e}}^{{}^{(O_2)}}_{ { \mathfrak{t}} }(\widetilde{f}_{{}_{(12)}}(\sigma))} = \overline{\,\widetilde{ \mathfrak{e}}^{{}^{(O_1)}}_{f^{{}^{(21)}}( { \mathfrak{t}} )}(\sigma)\,}= \widetilde{ \mathfrak{e}}^{{}^{(O_1)}}_{\;\;\overline{ f^{{}^{(21)}}({ \mathfrak{t}} )}}(\sigma)$.\\
Property (\ref{f12sigmastar}) is a direct consequence of (\ref{esigmalbaresigmastarl}) and (\ref{f21tbar}) :
$\forall \sigma\in { \mathfrak{S}}^{{}^{(O_1)}},\forall { \mathfrak{t}}\in { \mathfrak{T}}^{{}^{(O_2)}},\;\; \widetilde{ \mathfrak{e}}^{{}^{(O_1)}}_{{ \mathfrak{t}}}(\widetilde{f}_{{}_{(12)}}(\sigma^\star)) =
\widetilde{ \mathfrak{e}}^{{}^{(O_1)}}_{f^{{}^{(21)}}({ \mathfrak{t}})}(\sigma^\star) =\widetilde{ \mathfrak{e}}^{{}^{(O_1)}}_{\overline{f^{{}^{(21)}}({ \mathfrak{t}})}}(\sigma) =\widetilde{ \mathfrak{e}}^{{}^{(O_1)}}_{f^{{}^{(21)}}(\overline{\,{ \mathfrak{t}}\,})}(\sigma) = 
\widetilde{ \mathfrak{e}}^{{}^{(O_1)}}_{\overline{ \mathfrak{t}}}(\widetilde{f}_{{}_{(12)}}(\sigma)) =
\widetilde{ \mathfrak{e}}^{{}^{(O_1)}}_{{ \mathfrak{t}}}((\widetilde{f}_{{}_{(12)}}(\sigma))^\star)$.
\end{proof}

Before studying the preservation of orthogonality, we will firstly note some elementary results. First of all, we have
\begin{eqnarray}
\forall { \mathfrak{t}}_{{}_{(\Sigma,\Sigma^\star)}} \;\in \widetilde{ \mathfrak{T}}^{{}^{(O_2)}}_{{}_{disc}},&& f^{{}^{(21)}}({ \mathfrak{t}}_{{}_{(\Sigma,\Sigma^\star)}})\;\in \widetilde{ \mathfrak{T}}^{{}^{(O_1)}}_{{}_{disc}}.
\end{eqnarray}
Indeed, let us introduce $\kappa,\kappa'\in { \mathfrak{S}}^{{}^{(O_1)}}$ as follows
\begin{eqnarray}  
\kappa :=\bigsqcap{}_{{}_{ \mathfrak{S}}}\{\,\sigma\;\vert\; \widetilde{ \mathfrak{e}}^{{}^{(O_1)}}_{f^{{}^{(21)}}({ \mathfrak{t}}_{{}_{(\Sigma,\Sigma^\star)}})}(\sigma)=\textit{\bf Y}\;\}
\;\;\;\textit{\rm and}\;\;\;
\kappa' :=\bigsqcap{}_{{}_{ \mathfrak{S}}}\{\,\sigma\;\vert\; \widetilde{ \mathfrak{e}}^{{}^{(O_1)}}_{f^{{}^{(21)}}({ \mathfrak{t}}_{{}_{(\Sigma,\Sigma^\star)}})}(\sigma)=\textit{\bf N}\;\}
\end{eqnarray}
Using the surjectivity of $\widetilde{f}_{{}_{(12)}}$ (Lemma \ref{f12surjective}) and the fact that $\widetilde{f}_{{}_{(12)}}$ preserves infima (Lemma \ref{lemmaflawson}), we obtain
\begin{eqnarray}
\widetilde{f}_{{}_{(12)}}(\kappa)=\Sigma & \textit{\rm and} & \widetilde{f}_{{}_{(12)}}(\kappa')=\Sigma^\star.
\end{eqnarray}
${ \mathfrak{t}}_{{}_{(\Sigma,\Sigma^\star)}}\in  \widetilde{ \mathfrak{T}}^{{}^{(O_2)}}_{{}_{disc}}$ means that $(\Sigma \bowtie_{{}_{ \mathfrak{S}}} \Sigma^\star)$ and $\antiwidehat{\Sigma \Sigma'}{}^{{}^{ \mathfrak{S}}}$. Using the bijectivity of $\widetilde{f}_{{}_{(12)}}$ (Lemma \ref{f12surjective} and property (\ref{defchuinjectivitysigma})) and its order-preserving property (Lemma \ref{lemmaflawson}), with the explicit definition of the relations $\cdot \bowtie_{{}_{ \mathfrak{S}}}\!\!\!\!\cdot$ and $\antiwidehat{\cdot \;\;\cdot}{}^{{}^{ \mathfrak{S}}}$, we deduce that $\kappa \bowtie_{{}_{ \mathfrak{S}}}\!\!\!\!\kappa'$ and $\antiwidehat{\kappa \kappa'}{}^{{}^{ \mathfrak{S}}}$. As a result, $f^{{}^{(21)}}({ \mathfrak{t}}_{{}_{(\Sigma,\Sigma^\star)}})\;\in \widetilde{ \mathfrak{T}}^{{}^{(O_1)}}_{{}_{disc}}$. Moreover,
\begin{eqnarray} 
\kappa=\widetilde{f}^{-1}_{{}_{(12)}}(\Sigma) & \textit{\rm and} & \kappa'=\widetilde{f}^{-1}_{{}_{(12)}}(\Sigma^\star)=\kappa^\star.
\end{eqnarray}
As a result, we have
\begin{eqnarray}
f^{{}^{(21)}}({\mathfrak{t}}_{{}_{(\Sigma,\Sigma^\star)}} )={ \mathfrak{t}}_{{}_{\left(\widetilde{f}^{-1}_{{}_{(12)}}(\Sigma),\widetilde{f}^{-1}_{{}_{(12)}}(\Sigma^\star)\right)}}.
\end{eqnarray}

In the following, we will impose the following property of preservation of schemes :
\begin{eqnarray}
f^{{}^{(21)}}({ \mathfrak{U}}^{{}^{(O_2)}})&=&{ \mathfrak{U}}^{{}^{(O_1)}}.\label{theoremfpreservdisc}
\end{eqnarray}
We now check the following fundamental property

\begin{theoreme}{\bf [Preservation of the orthogonality of states]}\label{theorempreservesortho}
\begin{eqnarray}
\hspace{-2cm}\forall (\widetilde{f}_{{}_{(12)}},f^{{}^{(21)}})\in  { \bf {Sym}}\left[ ({ \mathfrak{S}}^{{}^{(O_1)}},{ \mathfrak{T}}^{{}^{(O_1)}},\widetilde{ \mathfrak{e}}^{{}^{(O_1)}})\rightarrow ({ \mathfrak{S}}^{{}^{(O_2)}},{ \mathfrak{T}}^{{}^{(O_2)}},\widetilde{ \mathfrak{e}}^{{}^{(O_2)}})\right],\forall \sigma_1,\sigma_2 \in { \mathfrak{S}}^{{}^{(O_1)}},\;\;\;\;
\sigma_1 \perp \sigma_2 \Rightarrow  \widetilde{f}_{{}_{(12)}}(\sigma_1) \perp \widetilde{f}_{{}_{(12)}}(\sigma_2).&&\;\;\;\;\;\;
\end{eqnarray}
%As an immediate consequence, we have $\forall \sigma \in { \mathfrak{S}}^{{}^{(O_1)}},\;\;\;\;\widetilde{f}_{{}_{(12)}}(\sigma^\star)=\widetilde{f}_{{}_{(12)}}(\sigma)^\star. $
\end{theoreme}
\begin{proof}
\begin{eqnarray}
\sigma_1 \perp \sigma_2 & \Leftrightarrow & \exists { \mathfrak{t}}\in { \mathfrak{U}}^{{}^{(O_1)}}\;\vert\; ( \widetilde{ \mathfrak{e}}_{ \mathfrak{t}}(\sigma_1)=\textit{\bf Y}  \;\;\;\;\textit{\rm and}\;\;\;\; \widetilde{ \mathfrak{e}}_{ \overline{\,{ \mathfrak{t}}\,}}(\sigma_2)=\textit{\bf Y})\nonumber\\
& \Leftrightarrow &
\exists { \mathfrak{u}}\in { \mathfrak{U}}^{{}^{(O_2)}}\;\vert\; 
( \widetilde{ \mathfrak{e}}_{f^{{}^{(21)}}({ \mathfrak{u}})}(\sigma_1)=\textit{\bf Y}  \;\;\;\;\textit{\rm and}\;\;\;\; \widetilde{ \mathfrak{e}}_{ \overline{\,f^{{}^{(21)}}({ \mathfrak{u}})\,}}(\sigma_2)=\textit{\bf Y})
\;\;\;\;\;\;\;\;\;\;\textit{\rm from eq. (\ref{defchusurjectivitysigma}) and property  (\ref{theoremfpreservdisc})}\nonumber\\
& \Leftrightarrow &
\exists { \mathfrak{u}}\in { \mathfrak{U}}^{{}^{(O_2)}}\;\vert\; 
( \widetilde{ \mathfrak{e}}_{{ \mathfrak{u}}}(f_{{}_{(12)}}(\sigma_1))=\textit{\bf Y}  \;\;\;\;\textit{\rm and}\;\;\;\; \widetilde{ \mathfrak{e}}_{ \overline{\,{ \mathfrak{u}}\,}}(f_{{}_{(12)}}(\sigma_2))=\textit{\bf Y})
\;\;\;\;\;\;\;\;\;\;\textit{\rm from eq. (\ref{defchumorphismsigma}) and Lemma \ref{chutbar} }\nonumber\\
& \Leftrightarrow & \widetilde{f}_{{}_{(12)}}(\sigma_1) \perp \widetilde{f}_{{}_{(12)}}(\sigma_2).
\end{eqnarray}
\end{proof}

\begin{lemme}
\begin{eqnarray}
\begin{array}{rcrclcrcrcl}
\varphi_L & :& { \mathcal{C}}({ \mathfrak{S}}_{{}_{pure}}) & \longrightarrow & { \mathfrak{S}} &\;\;\;\;\;\;\; & 
\varphi_D & :& { \mathfrak{S}} & \longrightarrow & { \mathcal{C}}({ \mathfrak{S}}_{{}_{pure}}) \\
& & { \mathfrak{c}} & \mapsto & \bigsqcap_{{}_{ \mathfrak{S}}} { \mathfrak{c}} & & & & \sigma & \mapsto & \underline{\{\sigma\}}  
\end{array}
\end{eqnarray}
satisfy the following Galois adjunction relation
\begin{eqnarray}
\forall { \mathfrak{c}}\in { \mathcal{C}}({ \mathfrak{S}}_{{}_{pure}}),\forall \sigma\in { \mathfrak{S}},&& (\,\varphi_L ({ \mathfrak{c}})\sqsubseteq_{{}_{ \mathfrak{S}}}\sigma \,)\Leftrightarrow (\, { \mathfrak{c}} \subseteq \varphi_D (\sigma)\,),
\end{eqnarray}
and
$\varphi_L \circ \varphi_D = id_{{}_{ \mathfrak{S}}}$.
\end{lemme}

\begin{MyDef}
Using the elements of a symmetry $(\widetilde{f}_{{}_{(12)}},f^{{}^{(21)}})\in  { \bf {Sym}}\left[ ({ \mathfrak{S}}^{{}^{(O_1)}},{ \mathfrak{T}}^{{}^{(O_1)}},\widetilde{ \mathfrak{e}}^{{}^{(O_1)}})\rightarrow ({ \mathfrak{S}}^{{}^{(O_2)}},{ \mathfrak{T}}^{{}^{(O_2)}},\widetilde{ \mathfrak{e}}^{{}^{(O_2)}})\right]$, we define the following maps
\begin{eqnarray}
\begin{array}{rcrclcrcrcl}
\underline{\widetilde{f}_{{}_{(12)}}} & :& { \mathcal{C}}({ \mathfrak{S}}^{{}^{(O_1)}}_{{}_{pure}}) & \longrightarrow & { \mathcal{C}}({ \mathfrak{S}}^{{}^{(O_2)}}_{{}_{pure}}) &\;\;\; & 
\underline{f^{{}^{(21)}}} & :& { \mathcal{C}}({ \mathfrak{S}}^{{}^{(O_2)}}_{{}_{pure}}) & \longrightarrow & { \mathcal{C}}({ \mathfrak{S}}^{{}^{(O_1)}}_{{}_{pure}}) \\
& & { \mathfrak{c}} & \mapsto & \bigvee \{\, \widetilde{f}_{{}_{(12)}}(\sigma) \;\vert\; \sigma\in { \mathfrak{c}}\,\} & & & & { \mathfrak{c}} & \mapsto & (  \varphi_D^{{}^{(O_1)}} \circ {f}^{\downarrow}_{{}_{(21)}} \circ \varphi_L^{{}^{(O_2)}})({ \mathfrak{c}})
\end{array}
\end{eqnarray}
Note that ${f}^{\downarrow}_{{}_{(21)}}$ has been defined in equation (\ref{definf21downarrow}). 
\end{MyDef}

\begin{lemme}
$(\underline{\widetilde{f}_{{}_{(12)}}},\underline{f^{{}^{(21)}}})$ forms a Galois connection. 
\end{lemme}
\begin{proof}
For any ${ \mathfrak{c}}_1\in { \mathcal{C}}({ \mathfrak{S}}^{{}^{(O_1)}}_{{}_{pure}})$ and ${ \mathfrak{c}}_2\in { \mathcal{C}}({ \mathfrak{S}}^{{}^{(O_2)}}_{{}_{pure}})$, we have
\begin{eqnarray}
(\,\underline{\widetilde{f}_{{}_{(12)}}}({ \mathfrak{c}}_1)\subseteq { \mathfrak{c}}_2\,) %&\Leftrightarrow &(\,\forall \sigma\in { \mathfrak{c}}_1,\; \{\widetilde{f}_{{}_{(12)}}(\sigma)\}\subseteq { \mathfrak{c}}_2\,)\nonumber\\
&\Leftrightarrow &(\,\forall \sigma\in { \mathfrak{c}}_1,\;  \varphi_L^{{}^{(O_2)}}({ \mathfrak{c}}_2)  \sqsubseteq_{{}_{{ \mathfrak{S}}^{{}^{(O_2)}}}}  \widetilde{f}_{{}_{(12)}}(\sigma)   \,)\nonumber\\
&\Leftrightarrow &(\,\forall \sigma\in { \mathfrak{c}}_1,\;  \widetilde{ \mathfrak{e}}^{{}^{(O_2)}}_{{ \mathfrak{t}}_{(\varphi_L^{{}^{(O_2)}}({ \mathfrak{c}}_2),(\varphi_L^{{}^{(O_2)}}({ \mathfrak{c}}_2))^\star)}}(\widetilde{f}_{{}_{(12)}}(\sigma))={\bf Y}   \,)\nonumber\\
&\Leftrightarrow &(\,\forall \sigma\in { \mathfrak{c}}_1,\;  \widetilde{ \mathfrak{e}}^{{}^{(O_1)}}_{{f}^{{}^{(21)}}({ \mathfrak{t}}_{(\varphi_L^{{}^{(O_2)}}({ \mathfrak{c}}_2),(\varphi_L^{{}^{(O_2)}}({ \mathfrak{c}}_2))^\star)})}(\sigma)={\bf Y}   \,)\nonumber\\
&\Leftrightarrow &(\,\forall \sigma\in { \mathfrak{c}}_1,\; \{\sigma\}\subseteq 
\underline{
\left\{ \bigsqcap{}_{{}_{{ \mathfrak{S}}^{{}^{(O_1)}}}}
\{\sigma'\in { \mathfrak{S}}^{{}^{(O_1)}}_{{}_{pure}}\;\vert\;
\widetilde{ \mathfrak{e}}^{{}^{(O_1)}}_{{f}^{{}^{(21)}}({ \mathfrak{t}}_{(\varphi_L^{{}^{(O_2)}}({ \mathfrak{c}}_2),(\varphi_L^{{}^{(O_2)}}({ \mathfrak{c}}_2))^\star)})}(\sigma)={\bf Y}  \}
\right\}
} \,)\nonumber\\
 &\Leftrightarrow &(\,\forall \sigma\in { \mathfrak{c}}_1,\; \{\sigma\}\subseteq (\varphi_D^{{}^{(O_1)}}\circ {f}^{\downarrow}_{{}_{(21)}} \circ \varphi_L^{{}^{(O_2)}})({ \mathfrak{c}}_2)\,)\nonumber\\
 &\Leftrightarrow &(\, { \mathfrak{c}}_1\subseteq \underline{f^{{}^{(21)}}}({ \mathfrak{c}}_2)\,)
\end{eqnarray}
\end{proof}

\begin{theoreme}
$\underline{\widetilde{f}_{{}_{(12)}}}$ is an injective map from ${ \mathcal{C}}({ \mathfrak{S}}^{{}^{(O_1)}}_{{}_{pure}})$ to ${ \mathcal{C}}({ \mathfrak{S}}^{{}^{(O_2)}}_{{}_{pure}})$ preserving suprema, mapping atoms to atoms and preserving orthogonality (i.e. $\forall { \mathfrak{c}}\in { \mathcal{C}}({ \mathfrak{S}}^{{}^{(O_1)}}_{{}_{pure}}),\; \underline{\widetilde{f}_{{}_{(12)}}}({ \mathfrak{c}}^{\underline{\perp}}) = (\underline{\widetilde{f}_{{}_{(12)}}}({ \mathfrak{c}}))^{\underline{\perp}}$). \\
It is then an ortho-morphism of Hilbert lattice (see \cite[Definition 5.12 and Definition 5.17]{STUBBE2007477}).
\end{theoreme}
\begin{proof}
$\underline{\widetilde{f}_{{}_{(12)}}}$ preserves atoms as an extension of $\widetilde{f}_{{}_{(12)}}$ from ${ \mathfrak{S}}^{{}^{(O_1)}}_{{}_{pure}}$ to ${ \mathcal{C}}({ \mathfrak{S}}^{{}^{(O_1)}}_{{}_{pure}})$. It is injective and preserves suprema as a left Galois adjunct. It preserves the orthogonality relation according to Theorem \ref{theorempreservesortho}. 
\end{proof}

\newpage

\section{Conclusion}

We {{}}{aim} to develop a new axiomatic approach {{}}{to} quantum theory and this article {{}}{is designed} as a first decisive step for this axiomatic program.  A precise semantic description of the space of preparations and of the associated 'mixed states' of the system {{}}{was} formulated.  This semantic {{}}{formalism} is based on a Chu space construction involving the set of preparations, the set of yes/no tests and an evaluation map with a three-valued target space. The values taken by this map are associated {{}}{with} counterfactual statements of the observer for a given yes/no test and a given prepared sample. The three values are interpreted in a possibilistic perspective, i.e., as 'certainly yes', 'certainly {{}}{no}' and '{{}}{maybe}'.  This domain structure on the target space led to an 'informational' interpretation {{}}{of} the set of preparations. The space of preparations {{}}{was} equipped with a notion of 'mixtures', expressed in terms of the meet operation on this poset.  From natural requirements about the inductive definition of states, it appeared that this Inf semi-lattice was also a pointed directed-complete partial order. Then, an 'Information Principle' {{}}{was} introduced {{}}{in} the form of two topological requirements on the space of states. Although new in its form, this principle is very standard in different quantum axiomatic {{}}{programs}. This basic set of axioms {{}}{was shown} to be sufficient to constrain the structure of the space of states to be a 'projective domain'.  The space of pure states was then basically identified in terms of maximal elements of this domain.\\
Then, the relation between yes/no tests and states {{}}{was} studied {{}}{from two perspectives}: 
using the notion of Chu duality and using the notion of 'measurement'.  Adopting the first perspective,  the notion of 'properties' of the system {{}}{was} defined and a 'property-state' {{}}{was} identified for any property, as in Piron's construction.  The second perspective emphasizes the recursive aspect of preparation {{}}{processes}. 
In order to identify a {{}}{subclass} of measurement operations corresponding to minimally disturbing measurements{{}}{,} it appeared necessary to clarify the notion of 'compatibility between measurements'.  The compatibility between two measurements was defined in terms of the existence of preparations {{}}{that simultaneously exhibited} the two corresponding properties as actual. This notion was used to define 'ideal first-kind measurements' and to characterize them as 'minimally disturbing measurements'. We finally proved the existence of a 'bi-extensional $Chu_{3}$ duality' between the space of minimally disturbing yes/no tests and the space of states. \\
The simultaneous ideal first-kind measurements of compatible properties was then studied and 'Specker's principle' was proved.  Using this result, we obtained a 'coherent domain' structure on the space of 'descriptions' formalizing the families of compatible properties used to define a state of the system.  An orthogonality relation was then defined on the space of states in terms of the class of discriminating yes/no tests.  We achieve the characterization of the domain structure on the space of states by requiring the existence of a scheme of discriminating yes/no tests necessary to the construction of an orthogonality relation on the space of states.  In generic examples, this last constraint is equivalent to the existence of an ortho-complementation on the projective domain defining the space of states.\\
Using the properties of the domain structure established on the space of states, we deduced that the set of ortho-closed subsets of pure states, equipped with the induced orthogonality relation, inherits a structure of Hilbert lattice.  This is the first part of our reconstruction theorem. \\
In Section \ref{sectionsymmetries}, we explored the properties of Chu morphisms with respect to previous notions. As a central result, a sub-algebra of the algebra of Chu morphisms, corresponding to 'symmetries' of the system, was defined.  {{}}{These symmetries appear to preserve} the class of minimally disturbing yes/no tests and the orthogonality of states.  The link between these Chu symmetries and the morphisms of Hilbert lattice, defined on the space of ortho-closed subsets of pure states, is finally emphasized. This is the second part of our reconstruction theorem.\\
Despite its self-contained character and the fact that it achieves to produce a reconstruction theorem from very basic premises for quantum theory, this paper requires more developments to be plainly satisfactory. First of all, it appears necessary to study the ways to recover a probabilistic formalism from a possibilistic one. Secondly, it appears necessary to pursue the construction to the case of compound systems by studying the tensor products structures on the domains appearing in the present paper. Endly, it seems unavoidable to study extensively the category of 'projective domains' extracted from our set of axioms.  These problems will be attacked in our forthcoming papers.

\end{document}